\documentclass[12pt,a4paper]{article}
\usepackage{jheppub}
\usepackage[utf8]{inputenc}
\usepackage{amsfonts}
\usepackage{slashed}
\usepackage{xcolor}

\usepackage[english]{babel}
\usepackage{subcaption}

\usepackage{changes}

\numberwithin{equation}{section}

\newcommand{\be}{\begin{equation}}
\newcommand{\ee}{\end{equation}}

\newcommand{\diff}{\text{d}}

\newcommand{\ns}{n_\text{s}}
\newcommand{\nnn}{n_\text{n}}

\usepackage[mathscr]{euscript}

\def\df{\mathrm{d}}

\title{Hydrodynamics and instabilities of relativistic superfluids at finite superflow}

\author[a,b]{Daniel Are\'an,}
\emailAdd{daniel.arean@uam.es}
\affiliation[a]{Instituto de F\'isica Te\'orica UAM/CSIC, Calle Nicol\'as Cabrera 13-15, 28049 Madrid, Spain}
\affiliation[b]{Departamento de F\'isica Te\'orica, Universidad Aut{\'o}noma de Madrid, Campus de Cantoblanco, 28049 Madrid, Spain}

\author[c]{Blaise Gout\'eraux,}
\emailAdd{blaise.gouteraux@polytechnique.edu}
\affiliation[c]{CPHT, CNRS, Ecole Polytechnique, Institut Polytechnique de Paris, Route de Saclay, 91128 Palaiseau, France}

\author[d]{Eric Mefford,}
\emailAdd{ericmefford@uvic.ca}
\affiliation[d]{Department of Physics and Astronomy, University of Victoria, Victoria, BC V8W 3P6, Canada}

\author[c]{Filippo Sottovia}
\emailAdd{filippo.sottovia@polytechnique.edu}

\date{\today}

\abstract{
We study the linear response of relativistic superfluids with a non-zero superfluid velocity. For sufficiently large superflow, an instability develops via the crossing of a pole of the retarded Green's functions to the upper half complex frequency plane. We show that this is caused by a local thermodynamic instability, i.e. when an eigenvalue of the static susceptibility matrix (the second derivatives of the free energy) diverges and changes sign. The onset of the instability occurs when $\partial_{\zeta}(n_s\zeta)=0$, with $\zeta$ the norm of the superfluid velocity and $n_s$ the superfluid density. The Landau instability for non-relativistic superfluids such as Helium 4 also coincides with the non-relativistic version of this criterion. We then turn to gauge/gravity duality and show that this thermodynamic instability criterion applies equally well to strongly-coupled superfluids. In passing, we compute holographically a number of transport coefficients parametrizing deviations out-of-equilibrium in the hydrodynamic regime and demonstrate that the gapless quasinormal modes of the dual planar black hole match those predicted by superfluid hydrodynamics.
}

\begin{document}

\maketitle

\newpage

\section{Introduction}\label{sec:intro}

Superfluids are quantum phases of matter whose most remarkable experimental manifestation is frictionless flow, \cite{landaulif,landau1980course9,chlub}. They arise when a global, continuous U(1) symmetry is spontaneously broken: in other words, an operator charged under the symmetry acquires a nonzero vacuum expectation value. The phase of this complex order parameter is related to the appearance of a new gapless mode in the spectrum, the Goldstone boson $\varphi$, directly responsible for frictionless flow.

The Goldstone mode transforms nonlinearly under U(1) gauge transformations, and so gauge invariance demands that the theory only depends on gradients of the Goldstone $\nabla_\mu\varphi$. This makes this degree of freedom topological: its winding can only be relaxed by topological defects, the vortices.
A modern understanding of superfluidity is in terms of an anomalous emergent higher-form symmetry, reflecting the conservation of winding of the Goldstone field, \cite{Delacretaz:2019brr,Armas:2023tyx}. The mixed `t Hooft anomaly with the original U(1) current sources a nonzero overlap of the charge current with a conserved quantity (the higher-form current) and in turn causes the conductivity to be infinite in the zero frequency limit, for much the same reason it is infinite in a phase with unbroken spatial translations. This is the origin of frictionless flow.

A finite density of the higher-form current corresponds to a non-trivial background winding and a nonzero supercurrent. As it turns out, increasing the background winding eventually triggers an instability and superfluidity is destroyed. Landau gave a famous argument to determine the onset of the instability in Helium 4, \cite{landau1980course9}. Besides the superfluid component, at nonzero temperature the current is also transported by a normal component made up of thermally excited `phonons', in other words quasiparticles with bosonic statistics and a dispersion relation $\epsilon_k$ linear at small wavenumbers $k$ in the superfluid rest frame. Assuming that the system enjoys Galilean invariance and boosting back to the normal fluid rest frame, the phonon energy becomes $\epsilon'_k=\epsilon_k+{\bf k}\cdot {\bf v_s}$, where $ {\bf v_s}$ is the superfluid velocity transporting superflow. Superfluidity is lost when it becomes energetically favorable to excite phonons, i.e. when $\epsilon'_k<0$. The critical velocity at which $\epsilon_k'$ becomes zero depends on the angle between the vectors ${\bf v_s}$ and ${\bf k}$ and is lowest when they are antiparallel. This gives Landau's criterion for the critical velocity
\begin{equation} \label{landaucrit}
v_\text{s} \leq v_c \equiv\min_k \frac{\epsilon_k}{k} \,.
\end{equation}
In a previous work \cite{Gouteraux:2022kpo}, a subset of us showed that the instability at large superflow can be recast as a local thermodynamic instability, expressed simply as
\begin{equation}
\label{instabcrit}
\frac{\partial^2p}{\partial v_s^2}\to\infty\quad\Rightarrow\quad\partial_{v_s}(n_s v_s)=0
\end{equation}
where $p$ is the pressure (equal to minus the free energy density), $n_s$ the superfluid density and $v_s$ the norm of the superfluid velocity, as we will define more precisely in Section \ref{sec:superfluid_review}. Computing the pressure for a non-relativistic system of bosonic thermal excitations with dispersion relation $\epsilon_k$ allows one to reproduce Landau's criterion \eqref{landaucrit}. 

The advantage of \eqref{instabcrit} is that it does not assume any particular boost symmetry,\footnote{In the absence of boosts, the divergence of the same second derivative of the pressure continues to signal the instability, but its precise expression now involves the normal fluid velocity \cite{Gouteraux:2022kpo} and an extra thermodynamic coefficient, \cite{armas2023ideal}.} or that the ground state is weakly-coupled. It also gives a formulation independent of the microscopic details of the system under consideration.

Gauge/gravity duality \cite{Maldacena:1997re,Witten:1998qj} allows one to construct strongly-coupled relativistic superfluids, \cite{Gubser:2008wz,Hartnoll:2008vx,Hartnoll:2008kx}. Holographic superfluids at finite superflow were studied in \cite{Herzog:2008he,Herzog:2010vz,Sonner:2010yx,Arean:2010wu,Bhattacharya:2011eea,Bhattacharya:2011tra,Herzog:2011ec,Amado:2013aea,Lan:2020kwn}. \cite{Amado:2013aea,Lan:2020kwn} observed that for a critical value of the superfluid velocity, a pole of the retarded Green's function crosses to the upper half plane, leading to perturbations exponentially growing with time. Here, we show that the critical velocity when this happens is precisely predicted by \eqref{instabcrit}.

This paper is structured as follows. In Section \ref{sec:superfluid_review}, we review relativistic superfluid hydrodynamics including first-order derivative corrections (earlier work on relativistic superfluid hydrodynamics includes \cite{Khalatnikov1,Khalatnikov2,Khalatnikov3,Khalatnikov4,Israel1,Israel2,Bhattacharya:2011eea}). In Section \ref{sec:linear}, working in the limit when the wavenumber and the superfluid velocity are collinear for simplicity, we describe the spectrum of linear, collective excitations and demonstrate \eqref{instabcrit}. In Section \ref{sec:holography}, we construct a holographic superfluid phase at finite superflow, and check that it becomes dynamically unstable when \eqref{instabcrit} holds. In passing, we compute all the thermodynamics of the state as well as the collinear transport coefficients, as well as check that the hydrodynamic dispersion relation for the gapless modes matches the location of the gapless quasinormal modes of the dual planar black hole, as expected.\footnote{A similar check at zero superflow was carried out in \cite{Arean:2021tks}.} We then conclude with some outlook. In Appendix \ref{numericsappendix}, we give details on the numerical calculations, while in Appendix \ref{probeAppendix}, we analyze the so-called probe limit where temperature and normal velocity fluctuations are frozen.

\section{\label{sec:superfluid_review}Review of superfluid hydrodynamics}

We follow the derivation in \cite{Banerjee:2016qxf} (see also \cite{Haehl:2014zda,Haehl:2015pja}), although we will choose a different frame which ultimately proves more convenient for our purposes.

We consider a $(d+1)$-dimensional manifold equipped with a metric $g_{\mu\nu}$ and a U(1) gauge field $A_\mu$, and impose that the theory is invariant under diffeomorphisms $\chi^\mu$ and U(1) gauge transformations $\Lambda$. Under these, the background fields transform as 
\begin{equation}
\delta_{\mathcal{X}}g_{\mu\nu}\equiv\mathcal L_{\chi}g_{\mu \nu}=\nabla_\mu\chi_\nu+\nabla_\nu\chi_\mu\,,\quad \delta_{\mathcal X}A_\mu\equiv\mathcal L_\chi A_\mu+\partial_\mu\Lambda_\chi=A^\nu\partial_\mu\chi_\nu+\chi^\nu\partial_\nu A_\mu+\partial_\mu\Lambda_\chi\,,
\end{equation}
where the set of gauge parameters is $\mathcal X=\{\chi^\mu,\Lambda_\chi\}$ and $\mathcal L_\chi$ represents the Lie derivative along $\chi$.

In thermal equilibrium, we can use the existence of a time-like Killing vector field $\beta^\mu$ to define our notion of temperature and fluid velocity
\begin{equation}
T\equiv \frac{T_0}{\sqrt{-\beta^\mu\beta_\mu}}\,,\quad u^\mu=\frac{\beta^\mu}{\sqrt{-\beta^\mu\beta_\mu}}\,,
\end{equation}
which transform appropriately under diffeomorphisms as a scalar and a vector, respectively.
$T_0$ is a normalization constant which does not play any role, and with our definition, the fluid velocity is timelike and $u_\mu u^\mu=-1$. We can also define a gauge and diffeomorphism invariant chemical potential as
\begin{equation}
\frac\mu{T}=\beta^\mu A_\mu+\Lambda_{\mathcal B}\,.
\end{equation}
The set of gauge parameters $\mathcal X$ is of course arbitrary, but it is advantageous to choose them to be $\mathcal B=\{\beta^\mu,\Lambda_{\mathcal B}\}$. Then, the variations of the background fields under $\mathcal B$ are
\begin{equation}
\delta_{\mathcal{B}}g_{\mu\nu}\equiv\mathcal L_{\beta}g_{\mu \nu}=\nabla_\mu\beta_\nu+\nabla_\nu\beta_\mu\,,\quad \delta_{\mathcal B}A_\mu\equiv\mathcal L_\beta A_\mu+\partial_\mu\Lambda_\mathcal{B}=A^\nu\partial_\mu\beta_\nu+\beta^\nu\partial_\nu A_\mu+\partial_\mu\Lambda_\mathcal{B}\,,
\end{equation}
and vanish in thermal equilibrium.

In a superfluid phase, we also define the gauge-covariant superfluid velocity $\xi^\nu$, which is built out of the Goldstone field and the external gauge field:
\begin{equation}
\xi_{\nu} \equiv \partial_\nu\varphi - A_\nu\,,\qquad \zeta^\mu=\Delta^{\mu\nu}\xi_\nu
\end{equation}
together with $\zeta^\mu$, the component of the superfluid velocity normal to the fluid velocity and the associated projector $\Delta^{\mu\nu}=g^{\mu\nu}+u^\mu u^\nu$.

The stress tensor $T^{\mu\nu}$ and the current $J^\mu$ of the theory can be computed by varying the background fields $g_{\mu\nu}$ and $A_\mu$ and computing the corresponding variation of the Lagrangian density:
\begin{equation}
\label{variationgeneratingfunc}
\delta \left(\sqrt{-g}\mathcal L\right)=\sqrt{-g}\left[\frac12\delta g_{\mu\nu} T^{\mu\nu}+\delta A_\mu J^\mu+K\delta\varphi\right]
\end{equation}
Here we have added a variation of $\varphi$, as $\varphi$ transforms under gauge transformation as 
\begin{equation} 
\delta_{\mathcal B}\varphi=\beta^\mu\partial_\mu\varphi+\Lambda_{\mathcal B}\,.
\end{equation}
Thus in superfluid phases, there is another `current' $K$, which as we will see obtains a constitutive relation similarly to $T^{\mu\nu}$ and $J^\mu$. In thermal equilibrium, $\Lambda_{\mathcal B}=\mu/T-\beta^\mu A_\mu$, so that we recover the familiar Josephson relation for the Goldstone field
\begin{equation}
\left.\delta_{\mathcal B}\varphi=0\right|_{th.eq.}\quad\Rightarrow\quad u^\mu\xi_\mu+\mu=0\,.
\end{equation}

Requiring that the variation \eqref{variationgeneratingfunc} does not depend on gauge parameters yields the conservation equations of the theory in the presence of background sources
\begin{equation}
\label{conseq}
\nabla_\mu T^{\mu\nu}=F^{\nu\mu}J_\mu+K\xi^\nu\,,\qquad \nabla_\mu J^\mu=K\,.
\end{equation}

With these equations in place, we now turn to the constitutive relations. They can be determined by requiring that the divergence of the entropy current is non-negative:
\begin{equation}
\nabla_\mu S^\mu\geq0\,,\qquad S^\mu=s u^\mu+\tilde s^\mu\,,
\end{equation}
where we have decomposed the entropy current into an ideal part where the normal fluid transports the entropy density $s$, and a non-ideal part $\tilde s^\mu$. This decomposition is based on the physical assumption that the superfluid component does not transport entropy at ideal level, but turns out to be correct once we impose that the ideal terms in the constitutive relations produce no entropy.

The pressure depends on all possible diffeomorphism and gauge invariant scalars, $p(T,\mu,\zeta^2)$, where we have chosen to work with $\zeta^2$ rather than $\xi^2$, which allows us to work in a convenient frame once we include derivative corrections. The pressure does not depend explicitly on the fluid velocity, as a state with a nonzero background fluid velocity can always be generated by a Lorentz boost from the state where the normal fluid is at rest. However, the norm $\zeta\equiv\sqrt{\zeta^2}$ of the superfluid velocity relative to the normal fluid velocity is a physical parameter. The variation of the pressure is
\begin{equation} \label{firstlaw}
\diff p = s\, \diff T + n \, \diff \mu - \frac{n_\text{s}}{2\mu} \, \diff (\zeta^2) 
\end{equation}
where $s$ and $n$ correctly turn out to be the entropy and charge densities as expected, while $n_\text{s}$ will be interpreted as the superfluid density, the fraction of the charge carried by the superfluid component. $n$, $s$ and $n_\text{s}$ all depend on $T$, $\mu$,  $\zeta^2$.

To determine the constitutive relations, it is more convenient to compute
\begin{equation}
\Delta\equiv \nabla_\mu S^\mu+\beta_\nu\left(\nabla_\mu T^{\mu\nu}-F^{\nu\mu}J_\mu-K\xi^\nu\right)+\frac{\mu}{T}\left(\nabla_\mu J^\mu-K\right)\geq0
\end{equation}
which can be shown to be non-negative off-shell as well \cite{Banerjee:2016qxf}, i.e. without needing to impose the equations of motion \eqref{conseq}. After some algebra, and with a little bit of hindsight, the constitutive relations are found to be\footnote{Here we limit ourselves to the canonical part of the entropy current, as the non-canonical part plays no role in physical observables, and is only non-trivial in our case at second order in derivatives. See \cite{Bhattacharyya:2013lha,Bhattacharyya:2014bha} for details and examples of the non-canonical contribution to the entropy current for charged, relativistic fluids.}
\begin{align}
T^{\mu\nu}=&(\epsilon+p)u^\mu u^\nu+p g^{\mu\nu}+2n_s u^{(\mu}\zeta^{\nu)}+\frac{n_s}{\mu}\zeta^\mu\zeta^\nu+\tilde T^{\mu\nu}\,,\label{eq:Tmunu}\\
J^\mu=&nu^\mu+\frac{n_s}{\mu}\zeta^\nu+\tilde J^\mu\label{eq:Jmu}\\
K=&T\Delta^{\mu\nu}\nabla_\mu\left(\frac{n_s}{\mu T}\zeta_\nu\right)+\tilde K\,, \label{Kconstrel}\\
\tilde S^\mu=&-\beta_\nu \tilde T^{\mu\nu}-\frac{\mu}{T}\tilde J^\mu+\frac{n_s}{\mu}\zeta^\mu\delta_{\mathcal B}\varphi\label{JScanconstrel}
\end{align} 
while
\begin{equation}
\label{Delta}
\Delta=-\tilde T^{\mu\nu}\delta_{\mathcal B}g_{\mu\nu}-\tilde J^\mu \delta_{\mathcal B}A_\mu-\tilde K\delta_{\mathcal B}\varphi\,.
\end{equation}
In passing we have also defined the energy density
\begin{equation}
\label{Gibbs}
\epsilon=sT+n\mu-p\,.
\end{equation}

Restricting for a moment to ordinary fluids, \eqref{Delta}
 shows that no entropy is produced when $\tilde T^{\mu\nu}$, $\tilde J^\mu$ are set to zero, i.e. when the stress tensor and charge current take their ideal forms. The structure of \eqref{Delta} and the requirement that $\Delta\geq0$ implies that derivative corrections to $T^{\mu\nu}$, $J^\mu$ should either drop out from $\Delta$, in which case they are called `non-dissipative', or contribute positively to $\Delta$, in which case they  are deemed `dissipative', \cite{Haehl:2014zda,Haehl:2015pja}. Further, these derivative corrections can be separated in those that are terms proportional to (derivatives of) $\delta_{\mathcal B}g_{\mu\nu}$ and $\delta_{\mathcal B}A_\mu$, which vanish in thermal equilibrium and are then called `non-hydrostatic', and terms which are not proportional to any of the $\delta_{\mathcal B}(\cdot)$ and so do not vanish in thermal equilibrium, which are called `hydrostatic'. The hydrostatic corrections capture deviations of the equilibrium state from homogeneity, and can be derived from writing derivative corrections to the pressure in the generating functional, \cite{Banerjee:2012iz,Jensen:2012jh,Bhattacharyya:2012xi}. By varying the external sources, one can then determine how they contribute to the constitutive relations of currents. Instead, the non-hydrostatic terms (either dissipative or not), are true non-equilibrium corrections. 

There is a subtlety with this derivative expansion related to the presence of a superfluid. Indeed, the constitutive relation for $K$ involves derivative mixing, coming from our choice of $\varphi$ as a dynamical field, \cite{Banerjee:2016qxf,Armas:2019sbe}. This choice forces us to take $\varphi\sim \mathcal O(\nabla^{-1})$ so that gradients of $\varphi$ contribute to the first law or constitutive relations at order $\mathcal O(\nabla^0)$. Then the first term in \eqref{Kconstrel} is actually an $\mathcal O(\nabla)$ correction, and positivity of entropy production implies that
\begin{equation}
\label{Ktildeconstrel}
\tilde K=\alpha\, \delta_{\mathcal B}\varphi+\mathcal O(\nabla).
\end{equation}
This type of derivative mixing is typical of theories with spontaneous symmetry breaking and can be avoided by trading $\varphi$ for a higher-form current and the associated anomalous conservation law stemming from the conservation of winding in the absence of topological defects, \cite{Delacretaz:2019brr,Armas:2021vku}. Then all currents are treated on the same footing. 

Here we will stick to the more familiar formulation of superfluid hydrodynamics keeping the Goldstone field as the dynamical field, and pay the corresponding price of having to contend with derivative mixing in the constitutive relations. Putting together \eqref{Kconstrel} and \eqref{Ktildeconstrel},
\begin{equation}
\label{deltaphiconstrel}
 \delta_{\mathcal B}\varphi=-\alpha^{-1}\left[K-T\Delta^{\mu\nu}\nabla_\mu\left(\frac{n_s}{\mu T}\zeta_\nu\right)\right]+\mathcal O(\nabla)
\end{equation}
which shows that the Josephson relation follows from the left-hand side at leading order in derivative, while first-order corrections are (possibly partially) captured by the right-hand side.

The upshot of writing \eqref{deltaphiconstrel} is that instead of parametrizing non-hydrostatic corrections to $K$ in terms of (derivatives of) $\delta_{\mathcal B}\varphi$, we can equivalently parametrize non-hydrostatic derivative corrections to $\delta_{\mathcal B}\varphi$ in terms of the combination $K-T\Delta^{\mu\nu}\nabla_\mu\left(\frac{n_s}{\mu T}\zeta_\nu\right)$, which also must vanish in thermal equilibrium. The advantage of doing so is that it avoids the proliferation of time derivatives at first order in gradients, which greatly facilitates the linear response analysis. This is also the reason why we chose to write the first law \eqref{firstlaw} in terms of $\zeta^2$ instead of $\xi^2$, as otherwise the combination in the right-hand side of \eqref{deltaphiconstrel} would have involved $\nabla_\mu[n_s \xi^\mu/(\mu T)]$, which generates time derivatives. This partial choice of frame corresponds to the ``modified phase frame" of \cite{Bhattacharya:2011eea}.
 
Before moving on to determining the non-ideal terms $\tilde T^{\mu\nu}$, $\tilde J^{\mu}$, we need to finish fixing our frame, i.e. make a specific choice for our definition of temperature, chemical potential and fluid velocity out of equilibrium. This requires a set of $d+2$ constraints. In this work, we choose to work with the Landau frame:
\begin{equation}
\label{LandauFrame}
u_\mu\tilde T^{\mu\nu}=0\,,\quad u_\mu J^\mu=0\,,
\end{equation}
which constrains the non-ideal corrections to $T^{\mu\nu}$ and $J^\mu$ to be orthogonal to the fluid velocity. In the Landau frame, 
\begin{equation}
\label{eq:DeltaLandau}
\Delta=-\tilde T_{TL}^{\mu\nu}\frac{\sigma_{\mu\nu}}{T}-\tilde T\frac{\nabla\cdot u}{dT}-\tilde J_\mu \Delta^{\mu\nu}\delta_{\mathcal B}A_\nu-\tilde\mu\left[\Delta^{\mu\nu}\nabla_\mu\left(\frac{n_s}{\mu T}\zeta_\nu\right)-\frac{K}T\right]\,,
\end{equation}
where we have decomposed $\tilde T^{\mu\nu}=\tilde T_{TL}^{\mu \nu}+\Delta^{\mu\nu}\,\tilde T/d$ into its traceless and tracefull pieces and denoted $-\tilde\mu$ the derivative corrections to $T\delta_{\mathcal B}\varphi$. $\sigma^{\mu\nu}$ is the shear tensor, which is the transverse, traceless combination of derivatives of the fluid velocity
\begin{equation}
\sigma^{\mu\nu}=\Delta^{\mu\alpha}\Delta^{\nu\beta}\nabla_{(\alpha}u_{\beta)}-\Delta^{\mu\nu}\frac{\nabla\cdot u}{d}\,.
\end{equation}

For relativistic superfluids and in the absence of parity violation, there are no hydrostatic corrections to the homogeneous pressure at first order in derivatives, \cite{Bhattacharyya:2012xi}. We then parametrize the non-hydrostatic corrections in the following way\begin{equation} \label{dissmat}
\begin{pmatrix}\tilde J^\mu \\ \tilde \mu\\ \tilde T_{TL}^{\mu \nu} \\ \tilde T\end{pmatrix} = 
-\begin{pmatrix} A^{ \mu \rho} & B^\mu & C^{\mu \rho \sigma} & D^\mu \\
E^\rho & F & G^{\rho \sigma} & H \\
I^{\mu \nu \rho} & K^{\mu \nu} & L^{\mu \nu \rho \sigma} & M^{\mu \nu}\\
N^{\rho}  & Q & R^{\rho \sigma} & S
\end{pmatrix} 
\begin{pmatrix} \Delta_\rho{}^{\nu}\delta_{\mathcal B}A_\nu \\\Delta^{\rho\sigma}\nabla_\rho\left(\frac{n_s}{\mu T}\zeta_\sigma\right) -\frac{K}T\\\frac1{T}\sigma_{\rho \sigma}  \\ \frac{1}{d\,T}\nabla_\rho u^\rho\end{pmatrix}
\end{equation}
This matrix must be positive (semi)definite in order to achieve positivity of entropy production. In particular, its diagonal elements must be nonnegative. Its elements can be parametrized as follows:
\begin{align}\label{Aons}
	\begin{split}
		&A^{\mu \rho } = \alpha_A\,\Delta^{\mu \rho} + \beta_A \, \tilde{\zeta}^\mu \tilde{\zeta}^\rho\,,\qquad\;\;                                                                      
		B^\mu =\alpha_B  \zeta^\mu\,,                       \\ 
		&C^{\mu \rho \sigma} = \alpha_C \, \big( \tilde{P}^{\mu \sigma}\tilde{\zeta}^\rho+ \tilde{P}^{\mu \rho}\tilde{\zeta}^\sigma  \big) + \beta_C \, \zeta^\mu P^{\rho \sigma}\,, \\
		&D^\mu = \alpha_D \, \zeta^\mu\,,\quad
		E^\sigma = \alpha_E \, \zeta^\sigma \,,\quad 
		G^{\rho \sigma} = \alpha_G\, P^{\rho \sigma}\,, \\
		&I^{\mu \nu \rho}  = \alpha_I \, \big( \tilde{P}^{\mu \rho}\tilde{\zeta}^\nu+ \tilde{P}^{\nu \rho}\tilde{\zeta}^\mu  \big) + \beta_I \, \zeta^\rho P^{\mu \nu}\,, \\
		&K^{\mu \nu} = \alpha_K \, P^{\mu \nu}\,, \\
		&L^{\mu \nu  \rho \sigma} = \alpha_L \, P^{\mu \nu}P^{\rho \sigma} + \beta_L \,\Delta^{\mu \rho}\Delta^{\nu \sigma} + \gamma_L \, \big( \zeta^\mu \tilde{\zeta}^\sigma \tilde{P}^{\nu \rho}+\zeta^\nu \tilde{\zeta}^\sigma \tilde{P}^{\mu \rho}\big)\,, \\
		&M^{\mu \nu} = \alpha_M \, P^{\mu \nu}\,,\quad
		N^\rho = \alpha_N \, \zeta^\rho\,,\quad
		R^{\rho \sigma} = \alpha_R \, P^{\rho \sigma} \\
		&F = \alpha_F\,, \quad H = \alpha_H\,, \quad Q = \alpha_Q\,, \quad S = \alpha_S \,,
	\end{split}
\end{align}
where $\tilde{\zeta}^\mu\equiv\zeta^{\mu}/\sqrt{\zeta^\nu \zeta_\nu}$, $P^{\mu \nu} \equiv  \tilde{\zeta}^\mu \tilde{\zeta}^\nu - \Delta^{\mu \nu}/d$ is a traceless, transverse tensor and $\tilde{P}^{\mu \nu}\equiv \Delta^{\mu \nu} - \tilde{\zeta}^\mu \tilde{\zeta}^\nu$ is the projector onto the subspace orthogonal to both $u^\mu$ and $\zeta^\mu$. There are a total of 21 non-hydrostatic parameters. The diagonal elements of this matrix are all dissipative, together with the symmetric part of the off-diagonal elements, which is a total of 14 non-hydrostatic dissipative parameters. The 7 parameters left in the anti-symmetric parts of the off-diagonal elements are on the other hand non-dissipative, \cite{Banerjee:2016qxf}.

Onsager reciprocity requires the matrix appearing in (\ref{dissmat}) to be symmetric, which sets all non-dissipative parameters to zero:
\be
\alpha_B = \alpha_E \quad \alpha_C = \alpha_I \quad \beta_C = \beta_I \quad \alpha_D = \alpha_N \quad \alpha_G = \alpha_K \quad \alpha_H = \alpha_Q \quad \alpha_M = \alpha_R 
\ee
We thus obtain a total of 14 dissipative parameters, which is consistent with \cite{Bhattacharya:2011tra}. 

Note that in the limit of vanishing background superfluid velocity $\zeta^\mu \to 0$ the unit norm vector $\tilde{\zeta}^\mu$ is not well-defined \cite{Bhattacharya:2011eea}, so that one must set $\beta_A=\alpha_C=\alpha_G=\alpha_L=0$, while other transport coefficients multiplying terms non-linear in $\zeta^\mu$ (such as eg $\alpha_B$) also drop out.  After imposing Onsager invariance, this means that the only non-vanishing dissipative contributions to the constitutive relations will be parametrized by a total of five coefficients: $\alpha_A$, $\beta_L$, $\alpha_F$, $\alpha_H$ and $\alpha_S$.

Imposing conformal invariance (as we will be interested in doing for holographic superfluids) requires the stress tensor to be traceless, which in turn sets $\alpha_N=\alpha_D = \alpha_Q =\alpha_H= \alpha_R =\alpha_M= \alpha_S = 0$, leaving us with a total of 10 dissipative coefficients after imposing Onsager reciprocity, \cite{Bhattacharya:2011tra}. In the limit of zero background superfluid velocity this set is further reduced to three coefficients: $\alpha_A$, $\beta_L$ and $\alpha_F$.

\section{Linear response}\label{sec:linear}

We now turn to the study of linear response in the presence of a nonzero background superfluid velocity. That is, we perturb around a state where the spatial component of the superfluid velocity has a nonzero background value $\partial^i\varphi=\zeta^i+\delta\partial^i\varphi e^{-i\omega t+i k_j x^j}$. The temperature and the chemical potential are perturbed similarly, while the spatial component of the normal fluid velocity is $u^i=\delta u^i e^{-i\omega t+i k_j x^j}$. For simplicity, we work in the collinear limit $\boldsymbol{\zeta} // \mathbf{u} // \mathbf{k} // \hat{e}_x$, where all these vectors are aligned along the same direction $\hat{e}_x$. We will henceforth drop the subscript on $\zeta_x = \zeta$ when discussing the magnitude of the superfluid velocity.

We choose to work in the grand-canonical ensemble, holding fixed the temperature and chemical potential, which couple to the energy and charge current densities $T^{tt}$ and $J^t$, respectively. The other sources are $u^i$ the spatial component of the normal velocity, which couples to the momentum density $T^{ti}$. By construction, the Goldstone field couples to $K$. As we saw in the previous Section, in thermal equilibrium, $K=T\Delta^{\nu\lambda}\nabla_\nu[(n_s\zeta_\lambda)/(\mu T)]$. It will be more convenient to consider $\delta\partial^x\varphi$ as a fluctuation, which as a consequence couples to $T\delta h^x\equiv T\delta[(n_s \zeta^x)/(\mu T)]$ after integrating by parts the previous relation. This defines a vector of sources $\delta s^A=\{T\delta (\mu/T),\delta T/T,\delta u^x,T\delta h^x\}$, which is conjugate to the vector $\delta O^A=(\delta J^t,\delta T^{tt},\delta T^{tx},\delta\partial^x\varphi)$. 

The linear equations of motion with external sources turned off take the form 
\begin{equation} \label{eomsM}
-i\omega \, \begin{pmatrix} \delta J^t \\ \delta T^{tt} \\ \delta T^{tx} \\ \delta \partial^x\delta\phi \end{pmatrix} +M \cdot \begin{pmatrix}T \delta( \mu/T) \\ \delta T/T \\ \delta u^x \\ T\delta h^x \end{pmatrix} = 0
\end{equation}
with
\small{
\begin{equation}
 M=\begin{pmatrix} \sigma_0\,k^2 & ik\frac{ \zeta n_\text{s}}{\mu} & ikn + \zeta\zeta_6\, k^2 & ik +k^2\zeta\zeta_2 \\
ik\frac{ \zeta n_\text{s}}{\mu}  &2 ik \zeta n_\text{s}& ik\left(sT+n\mu+\frac{n_\text{s}\zeta^2}{\mu}\right) &i k\mu\\
ikn + \zeta\zeta_6\, k^2 & ik\left(sT+n\mu+\frac{n_\text{s}\zeta^2}{\mu}\right) &2n_\text{s} \zeta \,ik  +(\eta+\zeta_b)\,k^2 & ik\zeta +k^2\zeta_1 \\
ik+k^2\zeta\zeta_2 &ik\mu & ik\zeta +k^2\zeta_1  & k^2\zeta_3
\end{pmatrix} 
\end{equation}} 
where 
\begin{equation}
\sigma_0=\frac{\alpha_A+\beta_A}{T}\,,\quad\zeta_1=\frac{\alpha_G+\alpha_H}{2T}\,,\quad\zeta_2=\frac{\alpha_B}T\,,\quad \zeta_3=\frac{\alpha_F}{T}\,,\quad \zeta_6=\frac{(\alpha_D+\beta_C)}{2T}\,,
\end{equation}
and
\begin{equation}
\eta=\frac{\beta_L+\zeta\gamma_L}{2T}\,,\quad \zeta_b=\frac{\alpha_S+\alpha_L+\alpha_M-2\zeta\gamma_L}{4T}\,.
\end{equation}
Here we have labelled the bulk viscosity, $\zeta_b$, so as not to confuse it with the spatial superfluid velocity $\zeta$. Interestingly, at nonzero superfluid velocity, the bulk viscosity is non-vanishing even when conformal invariance is imposed.

The dissipative coefficients can be extracted from the retarded Green's functions through the following Kubo relations:
\begin{equation} \label{kubo}
\begin{split}
\sigma_0 &= -\lim_{\omega \to 0, k\to 0} \, \frac1{\omega}\, \text{Im} \,G^R_{J^xJ^x}(\omega, k) \\
 \zeta_1 &=-\lim_{\omega \to 0, k\to 0} \, \frac{1}{k\omega}\, \text{Im} \,G^R_{T^{xx}\xi^x}(\omega, k) \\
\zeta \, \zeta_2 &= -\lim_{\omega \to 0, k\to 0} \, \frac{1}{k}\, \text{Im} \,G^R_{J^x\xi^x}(\omega, k) \\
\zeta_3 &= -\lim_{\omega \to 0, k\to 0} \, \frac{\omega}{k^2}\,  \text{Im} \,G^R_{\xi^x \xi^x}(\omega, k) \\
\zeta\, \zeta_6 &=- \lim_{\omega \to 0, k\to 0} \, \frac{1}{\omega}\, \text{Im}\, G^R_{J^xT^{xx}}(\omega, k) \\
\eta+\zeta_b &=- \lim_{\omega \to 0, k\to 0} \, \frac1{\omega}\, \text{Im} \,G^R_{T^{xx}T^{xx}}(\omega, k)\\
\eta &=- \lim_{\omega \to 0, k\to 0} \, \frac1{\omega}\, \text{Im} \,G^R_{T^{xy}T^{xy}}(\omega, k)
\end{split}
\end{equation}
To compute the retarded Green's functions and the location of the hydrodynamic poles, we need to work out the matrix of static susceptibilities which relates the $O^A$ to the $s^A$
\begin{equation}
\delta O^A=\chi^{AB}\delta s_B\,,
\end{equation}
for which we find
\begin{equation}
\chi=\begin{pmatrix}
\chi_{nn} + \chi_{nh}^2\chi_{\xi\xi} & \chi_{n\epsilon} + \chi_{nh}\chi_{\epsilon h}\chi_{\xi\xi} & \frac{\zeta n_\text{s}}{\mu} &  \chi_{nh}\chi_{\xi \xi} \\
 \chi_{n\epsilon} + \chi_{nh}\chi_{\epsilon h}\chi_{\xi\xi} & \chi_{\epsilon\epsilon}+\chi_{\epsilon h}^2\chi_{\xi \xi} & 2\zeta n_\text{s} &\chi_{\epsilon h} \\
  \frac{\zeta n_\text{s}}{\mu}  & 2\zeta n_\text{s} & \epsilon + p + \frac{n_\text{s}}{\mu} \zeta^2 & \mu \\
 \chi_{nh}\chi_{\xi \xi}  &  \chi_{\epsilon h}\chi_{\xi \xi}  &\mu & \chi_{\xi \xi}
\end{pmatrix}
\end{equation}
and where
\begin{equation}
\label{chixixi}
\chi_{\xi \xi} = \frac{\mu}{ \partial_{\zeta}\left( \zeta n_\text{s}\right)|_{T,\mu}}\,.
\end{equation}
Here we have often written the individual matrix elements in terms of the $\zeta$ derivatives rather than the $h$ derivatives, which will prove more direct to evaluate in holographic states. More explicitly,
\begin{equation}
\chi_{nh}\equiv\left.\frac{\partial n}{\partial\zeta}\right|_{T,\mu}
\end{equation}
Both the $\chi$ and $M$ matrices are Onsager symmetric, recalling that the eigenvalues under the time reversal symmetry of the $O^A$ are respectively $\{1,1,-1,-1\}$ and that $\xi,\zeta$ should be flipped under time reversal.

With this in hand, the retarded Green's functions can be computed through
\begin{equation}
\label{GRfromM}
G^R=-M^{-1}\cdot\left(i\omega-M\right)^{-1}\cdot\chi\,.
\end{equation}

It is also helpful to trade the energy density for the entropy density, which is conserved at linear level in fluctuations. This can be done using the first law
\begin{equation}
\delta \epsilon=T\delta s+\mu \delta n-\frac{n_\text{s}}{\mu}\delta(\xi^x-u^x)
\end{equation}
to combine the equations of motion \eqref{eomsM} and obtain the matrices $M$ and $\chi$ in the new basis $\delta O^A=(\delta J^t,\delta s,\delta T^{tx},\delta\partial^x\varphi)$ with sources $\delta s^A=\{\delta \mu,\delta T,\delta u^x,\delta h^x\}$. One can check that the row of the matrix $M$ corresponding to the $\delta s$ equation matches the linearized canonical entropy current given in \eqref{JScanconstrel}. 

The longitudinal sector contains four sound modes, $\omega = vk +i\Gamma k^2/2+\mathcal{O}(k^3)$, which are the zeroes of the denominator of retarded Green's functions computed with \eqref{GRfromM}. These modes reduce in the zero superfluid velocity case to the so-called first and second sound modes \cite{Landau, clark, putterman}.
In general, with a non-vanishing background superfluid velocity, these modes take complicated expressions which are not very helpful to write explicitly.

Clearly, $\chi_{\xi \xi}$ in \eqref{chixixi} diverges when 
\begin{equation}
\label{critrel}
 \partial_{\zeta}\left( \zeta n_\text{s}\right)|_{T,\mu}=0\,,\quad \zeta_c=-\frac{n_\text{s}}{ \left.\partial_{\zeta} n_\text{s}\right|_{T,\mu}}\,.
\end{equation}
As we showed in \cite{Gouteraux:2022kpo}, we expect a dynamical instability when this happens. Indeed, one can check that it is at this critical value that one of the sound modes becomes unstable (its imaginary part crosses onto the upper half plane while its real part changes sign).

Around the critical value \eqref{critrel}, one of the sound modes takes a particularly simple form. Setting $\zeta = \zeta_c+ \delta \zeta + \mathcal{O}(\delta \zeta^2)$ as before, we find a sound mode $\omega = vk +i\Gamma k^2/2$ which is linear in $\delta \zeta$:
\be \label{soundinst}
v =\nu \, \delta \zeta + \mathcal{O}(\delta \zeta^2)
\ee
\be 
\nu =\frac{\ns \, s^2}{2\, \mu \, \zeta_c \left( -s^2\,\chi_{nh} + s\, \chi_{sh}(n + \zeta_c \chi_{nh})-\zeta_c \,\nnn \,\chi_{sh}^2\right)} 
\ee
The attenuation constant $\Gamma$ can be written as:
\be \label{gammainst}
\Gamma = \frac{2 \,  \zeta_c\, \nu^2}{\mu \,\ns \,T^2\,s^4}\, \gamma_A \gamma_BM^{AB} \, \delta \zeta + \mathcal{O}(\delta \zeta^2) \, \text{,}
\ee 
with
\be
M^{AB} = \begin{pmatrix} \sigma_0 & \zeta_c  \,\zeta_6 &  \zeta_c \, \zeta_2 \\
\zeta_c \, \zeta_6 & \eta+\zeta_b & \zeta_1\\
\zeta_c \, \zeta_2 & \zeta_1 & \zeta_3
  \end{pmatrix}
\ee
and
\be
\gamma_A = \begin{pmatrix} 
\mu\, s\,\left(sT +\mu (n + \zeta_c\, \chi_{nh}) \right) - \mu \zeta_c \,\chi_{sh} \left(sT + 2\mu\, \nnn \right) \\
sT\mu\, \chi_{sh} \\
s^2\,\mu\, T \,\chi_{nh} + \ns s\zeta_c (n + \zeta_c\, \chi_{nh}) - 2\ns \nnn \,\zeta_c^2\,  \chi_{sh} -s\,T\,\mu\, n \,\chi_{sh}
\end{pmatrix} \, \text{.}
\ee
Most importantly, the matrix $M^{AB}$ is positive (semi)definite. This follows from positivity of entropy production, applied to matrix (\ref{dissmat}) in the collinear limit. Thus we see that for $\delta \zeta >0$ (i.e. $\zeta>\zeta_c$), the imaginary part of this mode becomes positive, thereby signaling a dynamical instability.
\section{Holographic superfluids}\label{sec:holography}

\begin{figure}[h!]
\centering
\includegraphics[scale=.4]{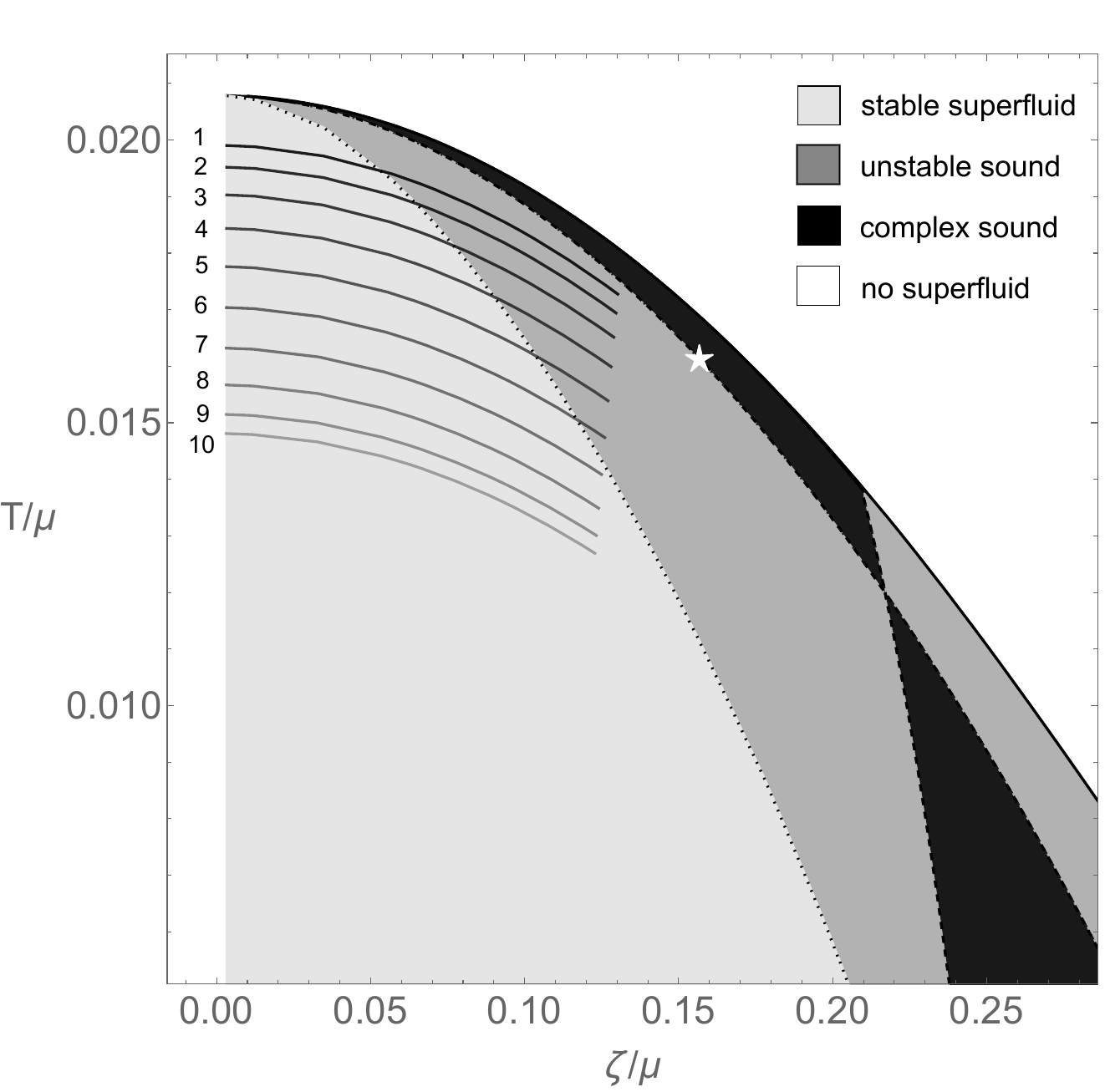}
\caption{\label{CutsThroughPhaseDiagram} The phase diagram for the holographic superfluid	we study (also reported in  \cite{Gouteraux:2022kpo}) as obtained from hydrodynamics and quasinormal modes. The curves on the upper left are slices through the phase diagram for which we report detailed hydrodynamic results. Here, we choose $Q=1$ where it is known that there is only one value of the condensate at fixed $\zeta/\mu$ and $T/\mu$ \cite{Sonner:2010yx}. Finally, the star marks the two-stream instability shown in Figure~\ref{twostreamfigure}.}
\end{figure}

To verify our hydrodynamic predictions, we turn to a class of numerically solvable holographic superfluids \cite{Gubser:2008px, Hartnoll:2008vx, Hartnoll:2008kx}. Holographic superfluids are strongly interacting $d+1$-dimensional quantum systems with a large $N$ number of colors that spontaneously break a global $U(1)$ symmetry. Via the gauge/gravity duality \cite{Maldacena:1997re,Witten:1998qj}, at leading order in $N$, these systems can be described by a $d+2$-dimensional classical gravitational action minimally coupled to a charged complex scalar field. A generic action for the system is given by
\begin{align}
\label{holographicaction1}
S_\text{bulk} = \int d^{d+2}x \sqrt{-g}\left[\frac{1}{16\pi G}(R-2\Lambda) - \frac{1}{4}F_{MN}F^{MN} - V(|\Psi|) - |D\Psi|^2 \right] \,.
\end{align}
Here $\Lambda = -d(d+1)/(2L_{AdS}^2)$ is the cosmological constant in units of the AdS radius, $L_{AdS}$, and $M=0,...,d+1$ runs over all bulk spacetime indices, whereas $\mu=0,...,d$ runs over only the boundary spacetime indices. $F_{MN}=\partial_M A_N-\partial_N A_M$ is the field strength for an abelian gauge field under which the complex scalar, $\Psi$, is charged,
\begin{align}
D_M\Psi = \nabla_M \Psi - i Q A_M \Psi.
\end{align}
In this work, we will only consider the simplest example of holographic superfluids where the scalar potential consists of just a mass term,
\begin{align}
V(|\Psi|) = \frac{m^2}{L_{Ads}^2}|\Psi|^2.
\end{align}

The first term in the action (\ref{holographicaction1}) is generically UV divergent. These divergences are regulated by introducing a boundary, $\partial\Sigma$, at large $r$ and adding local counterterms via $S_\text{bdy}$. $S= S_\text{bulk}+S_\text{bdy}$ is then regular as $r\to \infty$. The procedure leads to a well-behaved variational principle \cite{Balasubramanian:1999re,Skenderis:2002wp}. The choice of boundary action is dimension dependent and also depends on the choice of $m^2$. In what follows, we will restrict to $d=2$ and $m^2=-2$. The boundary action in this case is
\begin{align}
\label{boundaryaction}
S_\text{bdy} = \int_{\partial\Sigma}d^{3}x\sqrt{-\gamma}\left[\frac{1}{16\pi G}\left(2K-\frac{4}{L_{AdS}}-R^{(\gamma)}\right)-\frac{1}{L_{AdS}}|\Psi|^2\right]
\end{align}
Here, $\gamma_{MN}$ is the induced metric on the surface $\partial\Sigma$, $K_{MN}$ is the extrinsic curvature of the surface, $K$ is its trace, and $R^{(\gamma)}$ is the Ricci tensor associated with $\gamma_{MN}$. Without loss of generality, we can take $\Psi=\bar{\Psi}=\psi$. We furthermore set $G = 1/16\pi$ and $L_{AdS}=1$.

The equations of motion which follow from varying (\ref{holographicaction1}) are
\begin{align}
\begin{split}
R_{MN}-\frac{R}{2}g_{MN} - 3g_{MN} &= -\frac{1}{2}F_{MP}F^{P}_{\;\;N} +D_{(M}\Psi\bar{D}_{N)}\bar{\Psi}- \frac{1}{2}g_{MN}\biggl(\frac{1}{4}F^2 + V + |D\Psi|^2\biggr)\\
\nabla_M F^{MN}&=iQ\bigl(\bar{\Psi} D^N\Psi - \Psi \bar{D}^N\bar{\Psi}\bigr)\\
D_M(\sqrt{-g}D^M\Psi) &= -2\sqrt{-g}\Psi.
\end{split}
\end{align}

\begin{figure}[h]
\centering
\includegraphics[scale=.4]{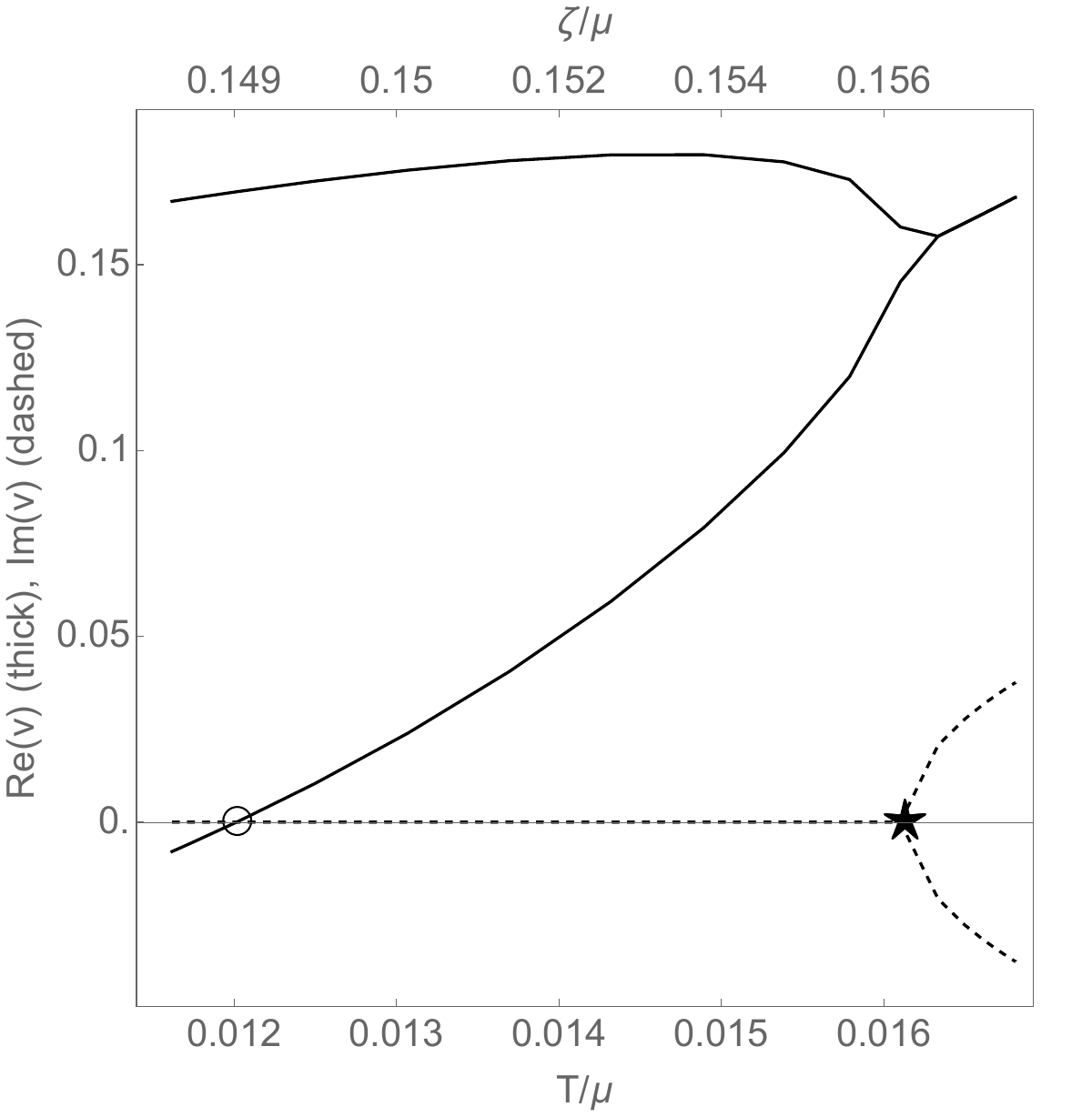}
\caption{\label{twostreamfigure} An example of the two types of instabilities that can appear at large enough superfluid velocities, illustrated in terms of the velocity of one of the hydrodynamic sound modes, $v\equiv \lim_{k\to 0}\omega/k$. The first, marked by an open circle, indicates the instability in~\eqref{critrel} where one of the velocities changes sign and is accompanied by a change in sign of the attenuation (here, we just show the velocity but in Figure~\ref{hydrocomparisonwithinstability} we also show the attenuations). The second, marked by a star, is the two-stream instability where the velocities become complex and one has a positive imaginary component. In general, this occurs when the first instability is already present, as demonstrated in the phase diagram of Figure~\ref{CutsThroughPhaseDiagram}.}
\end{figure}

The phase diagram of these holographic superfluids was constructed
in \cite{Gouteraux:2022kpo} and is shown in Figure~\ref{CutsThroughPhaseDiagram}. There one can see 
that the superfluid phase becomes dynamically unstable at the critical
superfluid velocity~\eqref{critrel}. Moreover, in some regions of the phase diagram if one increases the superfluid velocity beyond the critical value, a new instability can appear where two of the velocities of the sound modes pick up equal and opposite imaginary components and has been termed the ``two-stream" instability \cite{schmitt1,schmitt2,schmitt3}. We have illustrated this behavior in Figure~\ref{twostreamfigure}.

In the remainder of this section we will probe that phase diagram and carefully check that the hydrodynamic theory constructed in previous sections describes these holographic superfluids

\subsection{Thermodynamics and linear response from holography}
The equilibrium thermodynamics of holographic superfluids obeys the Landau-Tisza model of relativistic hydrodynamics described in Section \ref{sec:superfluid_review}, as was shown in \cite{Sonner:2010yx}. Here, we extend those results to include the leading order out of equilibrium transport by considering linearized fluctuations of the holographic solutions.

\begin{figure}[t!]
\centering
\includegraphics[scale=.35]{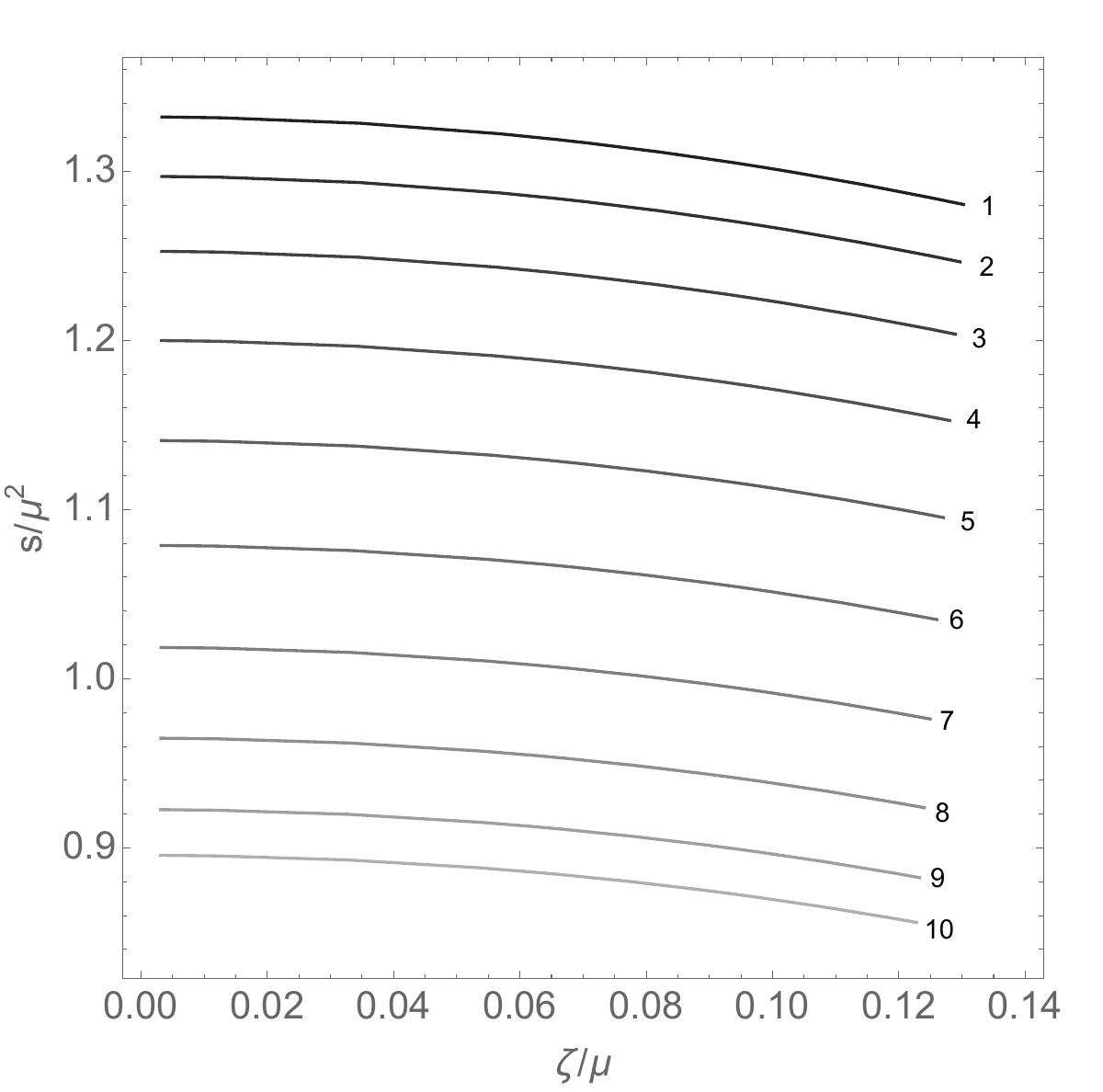}
\includegraphics[scale=.35]{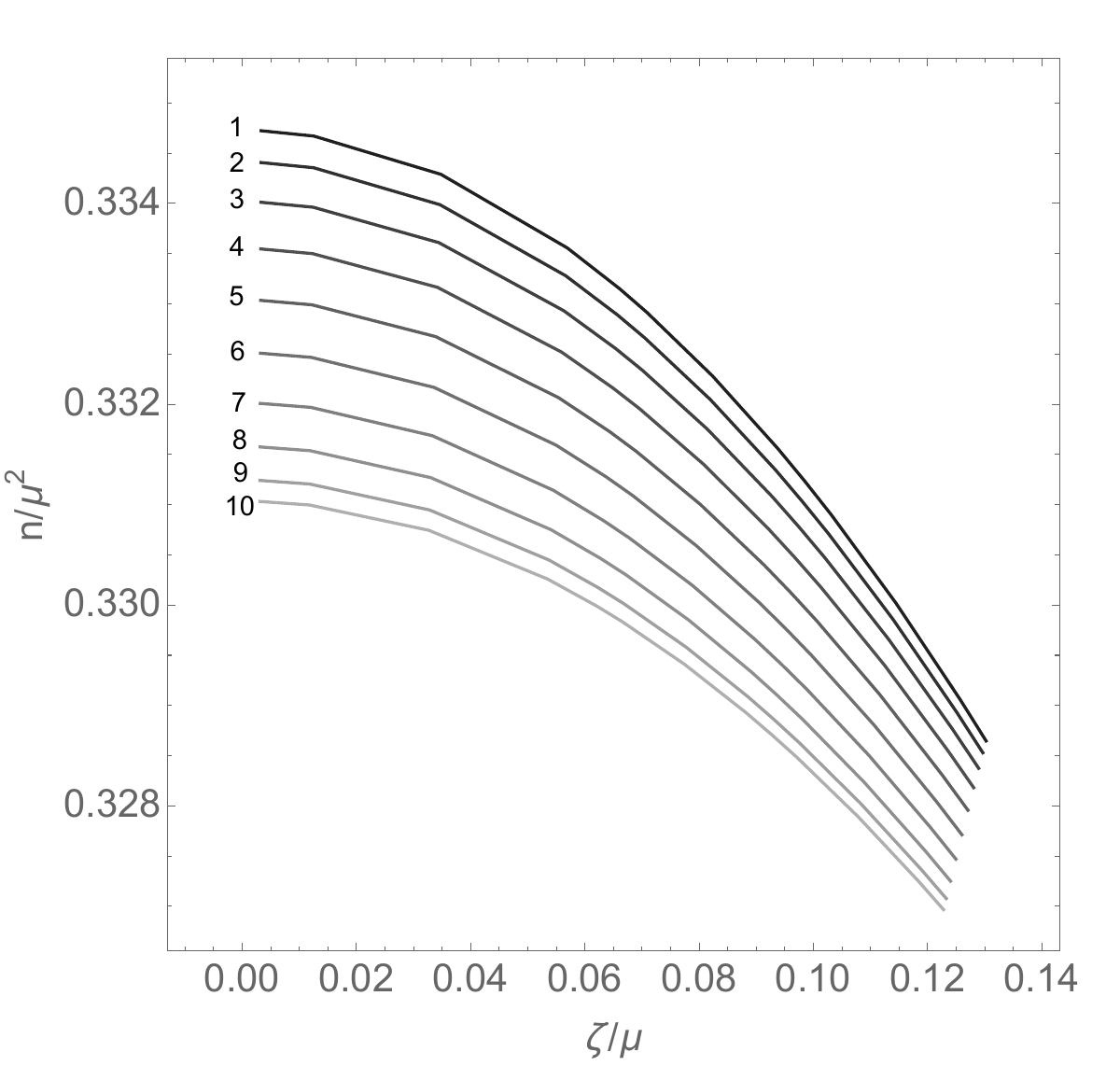}
\includegraphics[scale=.35]{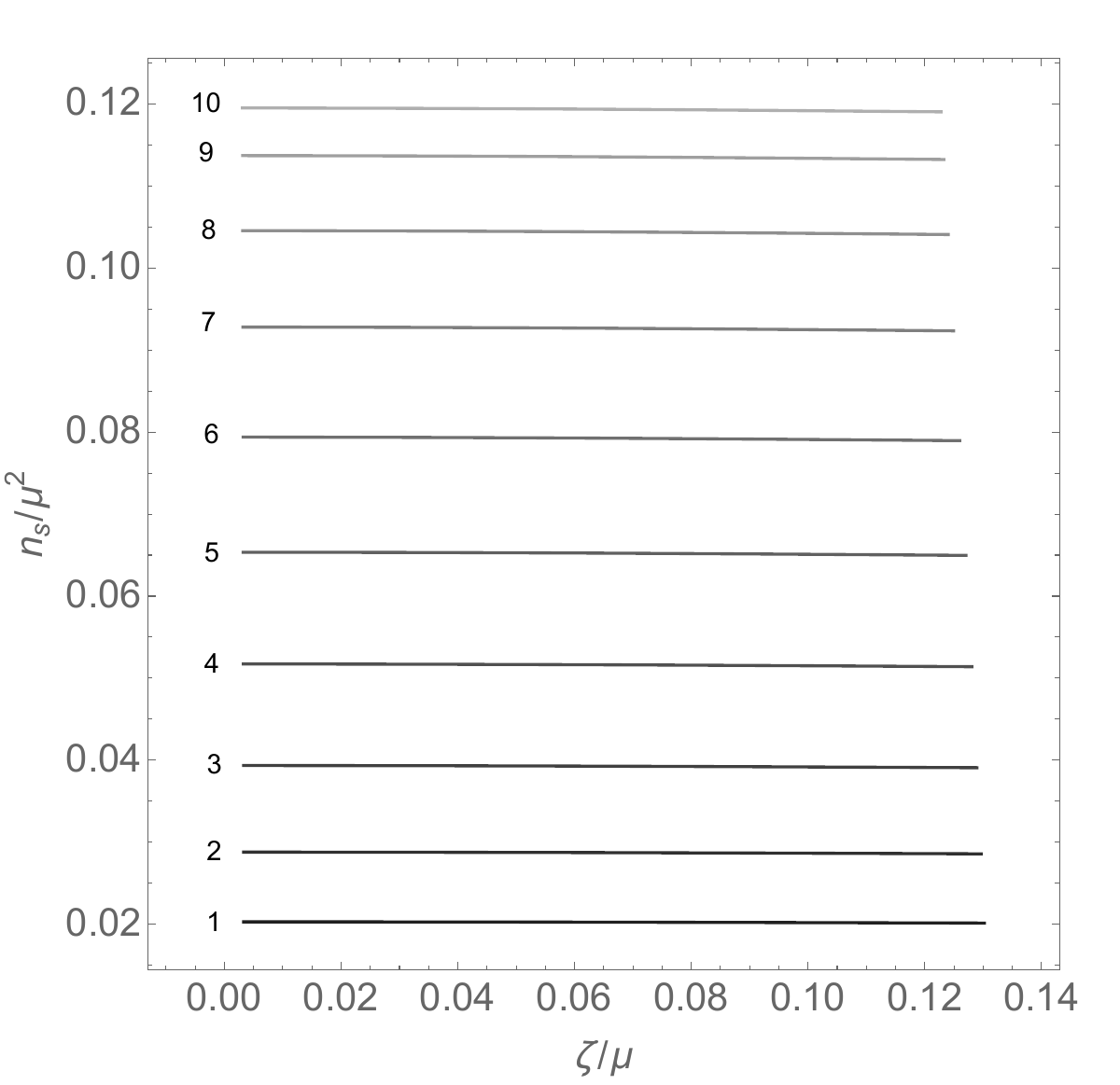}
\includegraphics[scale=.35]{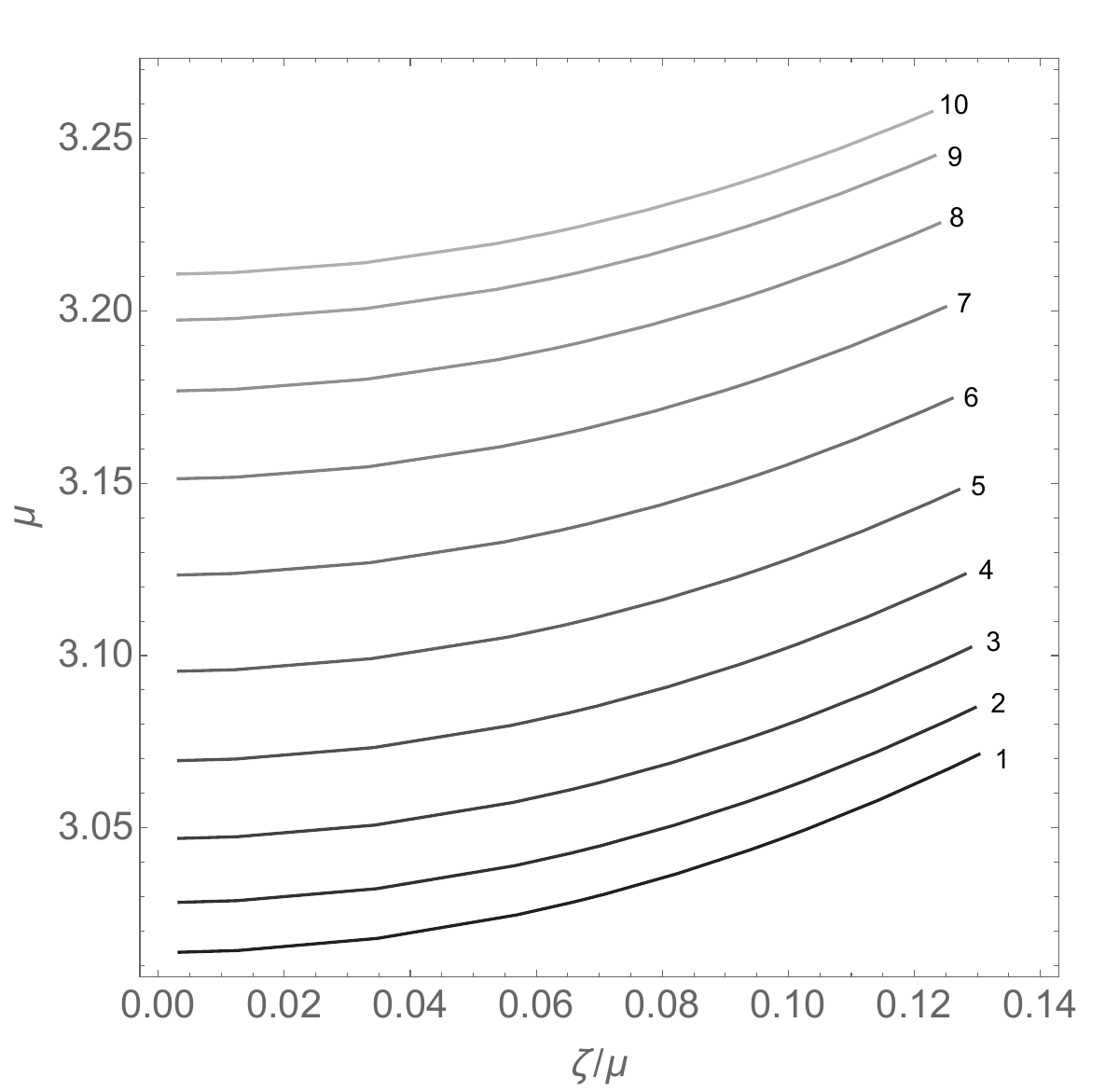}
\caption{\label{thermofigure1}Thermodynamic quantities along the curves 1-10 in Figure~\ref{CutsThroughPhaseDiagram} that are used to calculate the hydrodynamic dispersion relations. For convenience, in the last plot, we have include the values of $\mu$ that we use to construct numerical solutions when $r_h=1$.}
\end{figure}

We start by considering the equilibrium thermodynamics. Summarizing \cite{Sonner:2010yx}, variations of the on-shell action give access to the holographic stress tensor as well as the conserved U(1) current. This allows one to match to the non-tilded quantities in (\ref{eq:Tmunu}) and (\ref{eq:Jmu}) and obtain the quantities $\epsilon, p, n, s, n_s, \mu, \zeta$ and $T$. By considering a family of solutions parametrized by $\mu$, $\zeta$, and $T$, we have access to the entire matrix of static susceptibilities. A subset of these equilibrium quantities is illustrated in Figures~\ref{thermofigure1} and \ref{thermofigure2}.

We then consider linearized fluctuations of the gravitational and matter fields with plane wave dependence $e^{-i\omega t + ik_jx^j}$ which, on-shell, give access to the retarded Green's functions \cite{Son:2002sd}. Having obtained the retarded Green's functions via the holographic dictionary, we then use the Kubo formulae in (\ref{kubo}) to obtain the dissipative transport coefficients. A subset of the dissipative transport coefficients is illustrated in Figure \ref{dissipativefigure}.

In the absence of sources, the equations of motion for the linearized fluctuations only have solutions for specific values of $\omega$ and $k$. These solutions are called ``quasinormal modes" and the values of $\omega(k)$ for which the solutions exist are the quasinormal mode spectrum. At the same time, via the holographic dictionary, source-free solutions to the linearized equations of motion correspond to poles in the retarded Green's functions. Therefore, the gapless quasinormal modes, which have 
\begin{align}
\lim_{k\to 0}\omega(k) = 0, \nonumber
\end{align} 
can be identified with the spectrum of hydrodynamic modes \cite{kovstar}. 

Hence, to verify that holographic superfluids satisfy the theory of relativistic superfluid hydrodynamics including finite superflow and dissipation, as described in Section \ref{sec:superfluid_review}, we compare the numerically obtained spectrum of quasinormal modes to the prediction for the hydrodynamic spectrum using values for transport coefficients obtained via the holographic dictionary. A representative comparison is shown in Figures~\ref{fig:matching1}, \ref{fig:matching2}, \ref{fig:matching3}, \ref{fig:matching4}, where it is observed that the agreement is excellent.\footnote{A demonstration of the convergence of the matching between the quasinormal modes and the hydrodynamics with increasing grid size can be found in Figures~\ref{ConvergencePlots1} and \ref{ConvergencePlots2}.} 

In the main text, we consider the fully backreacted holographic superfluid. However, one can also consider decoupling the energy-momentum sector from the charge sector in a superfluid, which is equivalent to considering the case of a probe holographic superfluid. A convenient aspect of probe superfluids is that, for a $d=3$ dimensional superfluid, there exists an analytic solution \cite{Herzog:2010vz, Bhattacharya:2011eea}. This case provides an instructive introduction to the analysis of the main text, so we have included a discussion in Appendix \ref{probeAppendix}.

\subsection{Equilibrium thermodynamics}
In this Section, we discuss how to obtain the on-shell constitutive relations for the currents (again, working in $d=2$ with $m^2 = -2$). Choosing a superfluid velocity along the $x$ direction, we write the background metric as
\begin{align}
ds^2 = -dt^2 D(r) + B(r)dr^2 + C_{x}(r) dx^2 + C_{xt}(r)dt\,dx + C_{y}(r) dy^2
\end{align}
and the U(1) gauge field and charged scalar as
\begin{align}
A_Mdx^M = A_t(r)dt + A_x(r)dx,\quad \quad \Psi = \bar{\Psi} = \psi(r)
\end{align}
where $r$ is the AdS radial vector and we choose $r\to\infty$ to be the UV AdS boundary. The choice $\Psi = \bar{\Psi}$ fixes the phase of the superfluid $\varphi = 0$ in equilibrium. Given this ansatz for the coordinate dependence of the fields, we find that there are 5 second order and 2 first order equations for the 7 fields, so that to completely specify a solution we must impose 12 boundary conditions. The boundary conditions are specified as follows.

Finite temperature states of the superfluid are described in terms of black hole spacetimes. In Schwarzschild-like coordinates, this requires that there exists a black hole horizon at some radial coordinate $r_h$ where $D(r_h) =C_{xt}(r_h)= B^{-1}(r_h) = 0$. Near this surface, the metric reads
\begin{align}
ds^2 = \left[-4\pi T(r-r_h)(dt^2 - \frac{C_{xt}^h}{4\pi T} dt dx) + \frac{dr^2}{4\pi T(r-r_h)} + C_x^h dx^2 + \frac{s^2}{(4\pi)^2 C_x^h}dy^2 \right]\times[1+O(r-r_h)]
\end{align}
The matter fields are required to be well-behaved near the horizon---in particular, we require that $A_t(r_h) = 0$. Near $r=r_h$, then, the matter fields have the form, 
\begin{align}
A_t \to A_t^h(r-r_h)+O(r-r_h)^2,\quad A_x \to A_x^h + O(r-r_h),\quad \psi \to \psi_h + O(r-r_h).
\end{align}
Higher order terms in $r-r_h$ are determined in terms of the constants above. The vanishing of $D, B^{-1}, C_{xt}$ and $A_t$ at the black hole horizon constitutes 4 boundary conditions.

At large $r$, we require that the metric on a fixed $r$ hypersurface is conformal to Minkowski space,
\begin{align}
\lim_{r\to \infty} ds^2 = \frac{dr^2}{r^2} + r^2\biggl(-dt^2+dx^2+dy^2+\mathcal{O}(r^0)\biggr)
\end{align}
which fixes five boundary conditions. After fixing these boundary conditions, the equations of motion lead to an asymptotic form for the metric and matter fields
\begin{align}
\label{UVexpansion_equilibrium}
\begin{split}
D(r) &= r^2  - \frac{\epsilon}{3r}+O(r^{-2}),\quad 
B(r) =r^{-2},\quad 
C_x(r) = r^2 + \frac{\epsilon-p}{3r}+O(r^{-2}),\\
C_{xt}(r) &= -\frac{\zeta n_s}{3r}+O(r^{-2}),\quad
C_{y}(r)= r^{2}+\frac{p}{3r}+O(r^{-2}),\quad
A_t(r)= \mu - \frac{n}{r}+O(r^{-2}),\\
&A_x(r) = -\zeta + \frac{\zeta n_s}{\mu}\frac{1}{r} + O(r^{-2}),\quad
\psi(r) = \frac{\psi_s}{r}+\frac{\langle O^\psi\rangle}{2r^2} + O(r^{-3})
\end{split}
\end{align}
Note that the gauge invariant combination $\zeta = \partial_x\varphi -A_x$ fixes the sign of the leading falloff for $A_x$ when $\varphi = 0$. Here we have chosen a suggestive form for the coefficients of the subleading powers of $r$. Below, we will vary the on-shell action and observe that $\psi_s$ acts as a source for $\langle O^\psi \rangle$. One can also check that the remaining coefficients indeed correspond to the boundary theory quantities associated with the equilibrium hydrodynamics presented earlier. Interpreting $\langle O^\psi \rangle$ as the superfluid condensate, solutions with the boundary condition $\psi_s = 0$ describe spontaneous symmetry breaking. The final two boundary conditions can be chosen from the set of constants $\mu, \zeta, T, C^h_x, C^h_{xt}, A^h_t, A^x_t, \psi_h, s$. We will see that $\mu$ and $\zeta$ act as sources for $\rho$ and $n_s$, so it is natural to choose their values as the last two conditions. This is in line with their interpretation as intensive parameters in the thermodynamics. Finally, $r_h$ is determined by a choice of radial coordinate and can be set $r_h=1$ via a scale transformation $r \to r/r_h$. Ultimately, then, after fixing boundary conditions, the equation of state of a holographic superfluid is specified by two numbers, $p(T/\mu, \zeta/\mu)$.

The constitutive relations of the dual stress tensor and the U(1) current are obtained via variation of the renormalized on-shell action \cite{Balasubramanian:1999re,Skenderis:2002wp, Herzog:2009md, Kim:2016hzi, Arean:2021tks}. Using equations (\ref{holographicaction1}) and (\ref{boundaryaction}), the variation can be written

\begin{align}
\label{actionvariation}
\delta S_{ren}^{os} &=  \int_{\partial\Sigma} d^3x\sqrt{-\gamma}\biggl\{\delta\Psi \bigl[n^N(\partial_N+iq A_N)\bar{\Psi}-\bar{\Psi}\bigr]+\delta\bar{\Psi}\bigl[n^N(\partial_N-iq A_N)\Psi-\Psi\bigr]\nonumber\\
& \quad+n_M F^{MN}\delta A_N+\delta\gamma_{MN}\left[K^{MN}-\left(K+2+\frac{|\Psi|^2}{2}+\frac{R^{(\gamma)}}{2}\right)\gamma^{MN}-(R^{(\gamma)})^{MN}\right]\biggr\}.
\end{align}
Here $n_M$ is the vector normal to the surface $\partial\Sigma$, $\gamma_{MN}$ is the induced metric on this surface, $K_{MN}$ is its extrinsic curvature with $K$ the trace, and $R^{(\gamma)}_{MN}$ and $R^{(\gamma)}$ are the Ricci tensor and scalar associated with $\gamma_{MN}$.

From this expression, we find that
\begin{align}
\begin{split}
\langle T^{\mu\nu} \rangle \equiv \frac{2}{\sqrt{-\gamma}}\frac{\delta S_{ren}}{\delta\gamma_{\mu\nu}}&= \lim_{r\to\infty} r\biggl\{2\biggl[K^{\mu\nu} - (K+2)\gamma^{\mu\nu}\biggr]-|\psi|^2\gamma^{\mu\nu}\biggr\}\\
&=(\epsilon+p)u^\mu u^\nu + p \eta^{\mu\nu} + 2n_s u^{(\mu}\zeta^{\nu)} + \frac{n_s}{\mu}\zeta^\mu \zeta^\nu\\
\langle J^\mu \rangle \equiv \frac{1}{\sqrt{-\gamma}}\frac{\delta S_{ren}}{\delta A_\mu} &= \lim_{r\to\infty} r\biggl[n^MF_{M\mu}\biggr]\\
&=nu^\mu + \frac{n_s}{\mu}\zeta^\nu
\end{split}
\end{align}
matching the standard Landau-Tisza form \cite{Sonner:2010yx}. Of use are the identities
\begin{align}
\begin{split}
(S-S_{bdy})^{I}_{\text{on-shell}} &=\int dt d^2x \int_{r_h}^\infty dr\frac{d}{dr}\left[\frac{C_y'}{C_y}\sqrt{-g}g^{rr}\right]\\
(S-S_{bdy})^{II}_{\text{on-shell}}&= \int dt d^2x \int_{r_h}^\infty dr\frac{d}{dr}\left[\frac{C_y^2}{\sqrt{-g}}\frac{d}{dr}\biggl(\frac{\sqrt{-g}}{C_y^2}g^{rr}\biggr)+\sqrt{-g}g^{rr}F_{r\mu}A^\mu\right]. 
\end{split}
\end{align}
To arrive at these identities, one must use the equations of motion to rewrite the bulk action. Setting $(S-S_{bdy})^{I}_{\text{on-shell}}=(S-S_{bdy})^{II}_{\text{on-shell}}$ and evaluating for Euclidean time ($t = -i\tau$ with $\tau = \tau + \beta $), these enforce the thermodynamic relations 
\begin{align}
\epsilon+p = sT + \mu\rho,\quad\quad 2p-\epsilon = - \frac{\zeta^2}{\mu}n_s.
\end{align}
The latter of these two relations imposes the conformality condition $\langle T^\mu_{\;\;\mu} \rangle = 0$ and leads to $p$ being a function of $T/\mu$ and $\zeta/\mu$ instead of $T,\mu,\zeta$ independently. 

Variation of the Euclidean on-shell action yields the first law. To see this note
\begin{align}
-V_2\,\delta (\beta p) \equiv \delta(S_E - S_{E,bdy}) 
\end{align}
Using the asymptotics of the bulk fields, this variation can be written for fixed $\beta$ as \cite{Sonner:2010yx}
\begin{align}
-V_2\,\delta (\beta p) = \beta\, V_2 \biggl([\epsilon - \mu \,\rho]\,\delta r_h -\rho\,\delta\mu + \frac{n_s}{2\mu}\,\delta \zeta^2\biggr) .
\end{align}
After fixing $\beta$, we have the identity $\delta r_h = -(\delta T/T) r_h$ which then yields the first law
\begin{align}
\df p = s \,\df T + n \,\df \mu - \frac{n_s}{2\mu} \,\df \zeta^2\,.
\end{align}

\begin{figure}[t!]
\centering
\includegraphics[scale=.32]{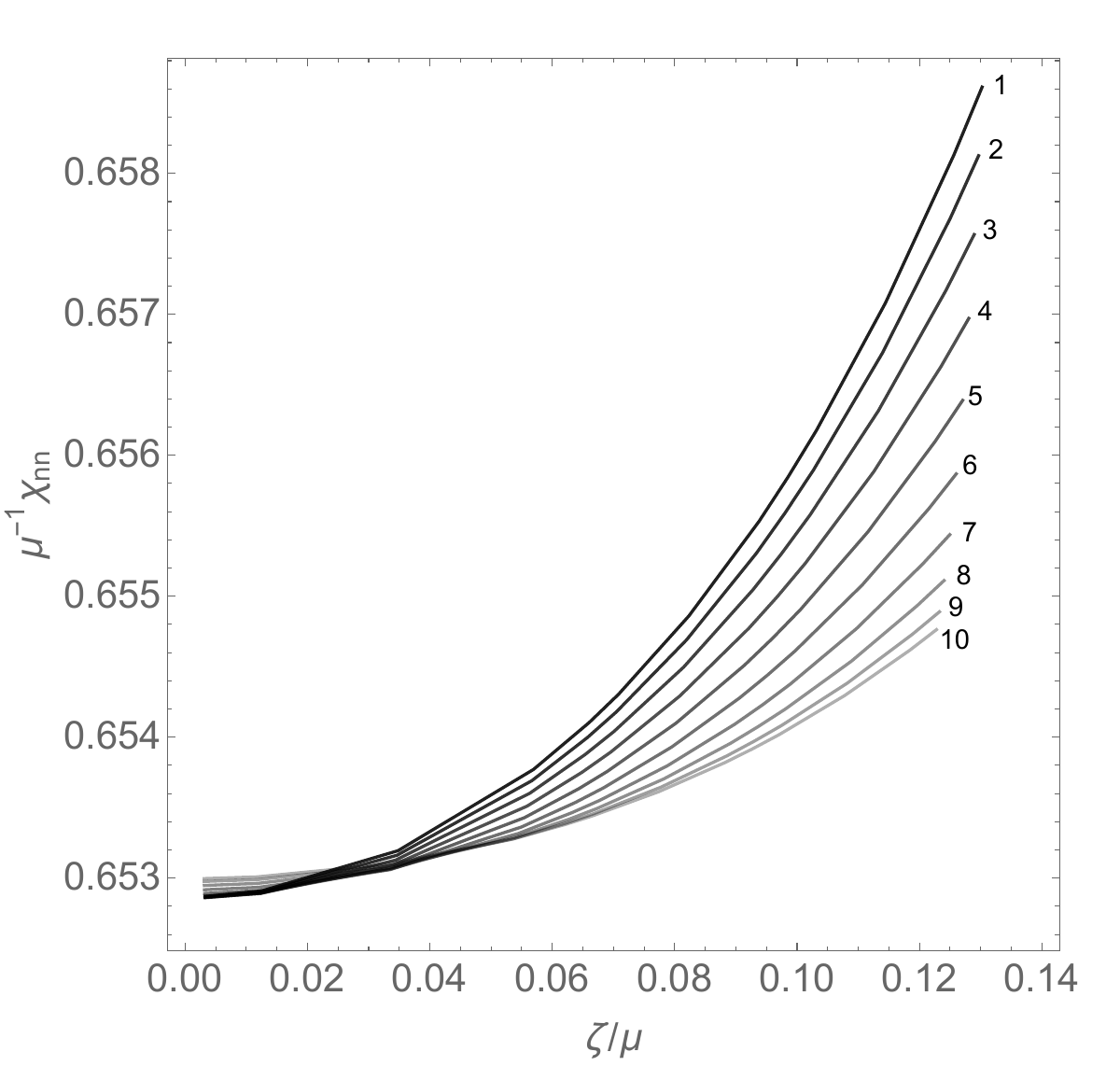}
\includegraphics[scale=.32]{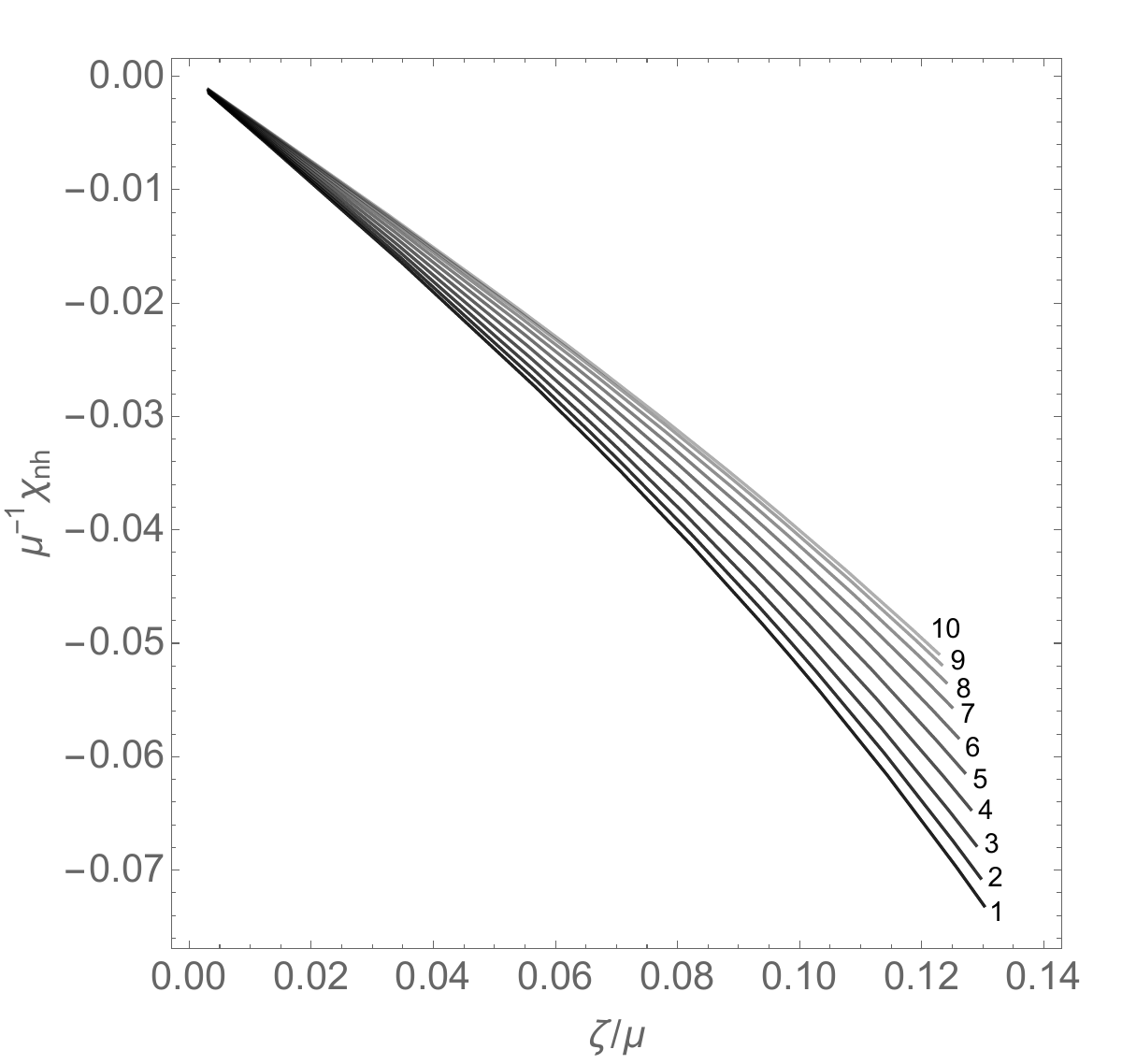}
\includegraphics[scale=.323]{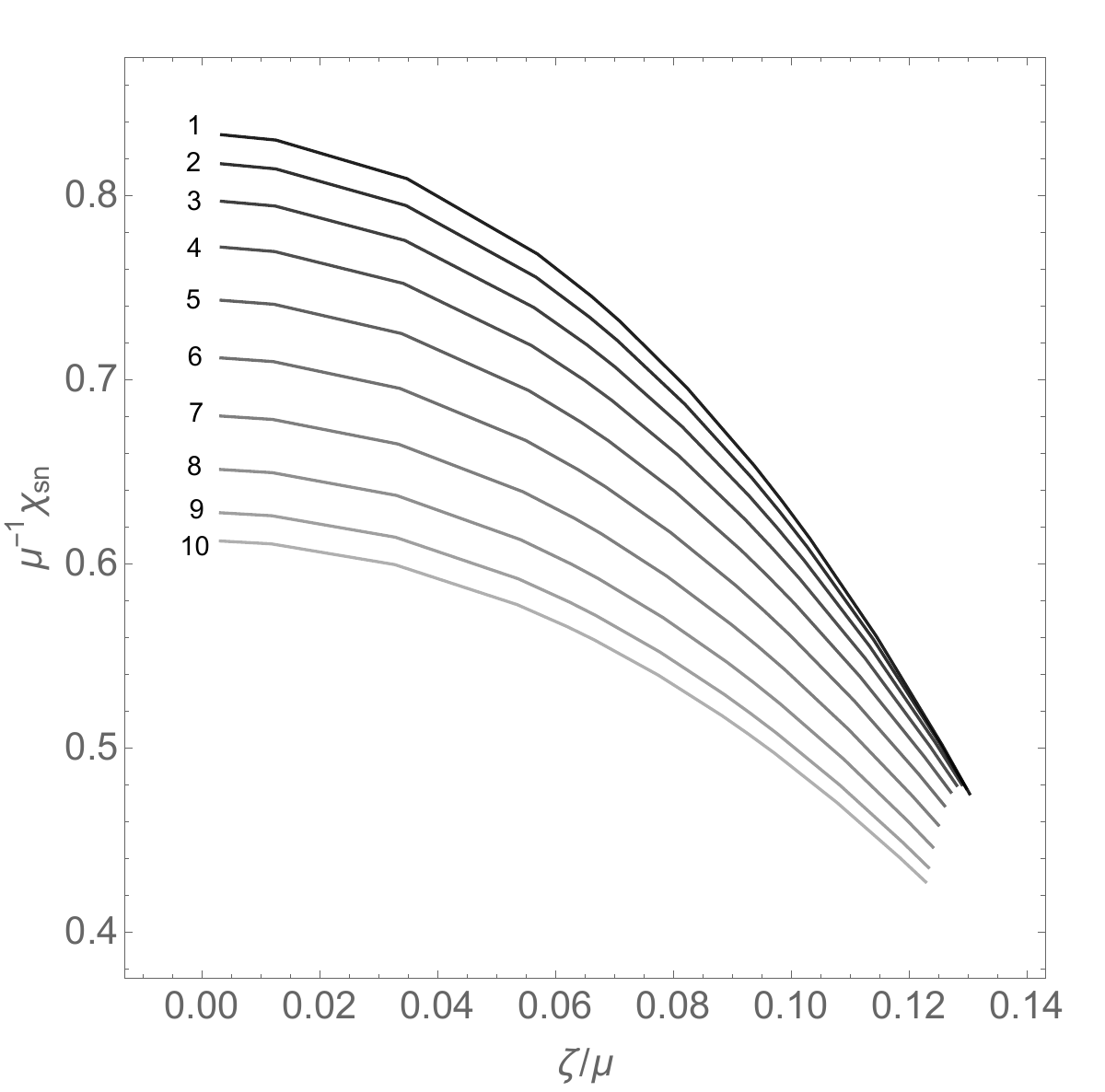}
\includegraphics[scale=.32]{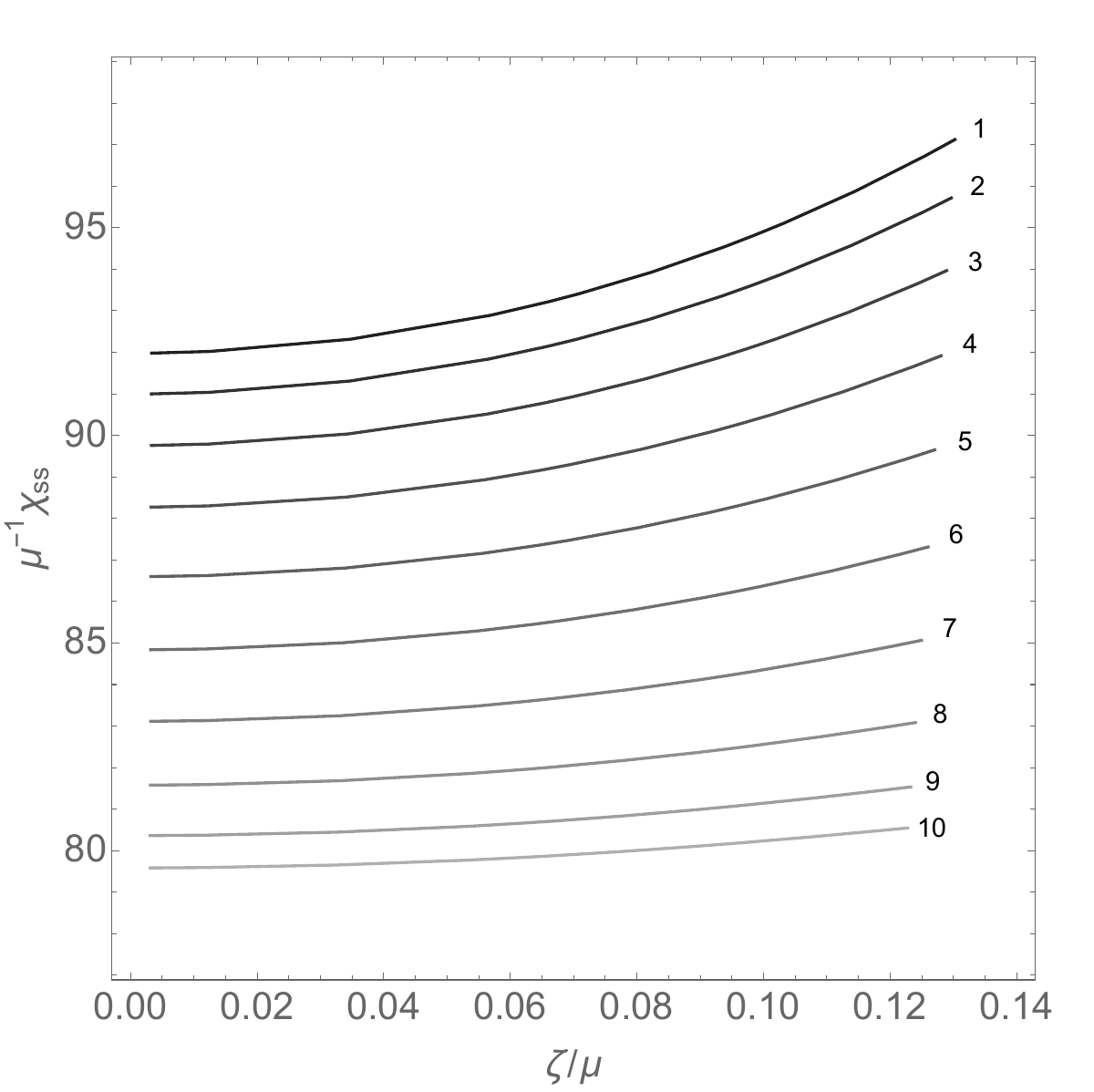}
\includegraphics[scale=.32]{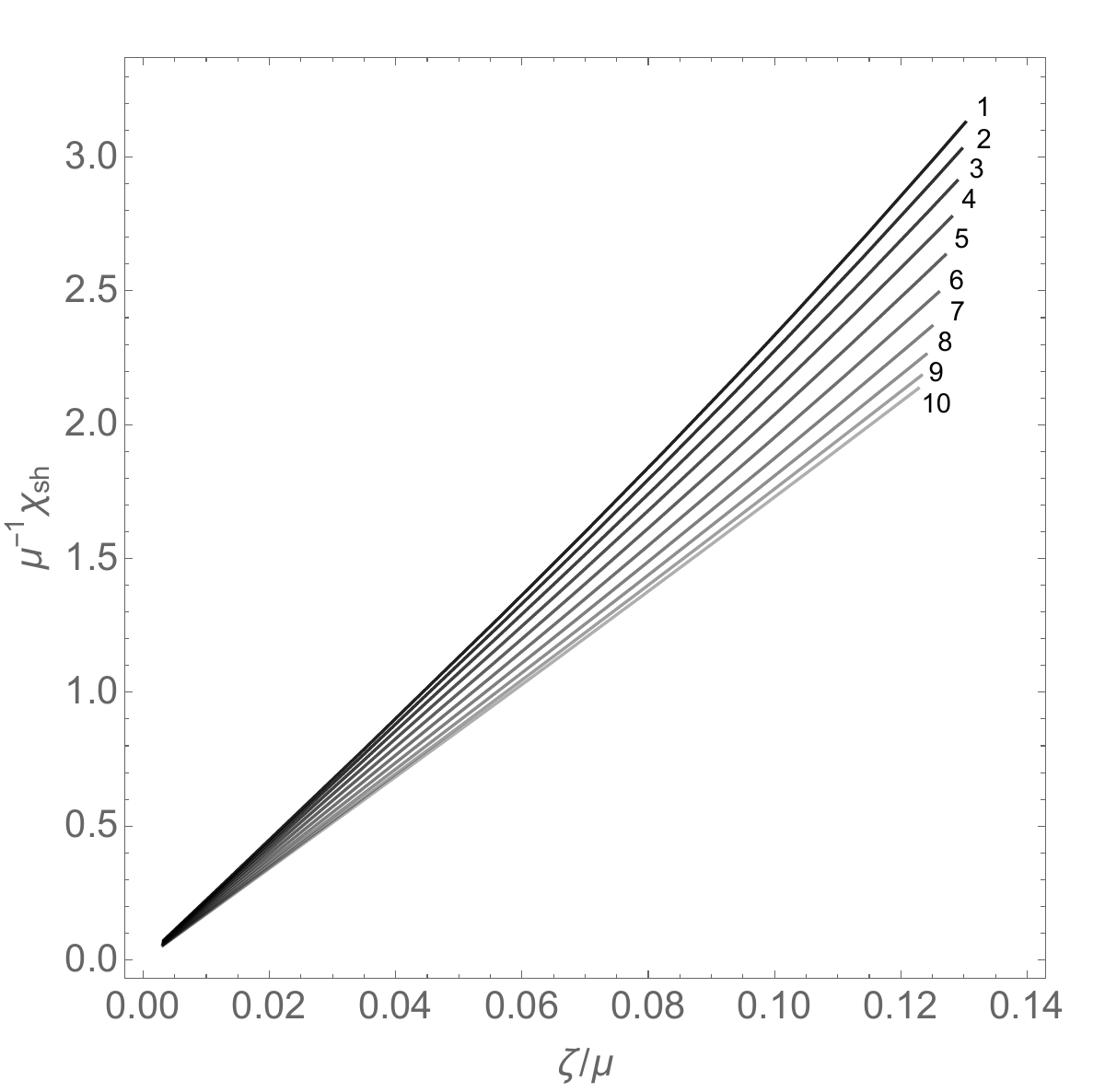}
\includegraphics[scale=.33]{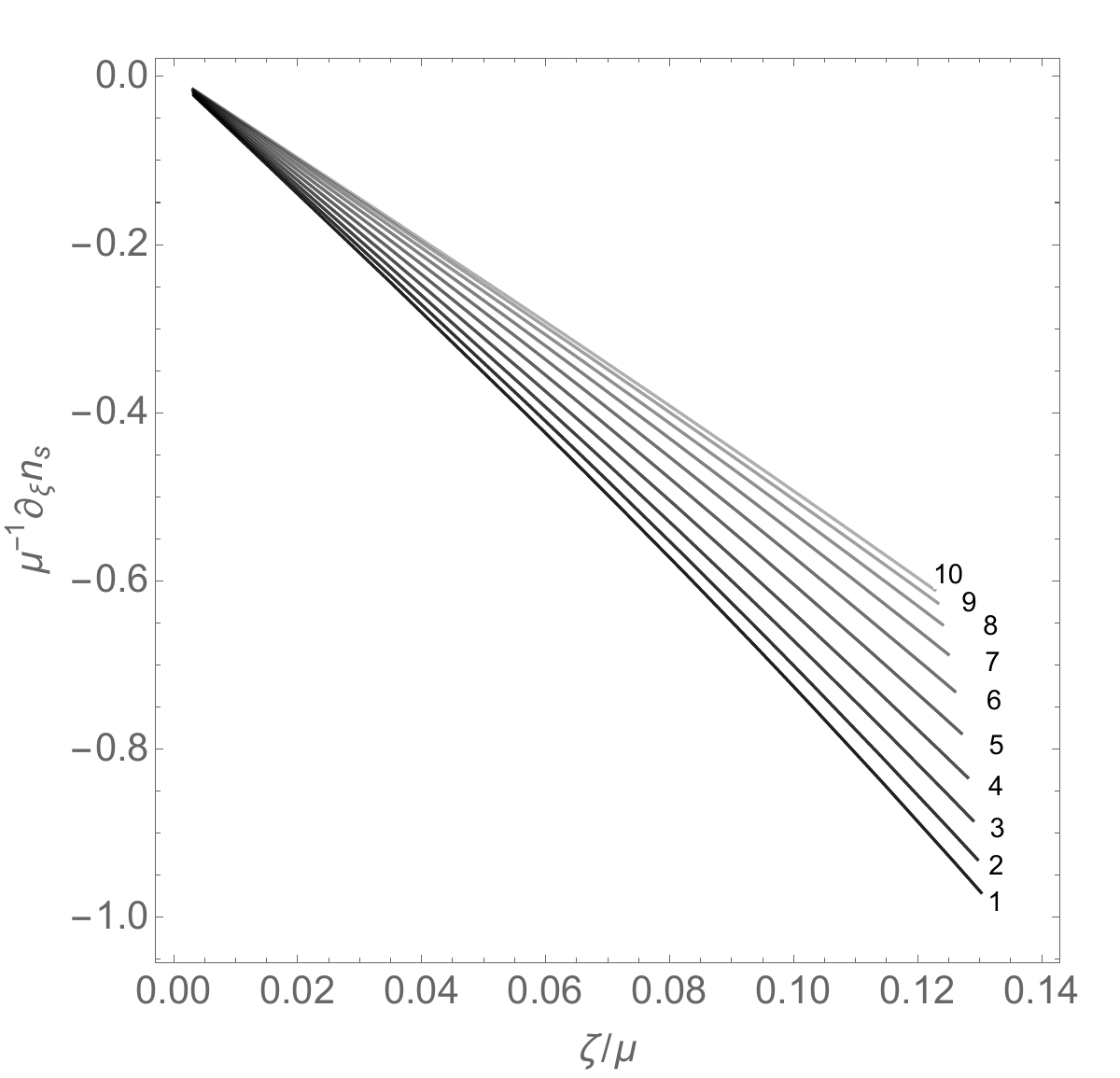}
\caption{\label{thermofigure2}Quantities appearing in the susceptibility matrix along the curves 1-10 in Figure~\ref{CutsThroughPhaseDiagram} that are used to calculate the hydrodynamic dispersion relations.}
\end{figure}

Using the Gibbs relation, we can establish
\begin{align}
\df \epsilon &= T\, \df s + \mu\, \df n + \frac{n_s}{2\mu}\, \df \zeta^2\,.
\end{align}
In Figures~\ref{thermofigure1} and \ref{thermofigure2}, we present results in the entropy ensemble. To compare to the susceptibilities used in Section~\ref{sec:linear}, one must perform a Legendre transform, leading to the identities
\begin{align}
\begin{split}
\chi_{\epsilon\epsilon} &= T^2\chi_{ss}+\mu T(\chi_{sn}+\chi_{ns})+\mu^2\chi_{nn}\bigr)\,,\\
\chi_{\epsilon n} = \chi_{n\epsilon} &= T\chi_{sn}+\mu  \chi_{nn}\,, \\
\chi_{\epsilon \zeta} = \chi_{\zeta \epsilon} & = T\chi_{sh}+\mu\chi_{nh} + \frac{n_s}{\mu}\zeta\,,
\end{split}
\end{align}
where
\begin{align}
\chi_{ss} = \frac{\partial s}{\partial T}\biggr|_{\mu,\zeta}, \;\; \chi_{sn} = \chi_{ns} = \frac{\partial n}{\partial T}\biggr|_{\mu,\zeta} =\frac{\partial s}{\partial \mu}\biggr|_{T,\zeta}, \;\; \chi_{nn} = \frac{\partial n}{\partial \mu}\biggr|_{T,\zeta}, \;\; \chi_{sh} = \frac{\partial s}{\partial\zeta}\biggr|_{\mu,T}, \;\; \chi_{nh} = \frac{\partial n}{\partial\zeta}\biggr|_{\mu,T}.
\end{align}

We have now specified how, given a $\mu$ and $\zeta$, we fully specify a bulk spacetime solution and how the subleading terms in (\ref{UVexpansion_equilibrium}) can be matched onto the parameters that appear in the equilibrium currents. By varying $\mu$ and $\zeta$, we can obtain an entire family of solutions and obtain the static susceptibility matrix for the holographic superfluid. In Figures~\ref{thermofigure1} and \ref{thermofigure2}, we plot the thermodynamic parameters for a wide range of $T/\mu$ and $\zeta/\mu$.

\subsection{Linearized fluctuations}
To identify the out of equilibrium constitutive relations, we consider linearized perturbations of the holographic superfluid. We work in radial gauge and consider plane wave perturbations with wavevector pointed along the superfluid velocity,
\begin{align}
	\begin{split}
&\delta \bar{\Psi} = (\sigma_r(r) - i\sigma_i(r))e^{-i(\omega t- k x)}\,,
\quad
\delta \bar{\Psi} = (\sigma_r(r) - i\sigma_i(r))e^{-i(\omega t- k x)}\,,\\
&\delta A_t = a_t(r) e^{-i(\omega t- k x)}\,, \qquad\quad
\delta A_x = a_x(r) e^{-i(\omega t- k x)},\quad\quad\;\;\,
\delta g_{tt} = -Dh_{tt}(r)e^{-i(\omega t- k x)},\\
%
&\delta g_{tx} = C_yh_{tx}(r)e^{-i(\omega t- k x)}\,,\quad
\delta g_{xx} = C_y h_{xx}(r)e^{-i(\omega t- k x)},\quad
\delta g_{yy} = C_y h_{yy}(r)e^{-i(\omega t- k x)} \,.
\label{eq:holflucts}
	\end{split}
\end{align}
For future reference, for fluctuations about the background state with $\Psi = \bar{\Psi}=\psi$,
\begin{align}
\delta\Psi = \delta \psi + iQ\psi \delta\varphi, \quad \delta\bar{\Psi} = \delta \psi - iQ\psi \delta\varphi.
\end{align}
Here, we can identify
\begin{align}
\sigma_r = 2\delta \psi, \quad \sigma_i = 2Q\psi \delta\varphi.
\end{align}
The equations of motion for these fluctuations are easily obtained but messy, so we omit their explicit expressions here. Nevertheless, we find that 4 of the equations are second order and 4 are first order, necessitating that we specify 12 boundary conditions. We now explain how to do so.

To describe the relaxation back to thermal equilibrium, we care about fluctuations which decay in time. Holographically, this is equivalent to imposing ingoing boundary conditions at the black hole horizon. This amounts to the following near horizon expansions, 
\begin{align}
\begin{split}
a_t(r) &= (r-r_h)^{i\omega/4\pi T}[a_{t}^h(r-r_h)+O(r-r_h)^2],\\
a_x(r) &= (r-r_h)^{i\omega/4\pi T}[a_x^h+O(r-r_h)],\\
\sigma_r(r) &=(r-r_h)^{i\omega/4\pi T}[\sigma_r^h+O(r-r_h)],\\
\sigma_i(r) &=(r-r_h)^{i\omega/4\pi T}[\sigma_i^h+O(r-r_h)],\\
h_{tt}(r) &=(r-r_h)^{i\omega/4\pi T}[h_{tt}^h(r-r_h)+O(r-r_h)^2],\\
h_{tx}(r) &=(r-r_h)^{i\omega/4\pi T}[h_{tx}^h(r-r_h)+O(r-r_h)^2],\\
h_{xx}(r) &=(r-r_h)^{i\omega/4\pi T}[h_{xx}^h+O(r-r_h)],\\
h_{yy}(r) &=(r-r_h)^{i\omega/4\pi T}[h_{yy}^h+O(r-r_h)] \,.
\end{split}
\end{align}
Not all of these constants are independent,
\begin{align}
\begin{split}
h_{tt}^h =&  \left(Q\psi_h\frac{8\pi T\bigl[-ik A_x^h-2(A_t^h)^2C_x^h\bigr]+i\omega\bigl[2A_t^h C_x^h + A_x^h C_{tx}^h\bigr]}{C_x^h(4\pi T-i\omega)(6\pi T-i\omega)}\right)\sigma_i^h\\
&+\left(s^2\frac{16\pi T\bigl[4k^2\pi T - (4\pi T+i\omega)Q^2(A_x^h)^2\psi_h^2\bigr]+\omega C_{tx}^h\bigl[-16\pi T k + \omega C_{tx}^h\bigr]}{256\pi^3(C_x^h)^3(4\pi T-i\omega)(6\pi T-i\omega)}\right)h_{xx}^h\\
&+\left(2\psi_h\frac{(A_x^h)^2Q^2(4\pi T+i\omega)- 2iC_x^h\omega}{C_x^h(4\pi T-i\omega)(6\pi T-i\omega)}\right)\sigma_r^h+\left(\frac{8\pi T(ikA_t^h + Q^2A_x^h\psi_h^2)-i\omega C_{tx}^hA_t^h}{C_x^h(4\pi T-i\omega)(6\pi T-i\omega)}\right)a_x^h \, , \nonumber
\end{split}
\end{align}
\begin{align}
\begin{split}
h_{tx}^h =& -\left(\frac{64\pi^3 T A_t^h}{s^2(4\pi T-i\omega)}\right)a_x^h + \left(i\frac{8\pi T k - \omega C_{tx}^h}{2C_x^h(4\pi T-i\omega)}\right)h_{xx}^h + \left(\frac{128 \pi^3 Q C_x^h A_x^h \psi_h}{s^2(4\pi T-i\omega)}\right)\sigma_i^h\,, \\
h_{yy}^h =& -\left(\frac{s}{4\pi C_x^h}\right)^2h_{xx}^h\,,\\
a_t^h =& \left(i\frac{8\pi T k - \omega C_{tx}^h}{2C_x^h(4\pi T-i\omega)}\right)a_x^h -\left(\frac{8\pi T Q \psi_h}{4\pi T-i\omega}\right)\sigma_i^h.
\end{split}
\end{align}
Fixing the radial dependence of the fluctuations to take an ingoing form amounts to 8 boundary conditions. In addition, we see that we are allowed to choose four independent constants, e.g. $\sigma_{i}^h, \sigma_r^h, g_{xx}^h$ and $a_x^h$. This fully specifies the solutions of the linearized equations of motion. 

In the UV, the fields have the expansion
\begin{align}
\begin{split}
a_t &= a_t^{(0)} + \frac{a_t^{(1)}}{r} + O(r^{-2}),\hspace{42pt}
a_x = a_x^{(0)} + \frac{a_x^{(1)}}{r} + O(r^{-2})\,,\\
\sigma_r &= \frac{\sigma_r^{(0)}}{r} + \frac{\sigma_r^{(1)}}{r^2}+O(r^{-3}),\hspace{42pt}
\sigma_i = \frac{\sigma_i^{(0)}}{r} + \frac{\sigma_i^{(1)}}{r^2}+O(r^{-3})\,,\\
h_{tt} & = h_{tt}^{(0)} + \frac{h_{tt}^{(2)}}{r^2} +\frac{h_{tt}^{(3)}}{r^3}+O(r^{-4})\,,\;\;
h_{tx}  = h_{tx}^{(0)} + \frac{h_{tx}^{(2)}}{r^2} +\frac{h_{tx}^{(3)}}{r^3}+O(r^{-4})\,,\\
h_{xx} & = h_{xx}^{(0)} + \frac{h_{xx}^{(2)}}{r^2} +\frac{h_{xx}^{(3)}}{r^3}+O(r^{-4})\,,\;\;
h_{yy}  = h_{yy}^{(0)} + \frac{h_{yy}^{(2)}}{r^2} +\frac{h_{yy}^{(3)}}{r^3}+O(r^{-4}) \,.
\end{split}
\end{align}
These expansions imply that the phase field behaves as
\begin{align}
\delta\varphi &= \frac{1}{Q\langle O^\psi \rangle}\bigl(\sigma_{i}^{(0)}r+\sigma_{i}^{(1)}+O(r^{-1})\bigr)\,.
\end{align}
There are potential terms at $\mathcal{O}(r^0)$ which can appear if $\psi_s\neq 0$, but we will never consider this case. Using this expansion for $\delta\varphi$ and (\ref{actionvariation}), we identify
\begin{align}
K = iQ\psi\biggl[n^N\partial_N(\bar{\Psi} -\Psi)+(\bar{\Psi} -\Psi)\biggr].
\end{align}
Expanding the fields near the boundary, we have 
\begin{align}
\delta K = - Q\langle O^\psi \rangle \sigma_i^{(0)}, \quad \delta \langle \varphi \rangle = \frac{\sigma_i^{(1)}}{Q\langle O\psi \rangle}.
\end{align}
Variations of the other currents are obtained similarly
\begin{align}
\label{deltaCurrentIdentities}
\begin{split}
\delta \langle J^t (\omega,k) \rangle &= -a_t^{(1)} +n\,h_{tt}^{(0)}-\frac{\zeta n_s}{\mu}\,h_{tx}^{(0)}\,,\\
\delta \langle J^x (\omega,k) \rangle &= a_x^{(1)} +\frac{\zeta n_s}{\mu}\,h_{xx}^{(0)}+n\,h_{tx}^{(0)}\,,\\
\delta \langle T^{tt} (\omega,k) \rangle &= -3h_{tt}^{(3)} + 2\zeta n_s\, h_{tx}^{(0)}- \epsilon\,h_{tt}^{(0)}-\langle O^\psi \rangle \sigma_r^{(0)}\,,\\
\delta \langle T^{tx} (\omega,k) \rangle &= -3h_{tx}^{(3)} - \zeta n_s\,(h_{xx}^{(0)}+h_{tt}^{(0)})-2p\, h_{tx}^{(0)}\,,\\
\delta \langle T^{xt} (\omega,k) \rangle &= -3h_{tx}^{(3)} - \zeta n_s\,(h_{xx}^{(0)}+h_{tt}^{(0)})-2p\, h_{tx}^{(0)}\,,\\
\delta \langle T^{xx} (\omega,k) \rangle &= 3 h_{xx}^{(3)} - 2\zeta n_s\, h_{tx}^{(0)} + (3p-2\epsilon) \, h_{xx}^{(0)} + \langle O^\psi \rangle \sigma_r^{(0)}\,, \\
\delta \langle T^{yy} (\omega,k) \rangle &= 3 h_{yy}^{(3)} - p\, h_{yy}^{(0)} + \langle O^\psi \rangle \sigma_r^{(0)}\,,\\
\delta \langle O^\psi (\omega,k) \rangle &= 2(\sigma_r^{(1)}+i\sigma_i^{(1)})\,,\\
\delta \langle \bar{O}^\psi (\omega,k)\rangle &= 2(\sigma_r^{(1)}-i\sigma_i^{(1)})\,.
\end{split}
\end{align}
The variation of the currents contain pieces that are proportional to the leading term in the asymptotic expansion for the fluctuations (i.e. terms with a superscript ``(0)"). These contribute contact terms in the response functions and are responsible for enforcing the Ward identities. To see this, we note that as $r\to \infty$, the equations of motion impose that
\begin{align}
\begin{split}
0=&\,3\mu\bigl(h_{tt}^{(3)}+h_{xx}^{(3)}+h_{yy}^{(3)}\bigr)-\zeta n_s\bigl(\zeta h_{xx}^{(0)}+2\mu h_{tx}^{(0)}\bigr) + 2\mu \langle O^\psi \rangle \sigma_r^{(0)},\\
0=&\,6\mu\bigl(\omega h_{tx}^{(3)}+k h_{xx}^{(3)}\bigr) +2\rho \mu \bigl(ka_t^{(0)}+\omega a_x^{(0)}\bigr) + 2iQ\mu\zeta\langle O^\psi\rangle\sigma_i^{(0)}+ \mu h_{tt}^{(0)}\biggl[k\bigl(p-2\epsilon\bigr)-\zeta n_s \omega\biggr] \\
& + \zeta n_s h_{xx}^{(0)}\bigl(2\zeta k + \mu \omega\bigr) + 2\zeta n_s h_{tx}^{(0)}\bigl(2\mu k + \zeta \omega\bigr) + \zeta n_s h_{yy}^{(0)}\bigl(\mu \omega - k \zeta\bigr)\,,\\
0=&\,6\mu\bigl(\omega h_{tt}^{(3)} - k h_{tx}^{(3)}\bigr) + 2\zeta n_s \bigl(k a_t^{(0)}+\omega a_x^{(0)}\bigr) - 2iQ\mu^2 \langle O^\psi \rangle \sigma_i^{(0)}-\zeta \mu n_s k h_{tt}^{(0)}\\
&+\mu h_{xx}^{(0)}\biggl[\omega\bigl(2\epsilon-p\bigr)+k\zeta n_s\biggr]+2\mu h_{tx}^{(0)}\biggl[2\zeta n_s \omega + k\bigl(p+\epsilon\bigr)\biggr]+\mu h_{yy}^{(0)}\biggl[\omega\bigl(p+\epsilon\bigr)-\zeta n_s k\biggr]\,,\\
0=&\,2\mu\bigl(\omega a_t^{(1)} + k a_x^{(1)}\bigr) -i2\mu Q\langle O^\psi \rangle \sigma_i^{(0)} + \bigl(h_{xx}^{(0)}-h_{tt}^{(0)}\bigr)\bigl(\omega \mu n + k \zeta n_s\bigr) \\
&-2h_{tx}^{(0)}\bigl(\omega \zeta n_s + k \mu n\bigr)-h_{yy}^{(0)}\bigl(\omega \mu n - k \zeta n_s\bigr)\,.
\end{split}
\end{align}
From these relations, one can see
\begin{align}
\begin{split}
0&=\eta_{\mu\nu}\delta \langle T^{\mu\nu} \rangle -g_{tt}^{(0)} \langle T^{tx}\rangle + 2g_{tx}^{(0)}\langle T^{tx} \rangle + g_{xx}^{(0)}\langle T^{xx} \rangle + g_{yy}^{(0)} \langle T^{yy}\rangle - \langle O^\psi \rangle \sigma_r^{(0)}\\
&=\delta \langle T^{\mu}_{\;\;\mu} \rangle - \delta\psi \langle O^\psi\rangle\,.
\end{split}
\end{align}
In the second line, we have identified $\delta\psi_s = \sigma_{r}^{(0)}$ as the fluctuation of the source for the condensate $\langle O^\psi \rangle$ which explicitly breaks conformal invariance. Furthermore, using the definition of $\delta J^\mu$ and $\delta K$ gives the variation of the Ward identity
\begin{align}
\delta\langle\nabla_\mu J^\mu-K\rangle = 0\,.
\end{align}
Finally, using the definitions of $\delta \langle T^{\mu\nu} \rangle$, we can check that
\begin{align}
\begin{split}
0&=\partial_\mu \delta T^{\mu\nu} + \delta \Gamma^{\mu}_{\mu\lambda}\langle T^{\lambda\nu}\rangle  + \delta \Gamma^{\nu}_{\mu\lambda}\langle T^{\mu\lambda}\rangle - \delta F^{\nu\mu}\langle J_\mu \rangle - \langle \xi^\nu \rangle \delta K - \langle O^\psi \rangle \partial^\nu\psi_s \\
&=\delta\langle\nabla_\mu T^{\mu\nu}  - F^{\nu\mu}J_\mu - K\xi^\nu - O^\psi \partial^\nu \psi_s\rangle.
\end{split}
\end{align}
As we did not consider operators which explicitly break conformal invariance in our hydrodynamic theory, the last term does not appear in (\ref{conseq}). By choosing $\delta\psi_s = 0$, our results will overlap with the hydrodynamics presented earlier.

While they are important for ensuring that the linearized fluctuations of the holographic superconductor respect the perturbations of the Ward identities, contact terms can differ between various hydrodynamic approaches to obtaining the retarded Green's functions, potentially introducing ambiguities into Kubo formulae.\footnote{The two main approaches are the canonical, i.e. Kadanoff-Martin approach \cite{KadanoffandMartin} which we outlined in Section \ref{sec:linear} and the variational approach \cite{Kovtun:2012rj}. Holographic computations follow the variational approach.} In particular, when computing Kubo formulae, we should always consider
\begin{align}
\tilde{G}^{R}_{AB}(\omega,q) = G^R_{AB}(\omega, q) - G^R_{AB}(\omega = 0, q)
\end{align}
which will give rise to the same transport coefficient independent of approach. We find that for this quantity, we can neglect the contact terms entirely and instead consider variations of the fields $\delta v^I$ defined by
\begin{align}
\begin{split}
\delta v^I &= \{-3h^{(3)}_{tt}, \; -3h^{(3)}_{tx}, \;3h^{(3)}_{xx}, \;3h^{(3)}_{yy} , \; -a_t^{(1)}, \;a_x^{(1)}, \; 2(\sigma_{r}^{(1)}+i\sigma_{i}^{(1)}), \; 2(\sigma_{r}^{(1)}-i\sigma_{i}^{(1)})\}^I
\end{split}
\end{align}
with respect to sources $\delta s^I$ defined by
\begin{align}
\delta s^I &= \{-\frac{h^{(0)}_{tt}}{2},\; h_{tx}^{(0)}, \; \frac{h_{xx}^{(0)}}{2}, \; \frac{h_{yy}^{(0)}}{2}, \; a_t^{(0)}, \; a_x^{(0)}, \; \sigma_{r}^{(0)}+i\sigma_{i}^{(0)}, \;\sigma_{r}^{(0)}-i\sigma_{i}^{(0)}\}^I.
\end{align}
In other words,
\begin{align}
\tilde{G}^R_{IJ} = -\frac{\delta v^I}{\delta s^J}.
\end{align}

\begin{figure}[h!]
\centering
\includegraphics[width=.4\textwidth]{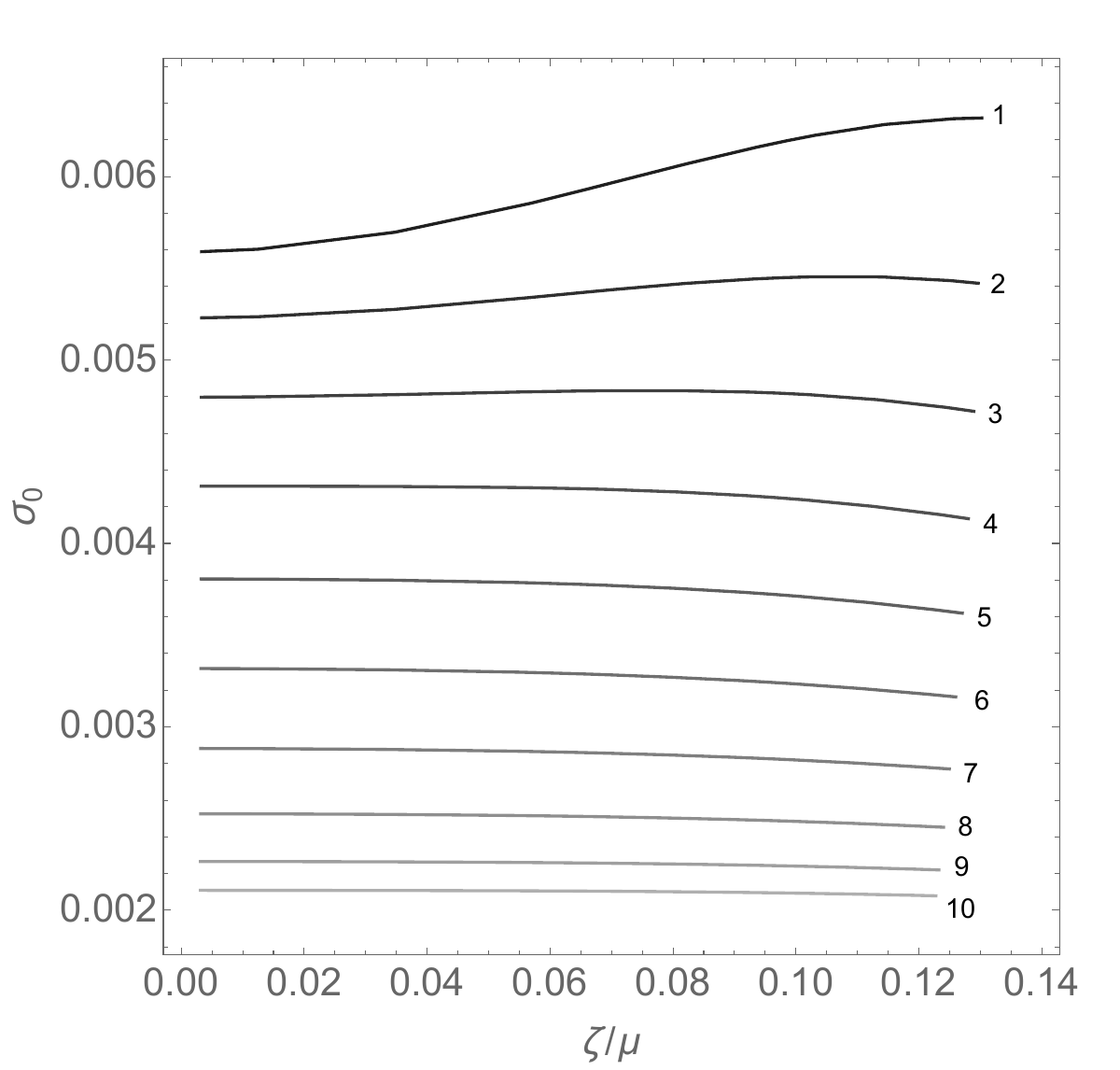}
\includegraphics[width=.41\textwidth]{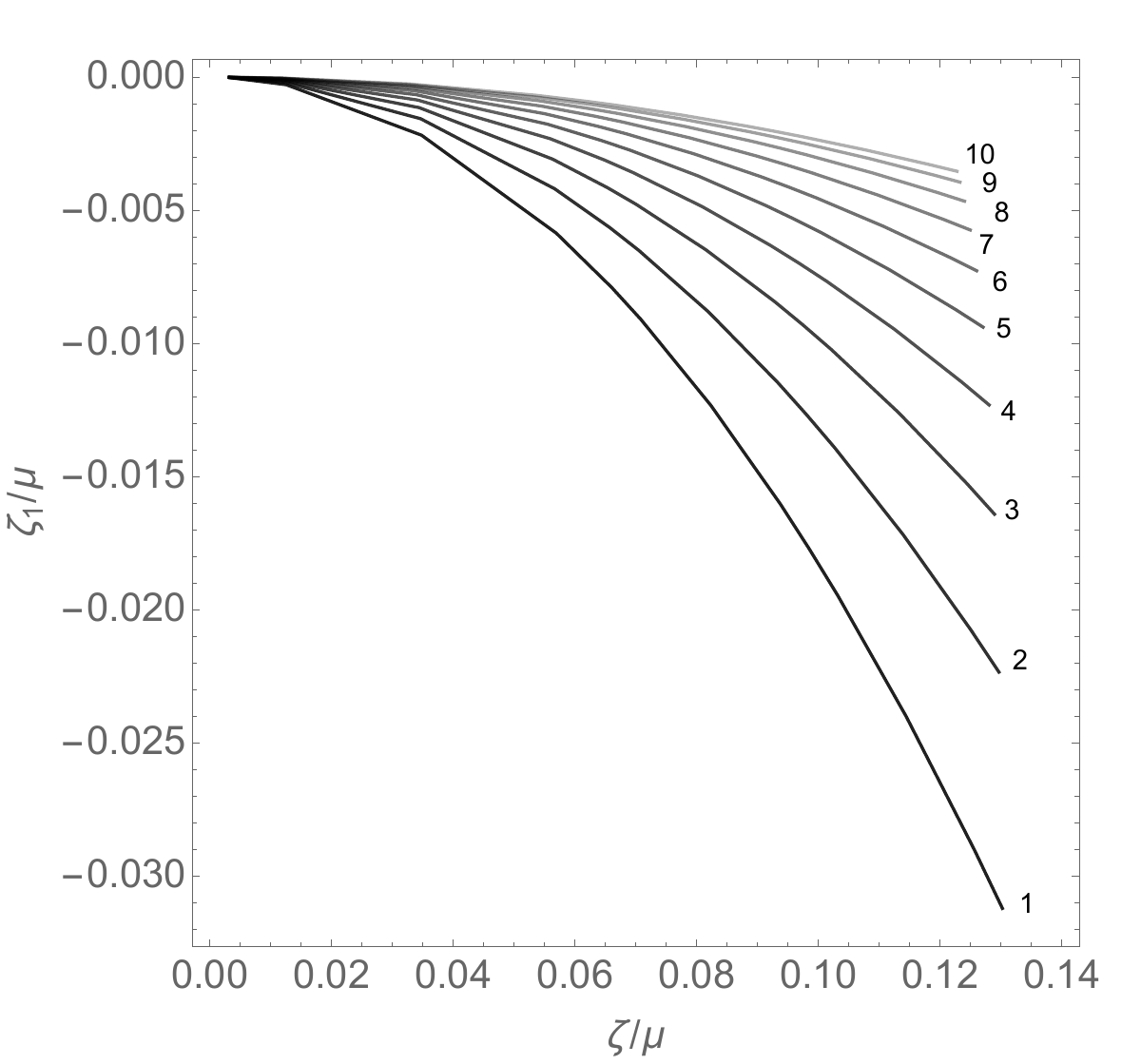}
\includegraphics[width=.42\textwidth]{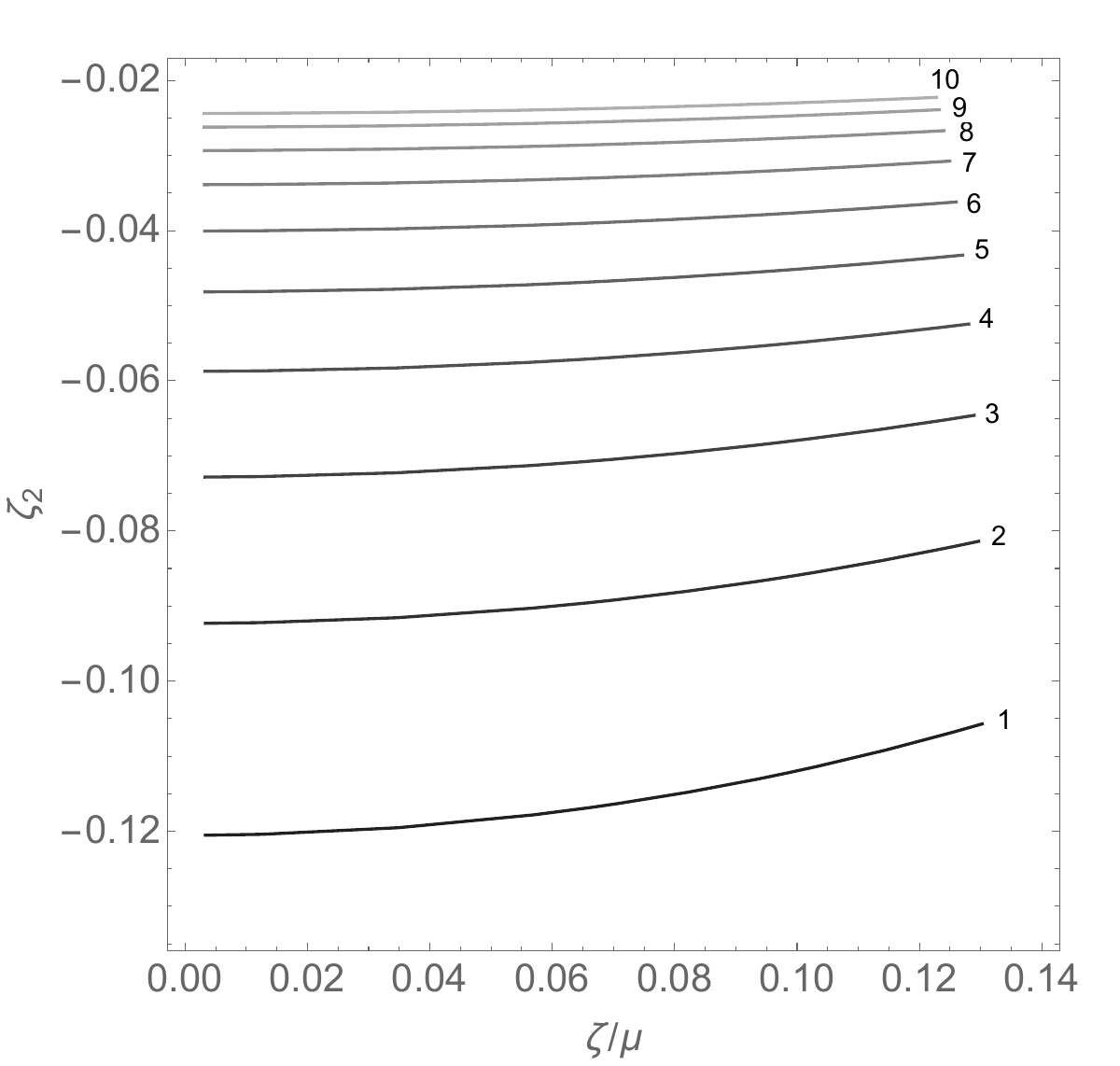}
\includegraphics[width=.4\textwidth]{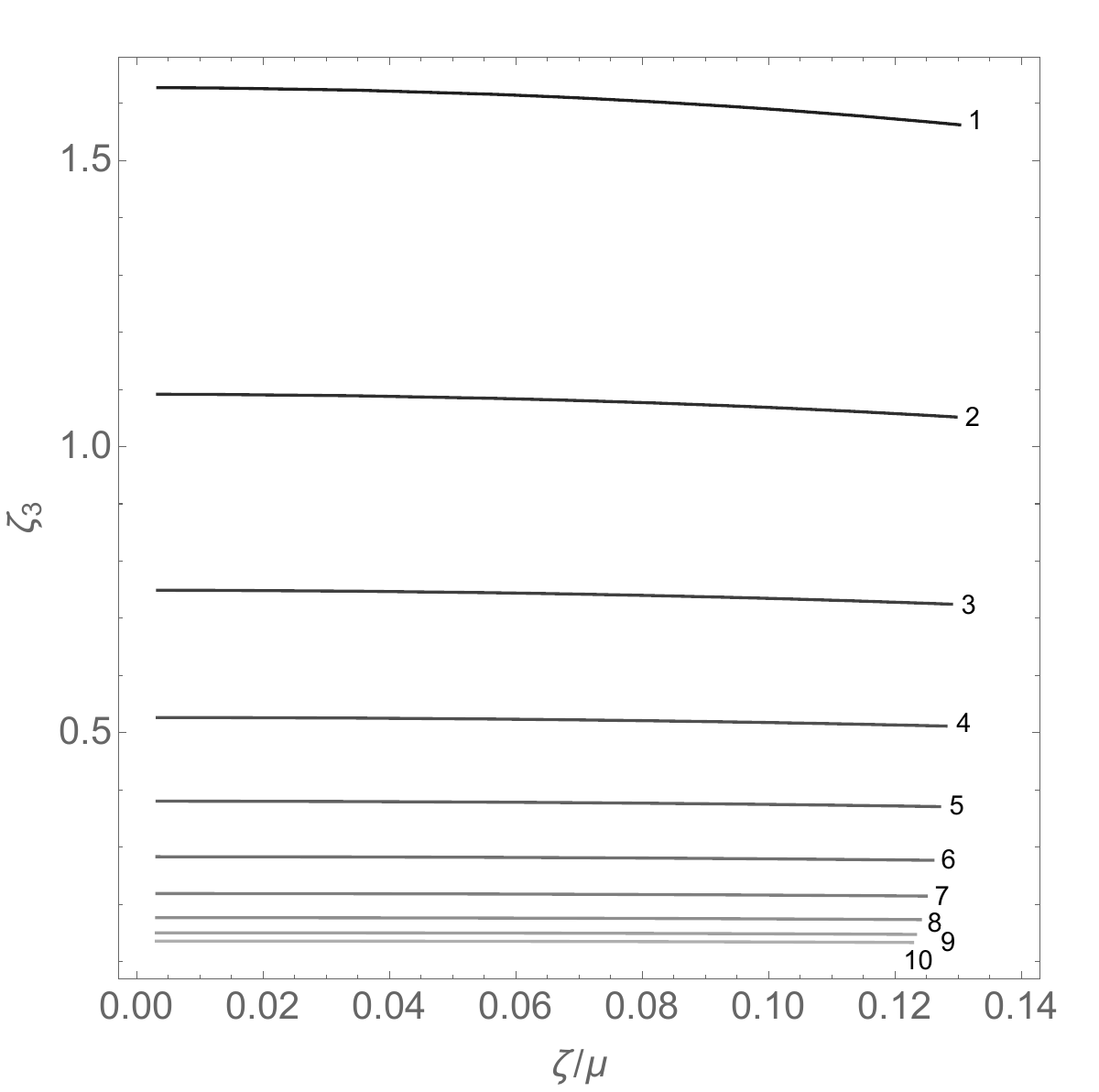}
\includegraphics[width=.44\textwidth]{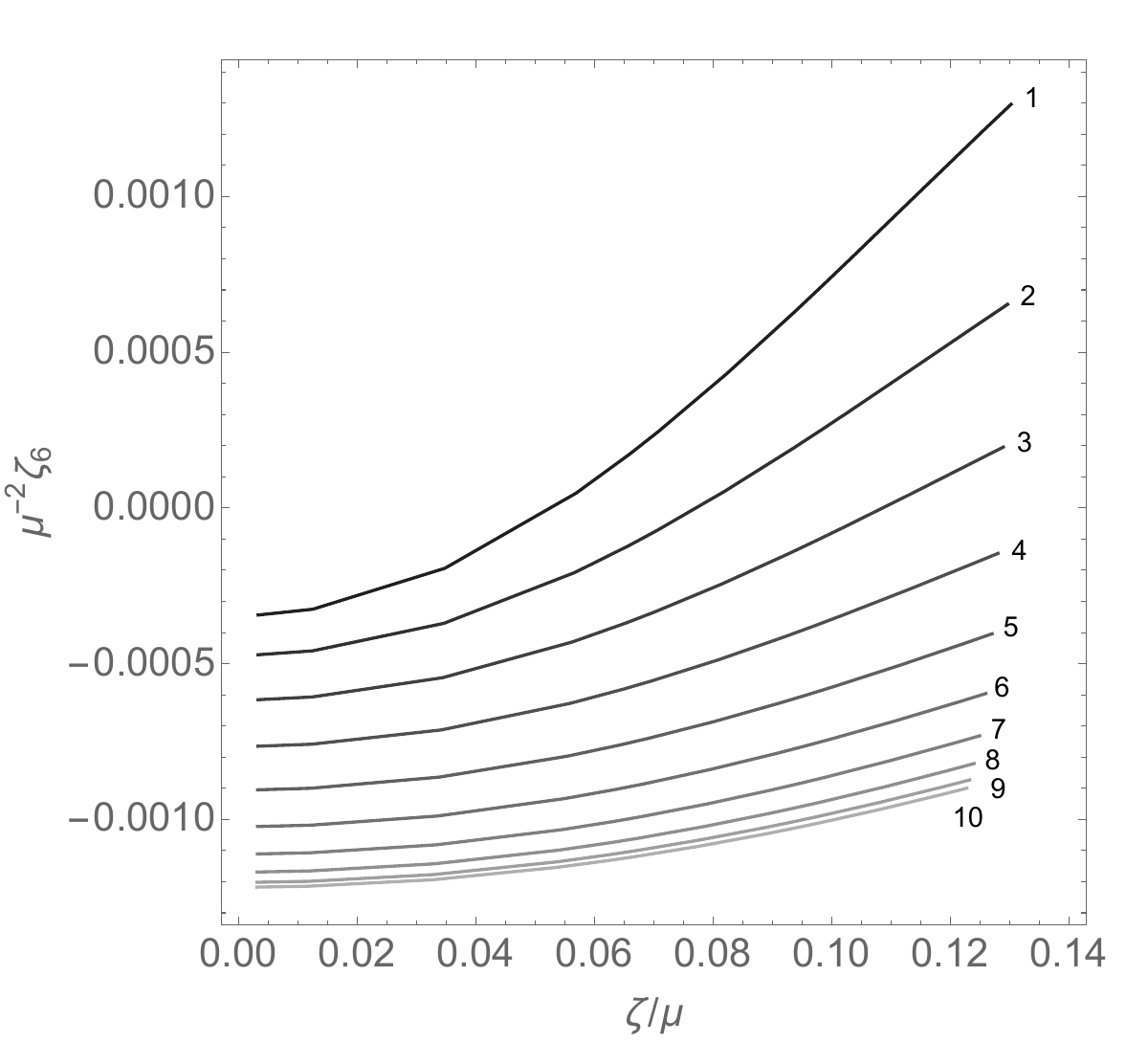}
\includegraphics[width=.42\textwidth]{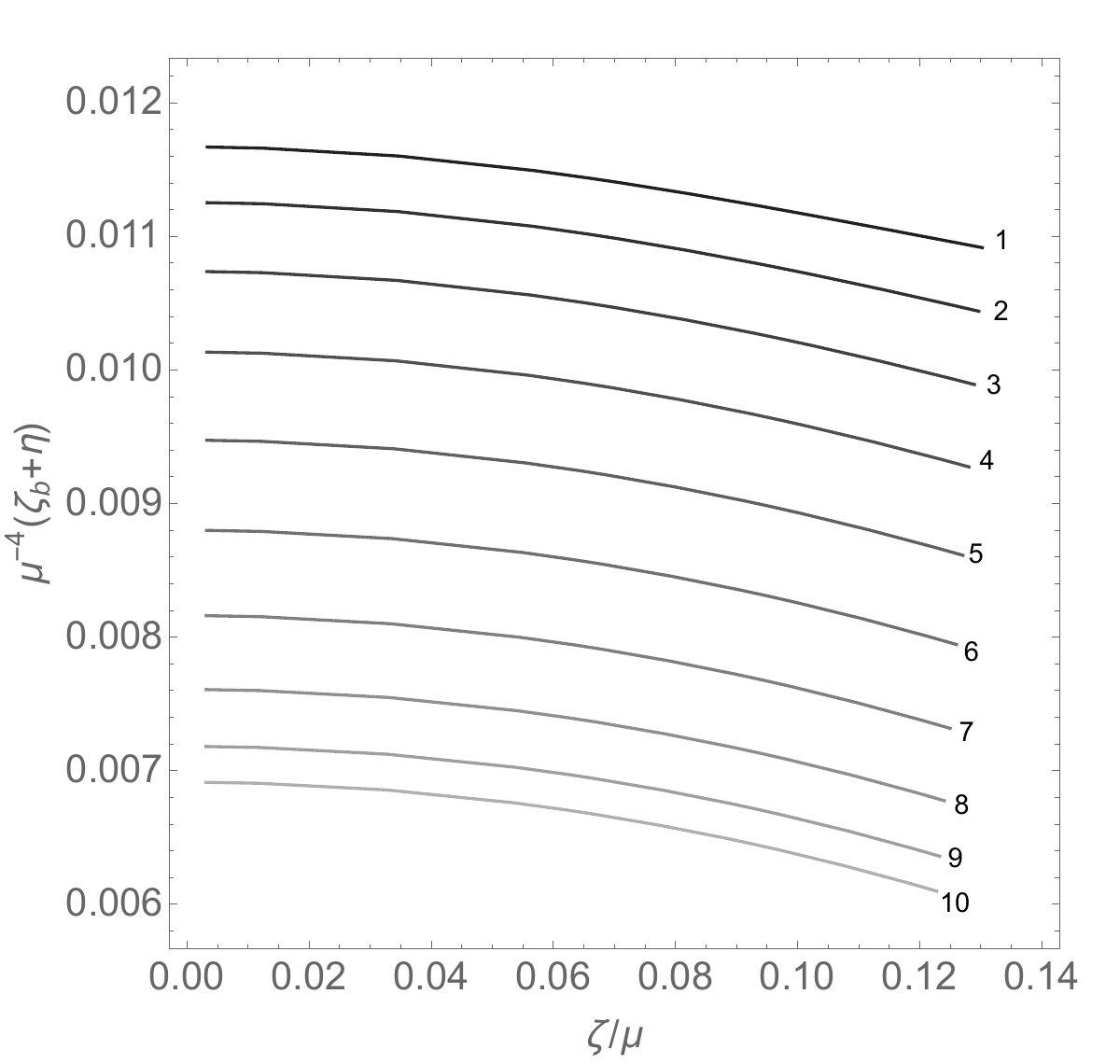}
\caption{\label{dissipativefigure}The dissipative quantities along the curves 1-10 in Figure~\ref{CutsThroughPhaseDiagram} that are used to calculate the hydrodynamic dispersion relations. Note that $\eta=s/4\pi$.}
\end{figure}

When working in radial gauge, we have shown that there are only four linearly independent solutions (specified by $a_x^h, \; g_{xx}^h, \; \sigma_r^h, \; \sigma_i^h$) with ingoing boundary conditions which is less than our eight sources, $\delta s^I$. In order to find the sufficient number of independent solutions, we supplement our ingoing solutions with the pure gauge solutions,
\begin{align}
\delta A_M &= \pounds_\beta A_M + \partial_M \Lambda, \quad \delta g_{MN} = \nabla_M \beta_N + \nabla_N \beta_M, \nonumber\\
\delta\Psi &= \beta^M\partial_M\psi+iQ\psi\Lambda, \quad \delta\bar{\Psi} = \beta^M\partial_M\psi - iQ\psi \Lambda.
\end{align}
which are defined in terms of four functions $\beta^M = \{\beta^{t}, \beta^x, 0, \beta^r\}^M$ and $\Lambda$. We take these to have the following behavior
\begin{align}
\beta^t = c_t e^{-i(\omega t- kx)}, \quad \beta^x = c_x e^{-i(\omega t- kx)}, \quad \beta^r = c_r/\sqrt{D(r)},\quad \Lambda = \lambda e^{-i(\omega t-kx)}.
\end{align}

The four pure gauge solutions and the four solutions specified by $\sigma_i^h, \sigma_r^h, a_x^h, g_{xx}^h$ form eight linearly independent solutions which we label $\vec{X}^I(r)$. Here, $I=1,...,8$ labels one of the solutions and $\vec{X}(r) = \{a_\mu(r), h_{\mu\nu}(r), \sigma_i(r), \sigma_r(r)\}$ is a vector of the linearized fluctuations corresponding to that solution (in other words, $\vec{X}^I(r)$ is an $8\times8$ dimensional object). Since we are only considering linearized fluctuations, a solution, $Y^I(r)$, corresponding to a particular configuration of sources $\delta s^I$ in the UV can be obtained by the linear transformation
\begin{align}
Y^I(r) = C_{IJ}(\delta s^I)X^J(r).
\end{align}
It is clear from this expression that the $C_{IJ}$ are functions of the sources. By considering solutions with all but one source non-vanishing $\delta s^I = \delta^I_J$, and identifying $\delta v^I$ by expanding the solutions near $r\to\infty$ we can then obtain
\begin{align}
\label{subtractedGreensFunctions}
\delta v^I = -\tilde{G}^R_{IJ}(\omega,k)\delta s^J.
\end{align}
The transport coefficients are then simply obtained by taking the appropriate limits of this expression, for instance
\begin{align}
\sigma_0 = - \lim_{\omega\to0,k\to 0}\frac{1}{\omega}\text{Im}\;\tilde{G}^R_{J_xJ_x}(\omega,k).
\end{align}
In Appendix \ref{numericsappendix} we give more details on the construction of the correlators and the numerical simulations leading to the results shown in this section.

In Figure \ref{dissipativefigure}, we plot the dissipative transport coefficients for a wide range of $T/\mu$ and $\zeta/\mu$. These results could also be obtained working directly at zero frequency, along the lines of \cite{Donos:2021pkk,Donos:2022www}.

Note that in addition to producing contact terms which enforce the Ward identities, the holographic retarded Green's functions also capture the non-linear physics of the bulk Einstein equations. It is only in the low energy and small wavelength limit, $\omega/T, k/T \lesssim 1$, where linear response is appropriate, that we expect the results to match the hydrodynamic expressions. 

\subsection{Quasinormal modes}

\begin{figure}[t!]

\centering
\begin{subfigure}{\textwidth}
\includegraphics[scale=.35]{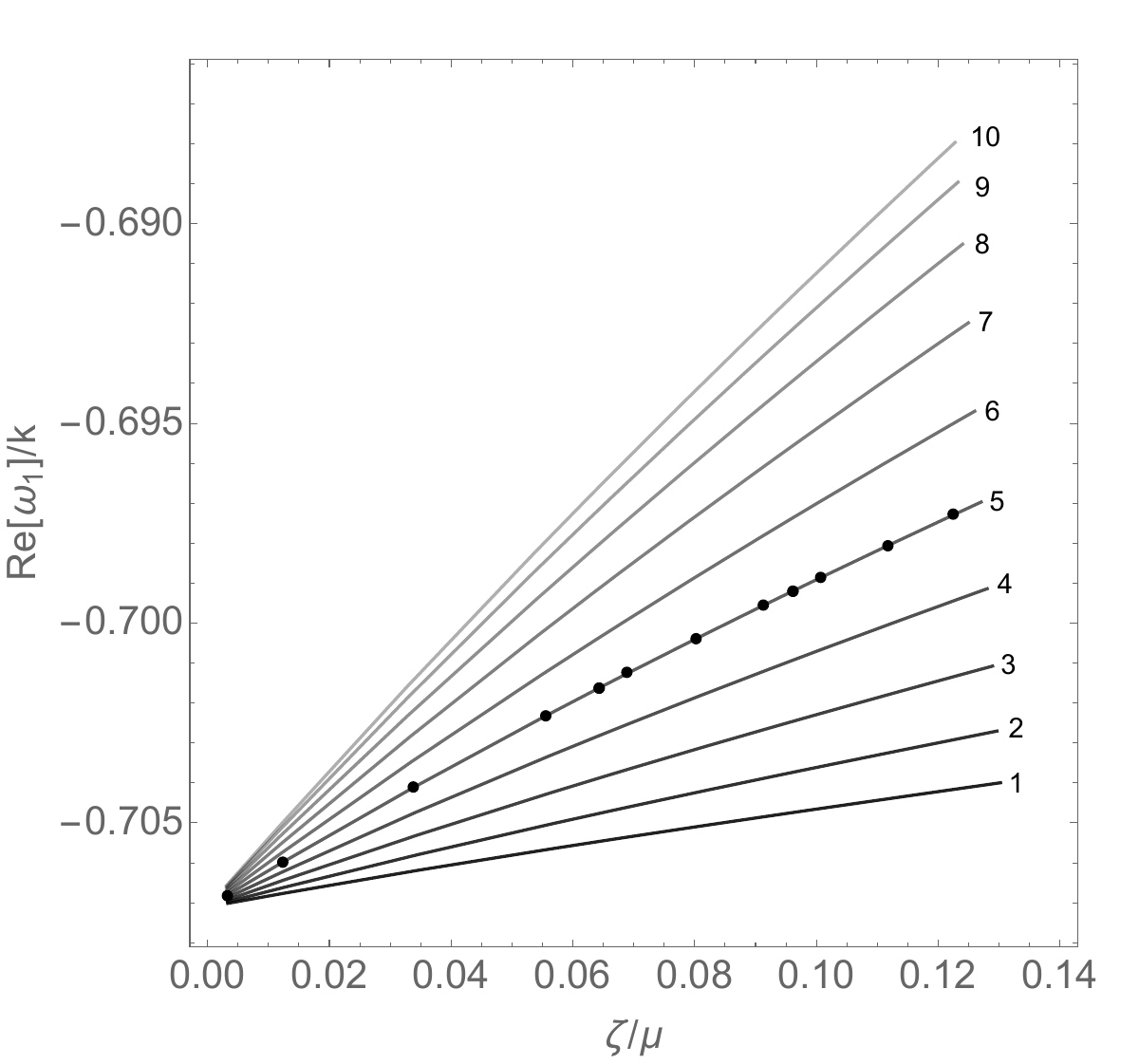}
\includegraphics[scale=.35]{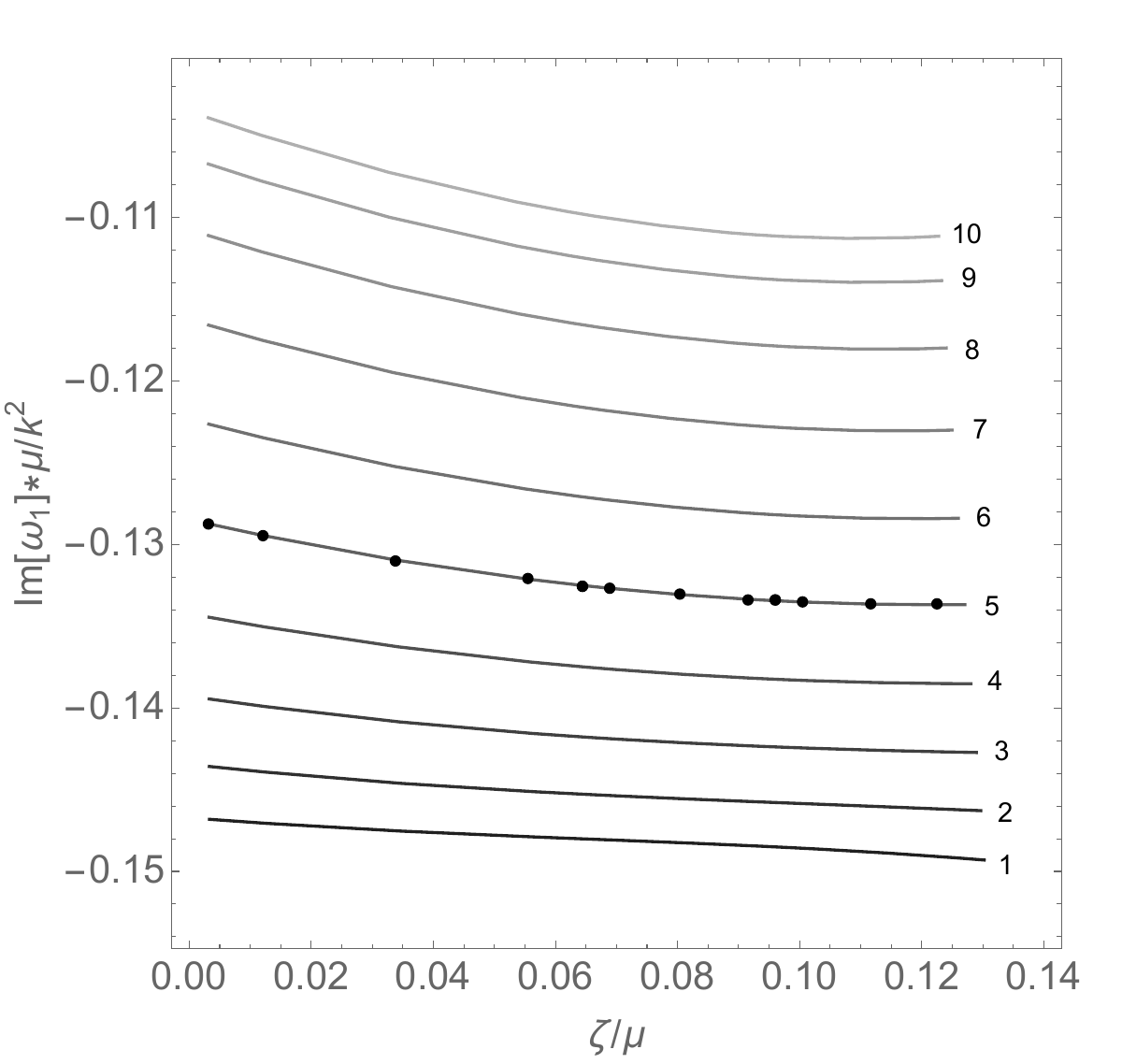}
\caption{\label{fig:matching1}}
\end{subfigure}
\hfill

\begin{subfigure}{\textwidth}
\includegraphics[scale=.35]{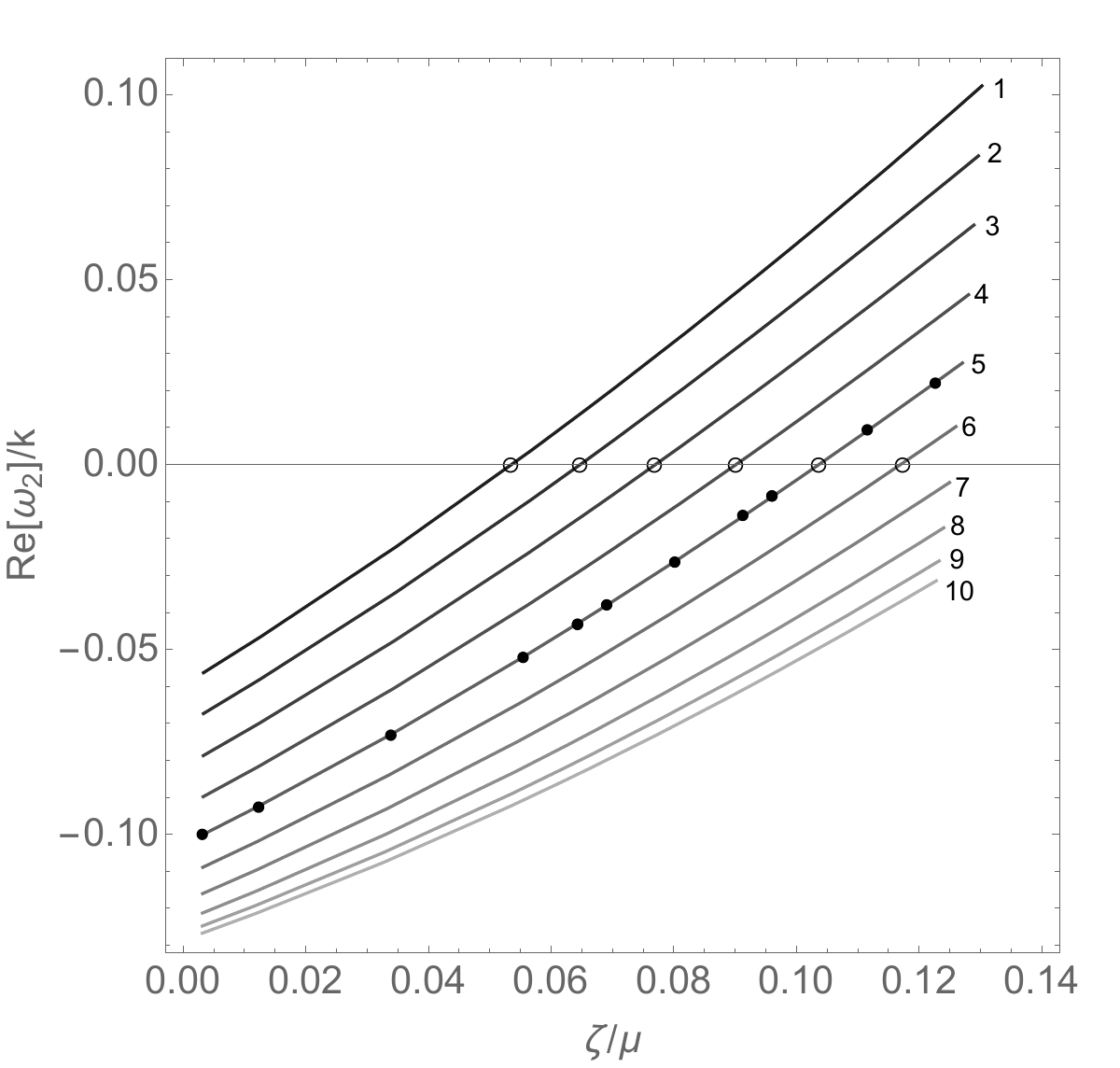}
\includegraphics[scale=.35]{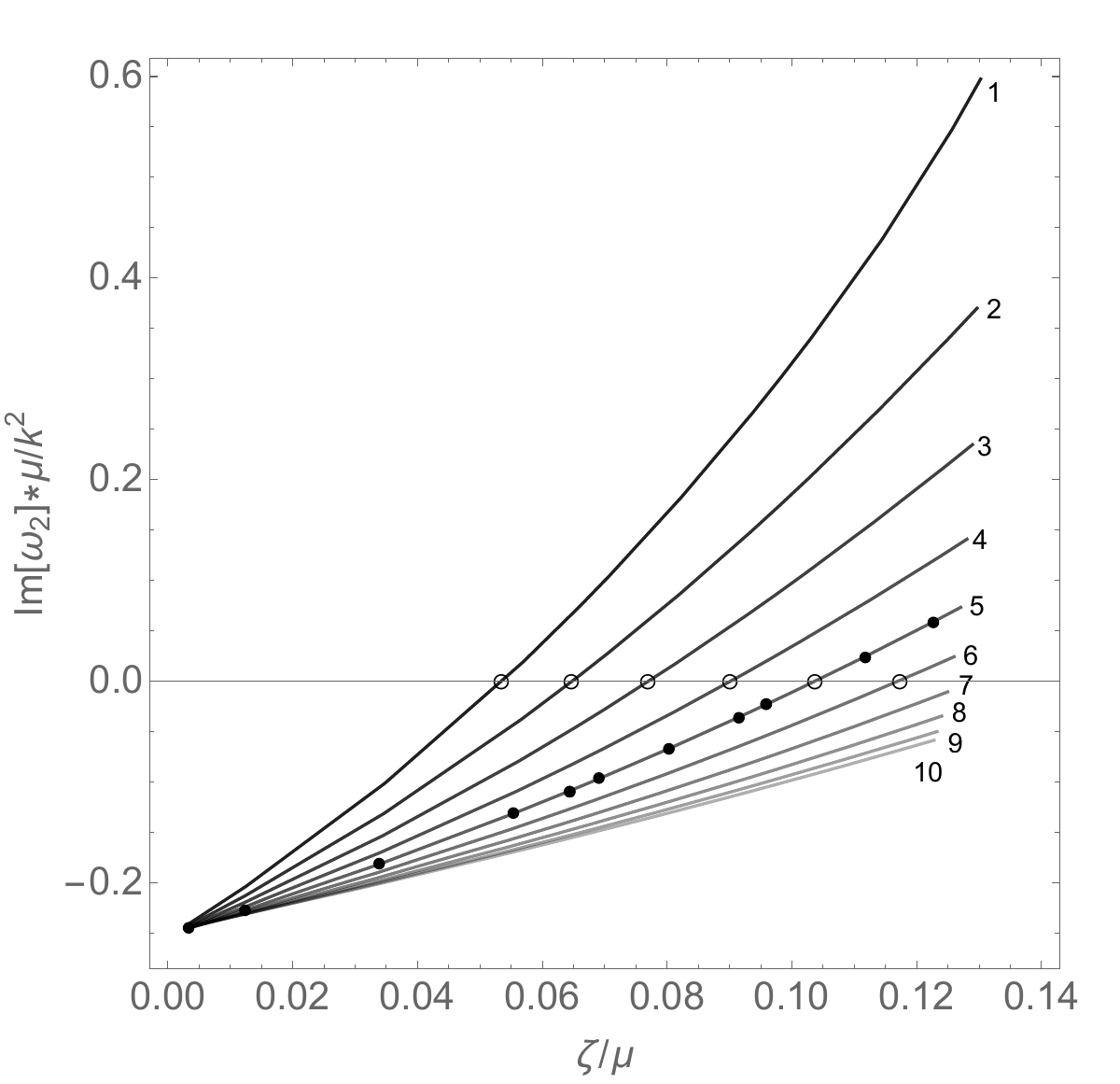}
\caption{\label{fig:matching2}}
\end{subfigure}

\caption{\label{hydrocomparisonwithinstability}Plots of the real and imaginary components of $\omega/\mu$ at $k/\mu = 3.2\times 10^{-7}$ obtained from hydrodynamics (lines) and from quasinormal modes (points) for one of the hydrodynamic sound modes. The color of the lines correspond to the curves in Figure~\ref{CutsThroughPhaseDiagram}. We compare the hydrodynamic results to the quasinormal mode results only on one curve for ease of reading. The mode which signals the instability is $\omega_2$. We also plot the critical value of $\zeta$ obtained from \eqref{critrel} in open circles and show that it coincides the sign change of the real and imaginary components of $\omega_2$.}
\end{figure}

\begin{figure}[t!]
\centering
\begin{subfigure}{\textwidth}
\includegraphics[scale=.35]{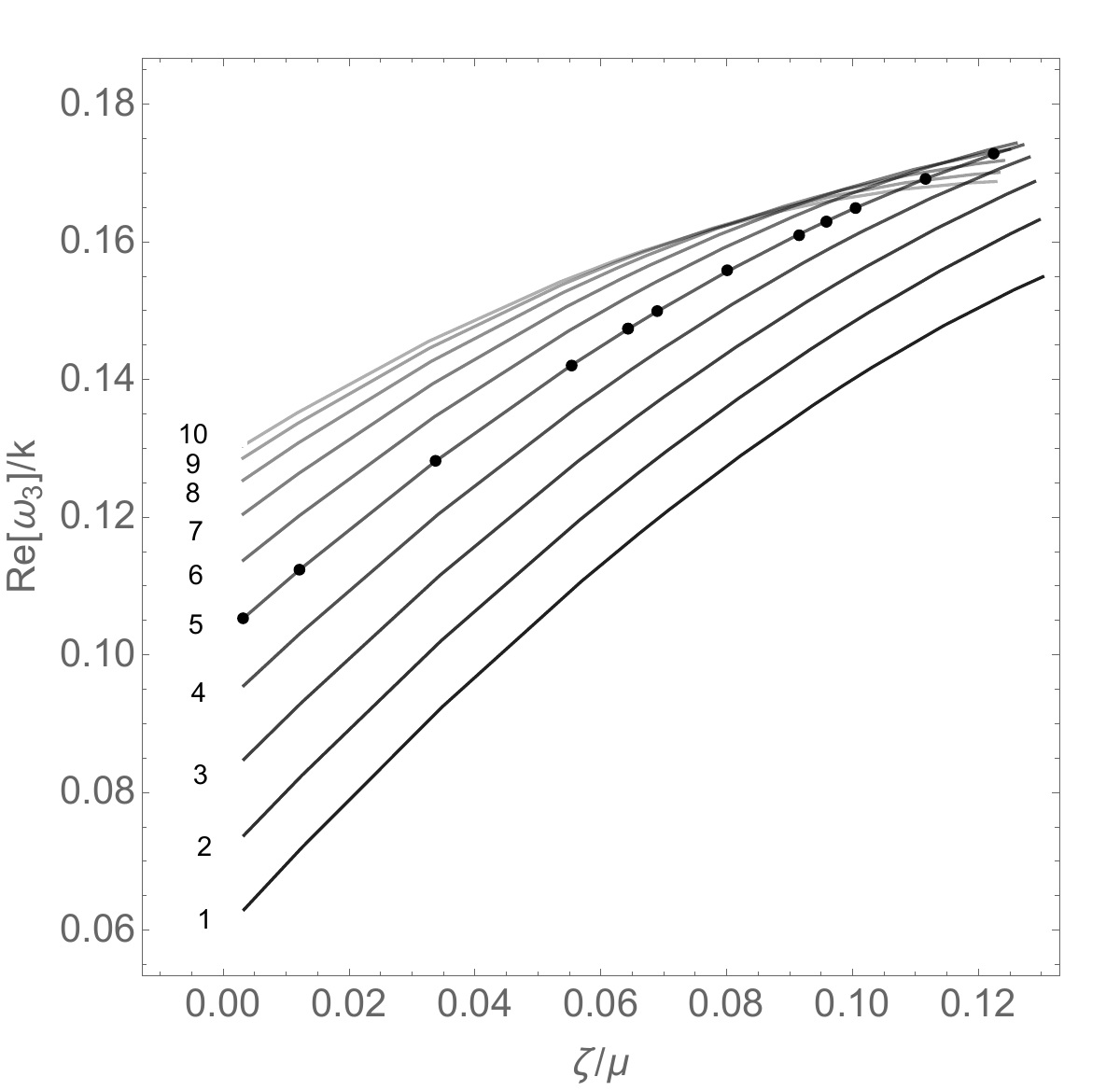}
\includegraphics[scale=.35]{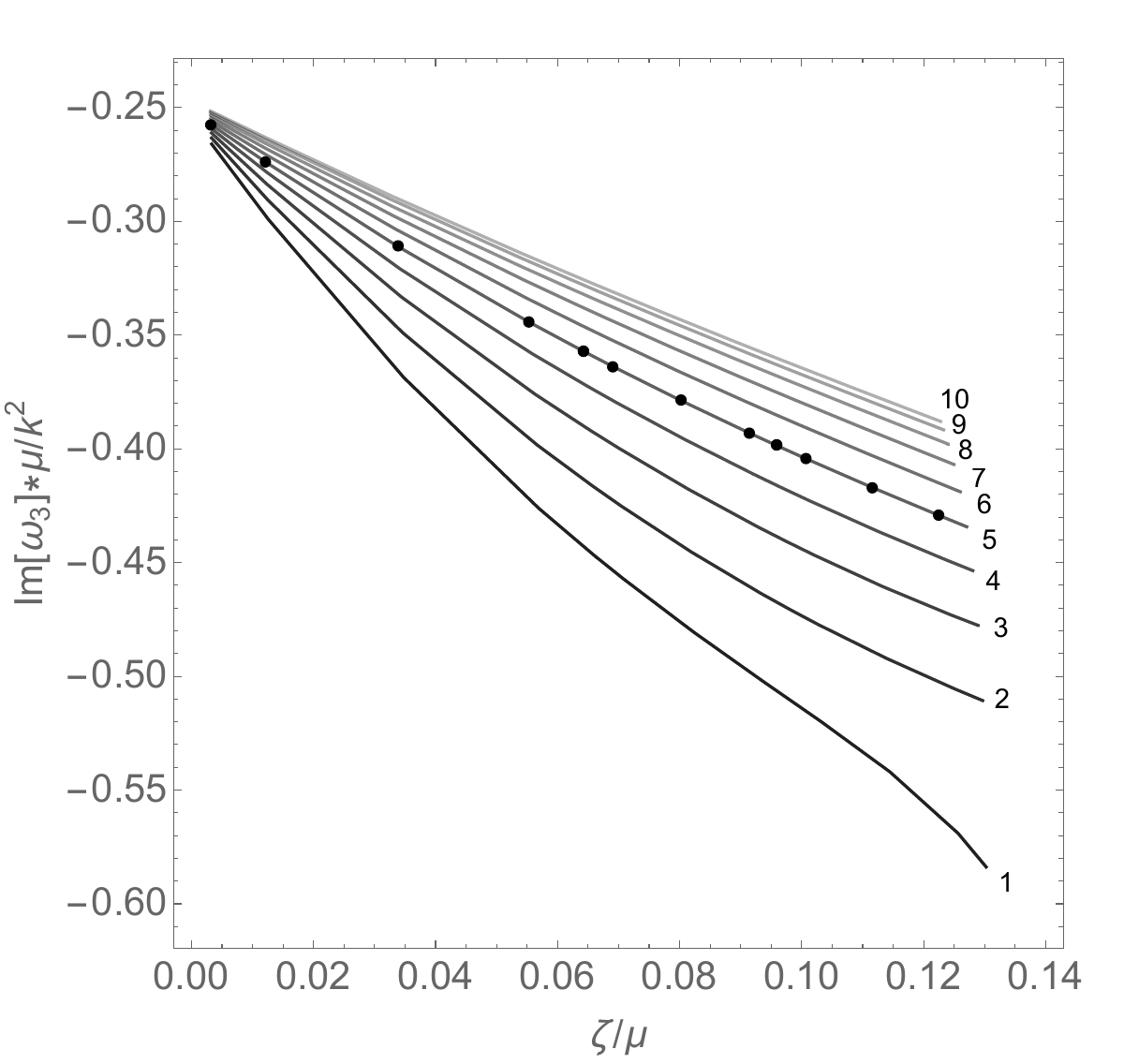}
\caption{\label{fig:matching3}}
\end{subfigure}
\hfill

\begin{subfigure}{\textwidth}
\includegraphics[scale=.35]{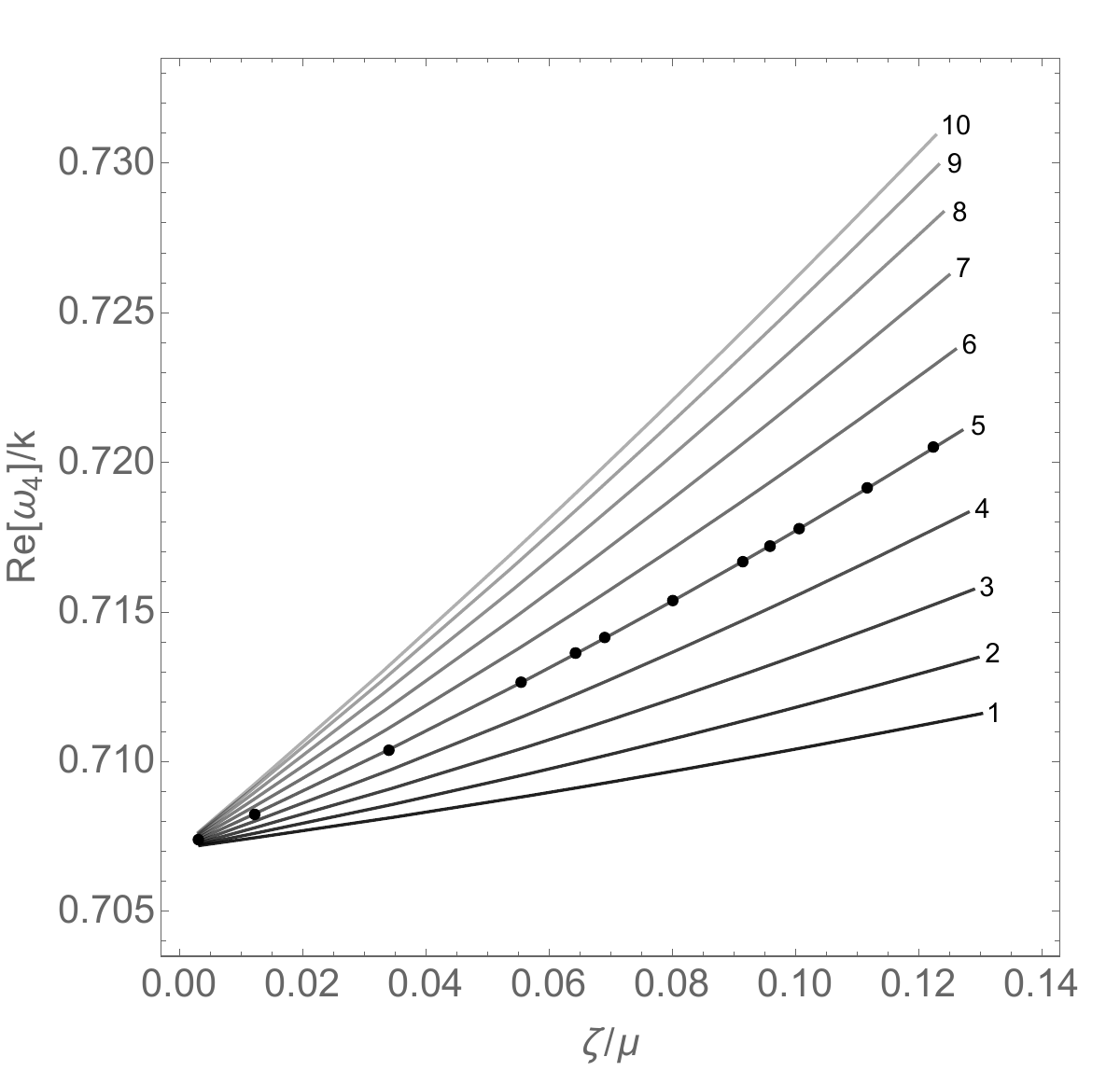}
\includegraphics[scale=.35]{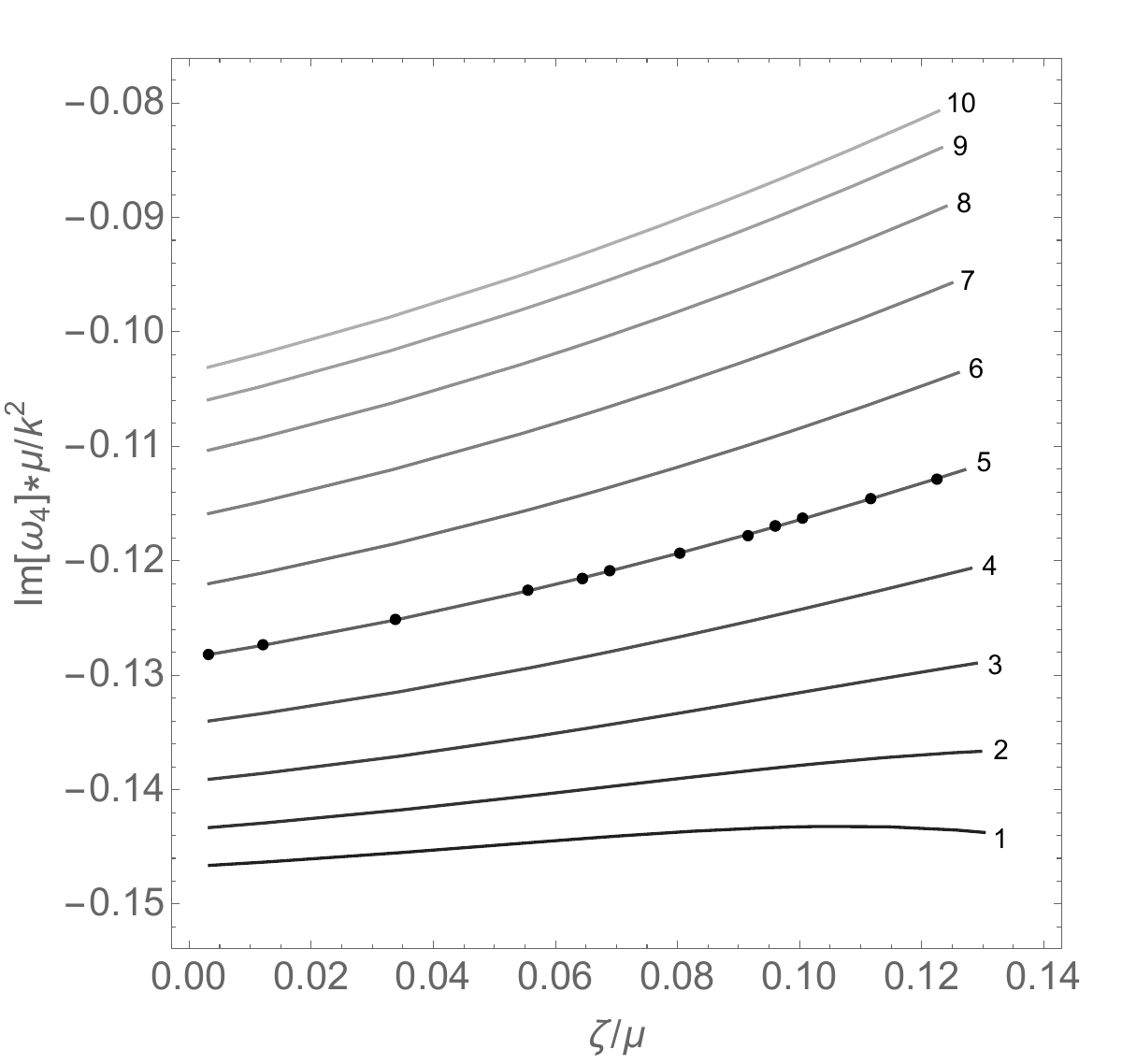}
\caption{\label{fig:matching4}}
\end{subfigure}

\caption{\label{matchingfigure}Plots of the real and imaginary components of $\omega/\mu$ at $k/\mu = 3.2\times 10^{-7}$ obtained from hydrodynamics (lines) and from quasinormal modes (points) for one of the hydrodynamic sound modes. The color of the lines correspond to the curves in Figure~\ref{CutsThroughPhaseDiagram}. We compare the hydrodynamic results to the quasinormal mode results only on one curve for ease of reading.}
\end{figure}

We have shown how to obtain the thermodynamic and dissipative transport coefficients via holography for holographic superfluids. Furthermore, by varying the quadratic on-shell action, we have seen how linearized fluctuations of the currents as obtained from holography manifestly satisfy the linearized Ward identities. It is therefore not surprising that, in the low energy and long wavelength limit where linear response is an appropriate approximation, we can obtain the dispersion relations of the hydrodynamic modes from a holographic calculation as well. 

The hydrodynamic modes are poles of the retarded Green's functions whose dispersion relation $\omega(k)$ vanishes as $k\to 0$. As we have illustrated that, up to contact terms, the retarded Green's functions are given by (\ref{subtractedGreensFunctions}), the poles of the retarded Green's functions are obtained when there exists a solution when $\delta v^I\neq 0$ but $\delta s^I = 0$ \cite{kovstar}. We have similarly argued that, for a given $\omega$ and $k$, there exist solutions $\delta v^I$ with arbitrary sources $\delta s^I$. However, setting $\delta s^I = 0$ will generically result in $\delta v^I = 0$ except for certain sets of $\omega$ and $k$ which are called quasinormal modes. These modes generically have complex frequencies, and, in the hydrodynamic limit where $k/T\lesssim1$, the modes which have $|\omega(k)|/T\lesssim 1$ are expected to have dispersion relations which agree with hydrodynamics.\footnote{The scale over which hydrodynamics is a good approximation can be reduced when there are dangerously irrelevant operators, see \cite{Grozdanov:2018fic,Davison:2022vqh}.}
As expected we have found four hydrodynamic sound modes whose frequencies we denote as $\omega_i$, $(i=1,\dots,4)$. In Figures~\ref{fig:matching1}, \ref{fig:matching2}, \ref{fig:matching3}, and \ref{fig:matching4}, we show that, indeed, the results agree very well. In particular, in Figure  \ref{fig:matching2} we show that the mode $\omega_2$ becomes unstable at the critical value of the superfluid velocity~\eqref{critrel} derived in the hydrodynamic theory. For comparison to previous results, e.g. \cite{Arean:2021tks}, we note that in the limit of zero superfluid velocity $\omega_1$ and $\omega_4$ coincide with superfluid first sound and $\omega_2$ and $\omega_3$ coincide with superfluid second sound.

The recipe for obtaining the hydrodynamic description of holographic superfluids is straightforward in principle, but due to the non-linear nature of the Einstein equations, solutions most often must be found numerically. In Appendix \ref{numericsappendix}, we outline the specific way in which we formulated this numerical recipe. 

\section{Discussion}

In this work, we have studied the instabilities of relativistic superfluids at finite superflow, reconciling the familiar formulation of such instabilities established long ago by Landau under the assumption of a weakly-coupled bosonic quasiparticle spectrum, \cite{landau1980course9}; and more recent results on strongly-coupled superfluids without quasiparticles obtained using gauge/gravity duality, \cite{Amado:2013aea}, which reported a dynamical instability triggered by the crossing to the upper complex frequency plane of a gapless pole of the retarded Green's functions. More precisely, we have showed that the instability is thermodynamic in nature, and caused by the divergence of the second derivative of the pressure with respect to the (relative) superfluid velocity, \eqref{instabcrit}. Assuming a bosonic quasiparticle spectrum, evaluating this quantity and checking when it diverges leads to the usual Landau criterion \eqref{landaucrit}, \cite{Gouteraux:2022kpo}. On the other hand, working in the hydrodynamic regime, we also showed that precisely when the thermodynamic derivative diverges, one of the hydrodynamic mode crosses to the upper half plane. This is exactly what happens in holographic superfluids, as we verified in this work. 

Thus, we have given a unified description of superfluid instabilities, which apply equally well at weak and strong coupling, and as shown in \cite{Gouteraux:2022kpo}, irrespective of any boost invariance. In passing, we have also demonstrated that the dispersion relation of the gapless poles of holographic retarded Green's functions matches the corresponding hydrodynamic predictions at sufficiently low wavenumbers, as expected.

We note that the criterion \eqref{instabcrit} as formulated makes no prediction for the endpoint of the instability. On the other hand, it applies irrespective of its specific mechanism, that is, whether it is caused by exciting rotons in Helium 4 or the nucleation of vortices in holographic superfluids as in \cite{Lan:2020kwn}. In fact, in \cite{Lan:2020kwn}, relaxation of the superfluid velocity through the nucleation of vortices was demonstrated when the initial state was prepared with a superfluid velocity located in the darker grey region of the phase diagram in Figure~ \ref{CutsThroughPhaseDiagram}, freezing normal velocity and temperature fluctuations. It would be interesting to include backreaction of these fluctuations, and to probe the black-colored region of the phase diagram. There, sound modes become complex (the `two-stream' instability discussed in \cite{schmitt1,schmitt2,schmitt3}), the second derivative of the pressure with respect to the superfluid velocity becomes positive again, while instead the heat capacity changes sign. 

Another interesting direction for future work is to study the fate of such instabilities at very low temperatures, when the superfluid is coupled to a quantum critical sector, \cite{Gouteraux:2019kuy,Gouteraux:2020asq}. In this regime, the dynamical instability may no longer be within the regime of validity of hydrodynamics, if it occurs for values of the superfluid velocity large compared to the temperature. If the critical superfluid velocity is not large compared to the chemical potential, it may still be captured by a zero temperature effective field theory \cite{Son:2002zn}, where the derivative \eqref{instabcrit} can still meaningfully be computed.

\subsection*{Acknowledgements}

The work of D.~A is supported through the grants CEX2020-001007-S and PID2021-123017NB-100, PID2021-127726NB-I00 funded by MCIN/AEI/10.13039/501100011033 and by ERDF ``A way of making Europe''.
The work of B.~G.~ and F.~S.~ is supported by the European Research Council (ERC) under the European Union's Horizon 2020 research and innovation programme (grant agreement No758759). The work of E.~M.~was supported in part by NSERC and in part by the European Research Council (ERC) under the European Union's Horizon 2020 research and innovation programme (grant agreement No758759). We all gratefully acknowledge Nordita's hospitality during the Nordita program `Recent developments in strongly-correlated quantum matter' where part of this work was carried out. Part of this work was performed at the Aspen Center for Physics, which is supported by National Science Foundation grant PHY-2210452. E.~M.~ and B.~G.~ also acknowledge the hospitality of the Kavli Institute for Theoretical Physics where part of this work was performed and which is supported in part by the National Science Foundation under Grants No.~NSF PHY-1748958 and PHY-2309135.
\appendix

\section{\label{numericsappendix} Details of the numerics}

\begin{figure}[t!]
\centering
\includegraphics[scale=.35]{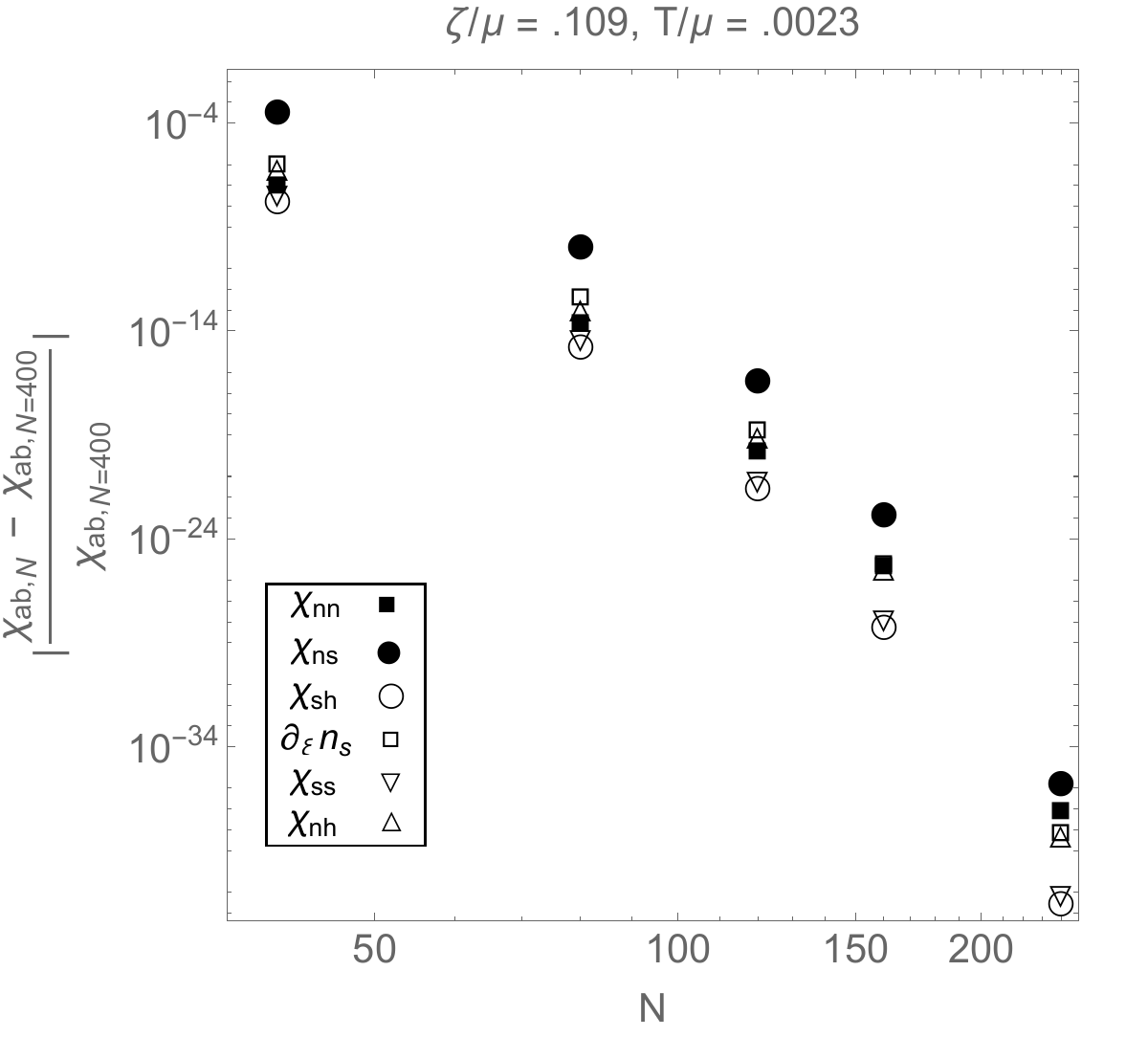}
\includegraphics[scale=.35]{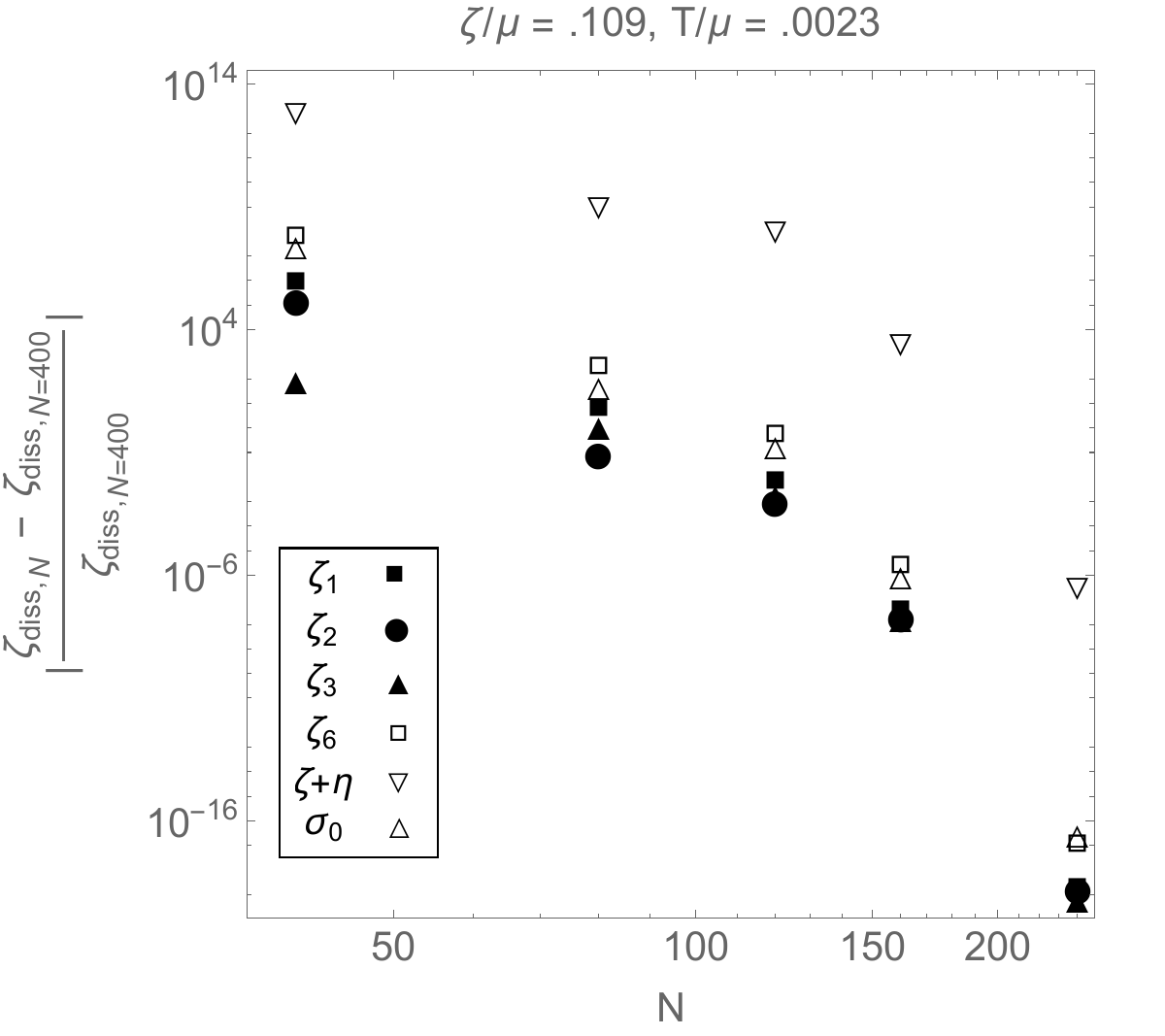}
\caption{\label{ConvergencePlots1}(Left) Comparison of the components of the susceptibility matrix as a function of grid size $N$ to the values at $N=400$ showing exponential convergence. (Right) Comparison of the dissipative coefficients as a function of grid size $N$ to the values at $N=400$ showing exponential convergence. The value of $\zeta+\eta$ converges slowest and is the largest source of error in the hydrodynamic prediction.}
\end{figure}

\begin{figure}[t!]
\centering
\includegraphics[scale=.32]{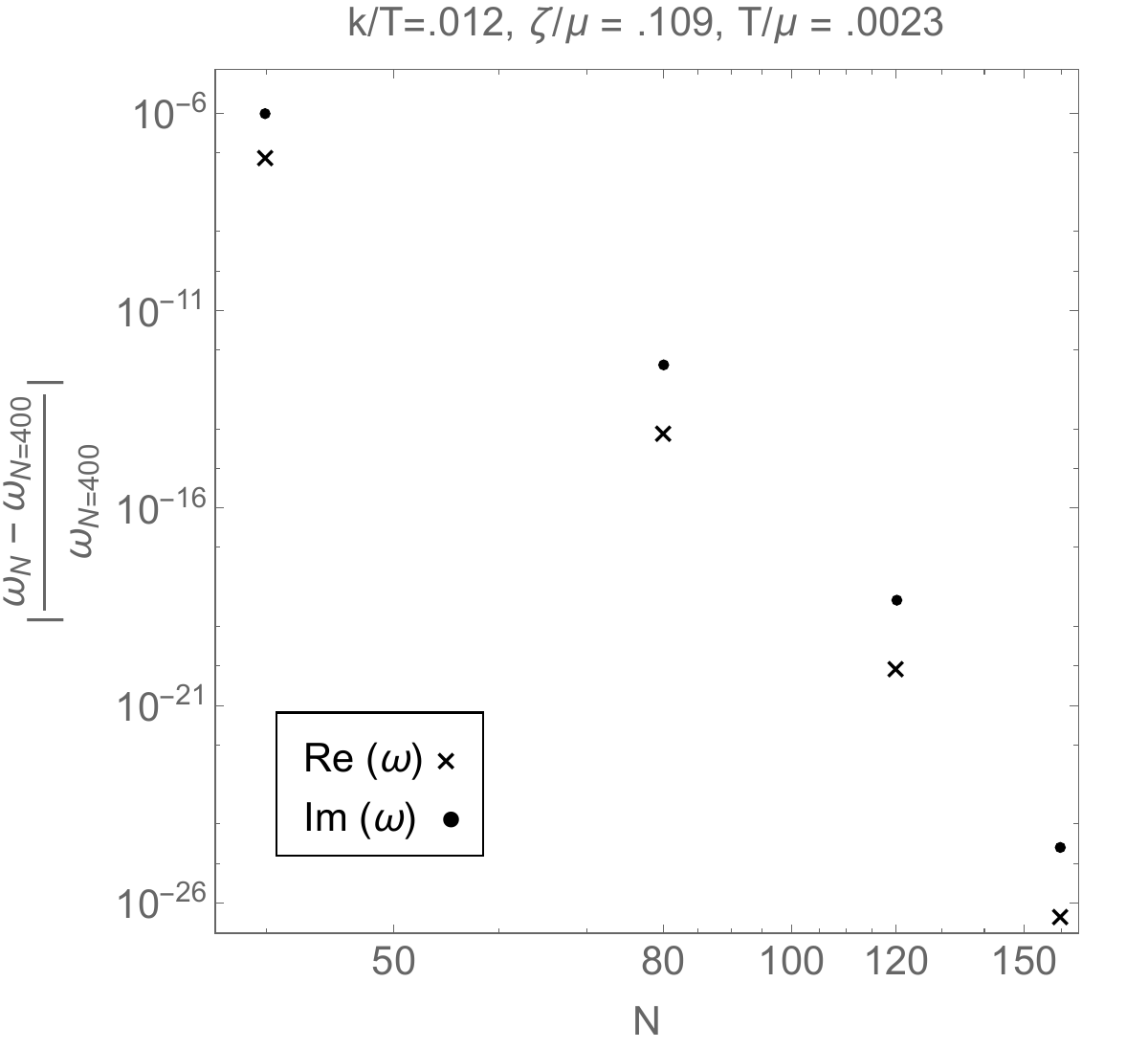}
\includegraphics[scale=.32]{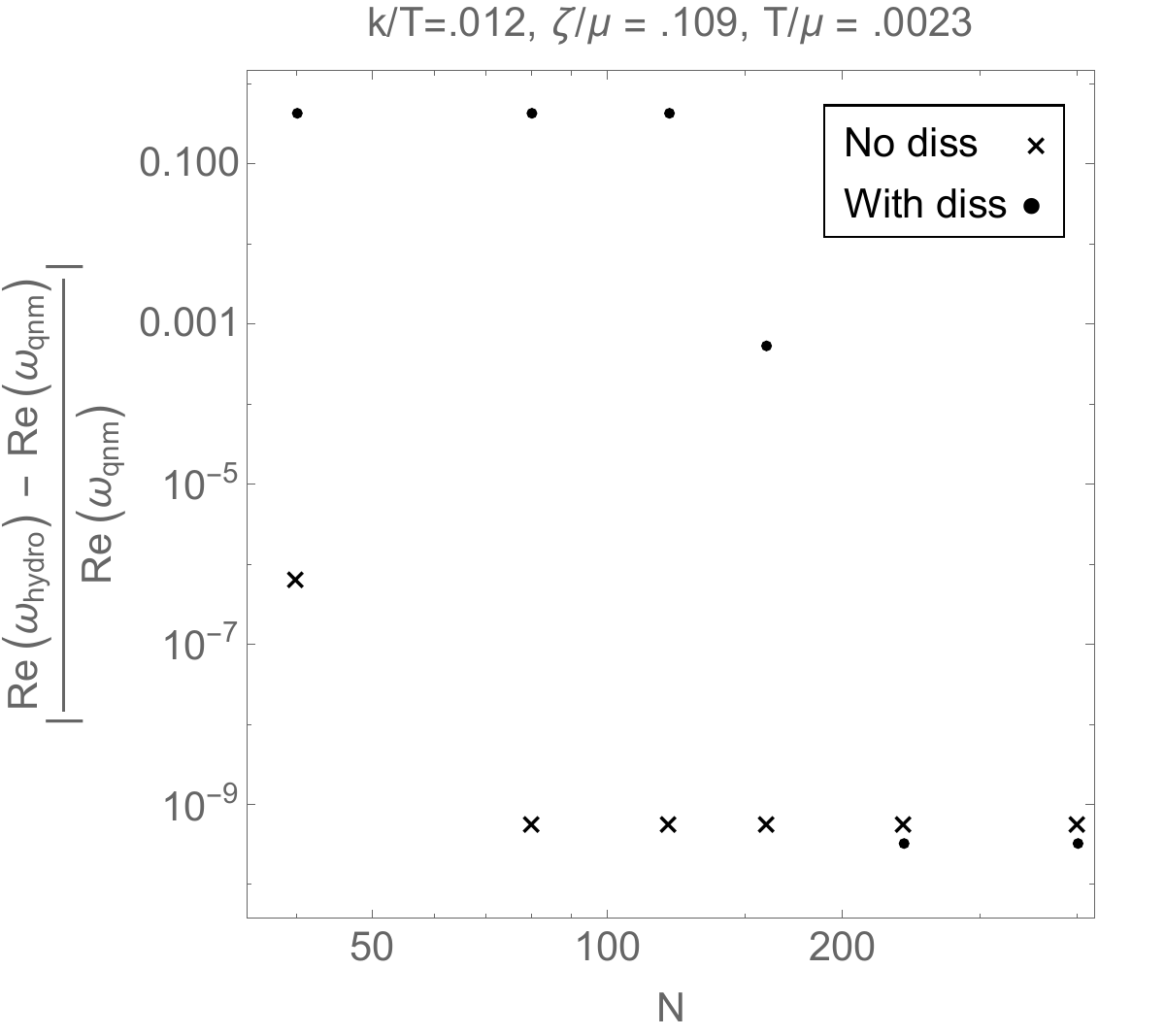}
\includegraphics[scale=.32]{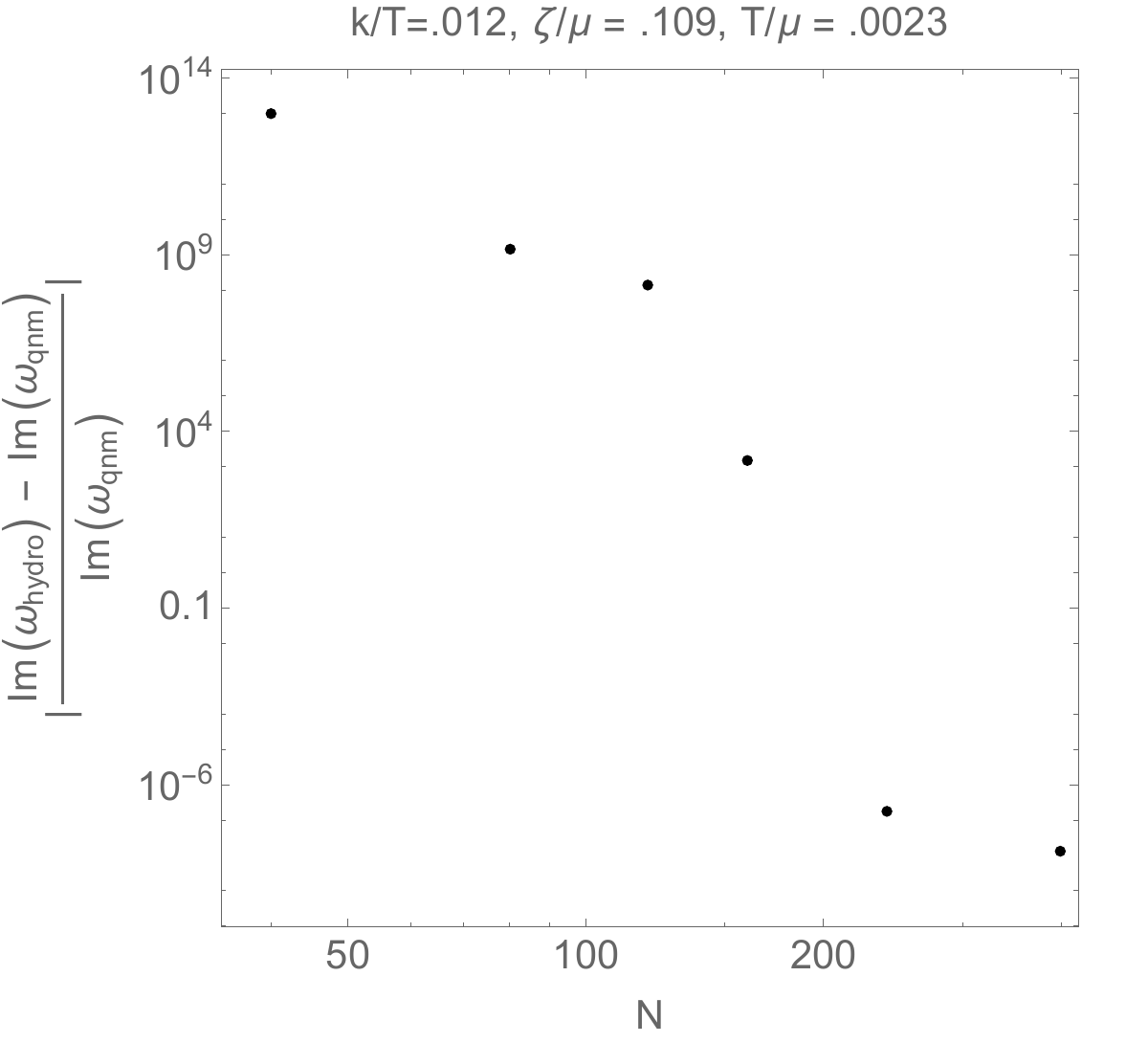}
\caption{\label{ConvergencePlots2}(Top left) A comparison of the real and imaginary components of one hydrodynamic mode as a function of grid size $N$ to the value at $N=400$. At $N=400$, $\omega/\mu = -(.0721477...)(k/\mu) - i(.0580989...) (k/\mu)^2+\mathcal{O}(k/\mu)^3$  (Top right) A comparison of the hydrodynamic prediction for the real part of this mode to the real part of the QNM as a function of grid size $N$. The points represent $\text{Re}(\omega_{\text{hydro}})$ including the value for the dissipative coefficients obtained using Kubo formulae at the same value of $N$ whereas the crosses represent the value for $\text{Re}(\omega_{\text{hydro}})$ when the dissipative coefficients are set to zero. The strong convergence in the absence of dissipative coefficients reflects the strong convergence of the susceptibility matrix in Figure \ref{ConvergencePlots1}. (Bottom) Comparison of the imaginary component of the hydrodynamic prediction to the imaginary component of the QNM as a function of grid size $N$ (the dissipative coefficients are non-zero).}
\end{figure}

\begin{figure}[t!]
\centering
\includegraphics[scale=.4]{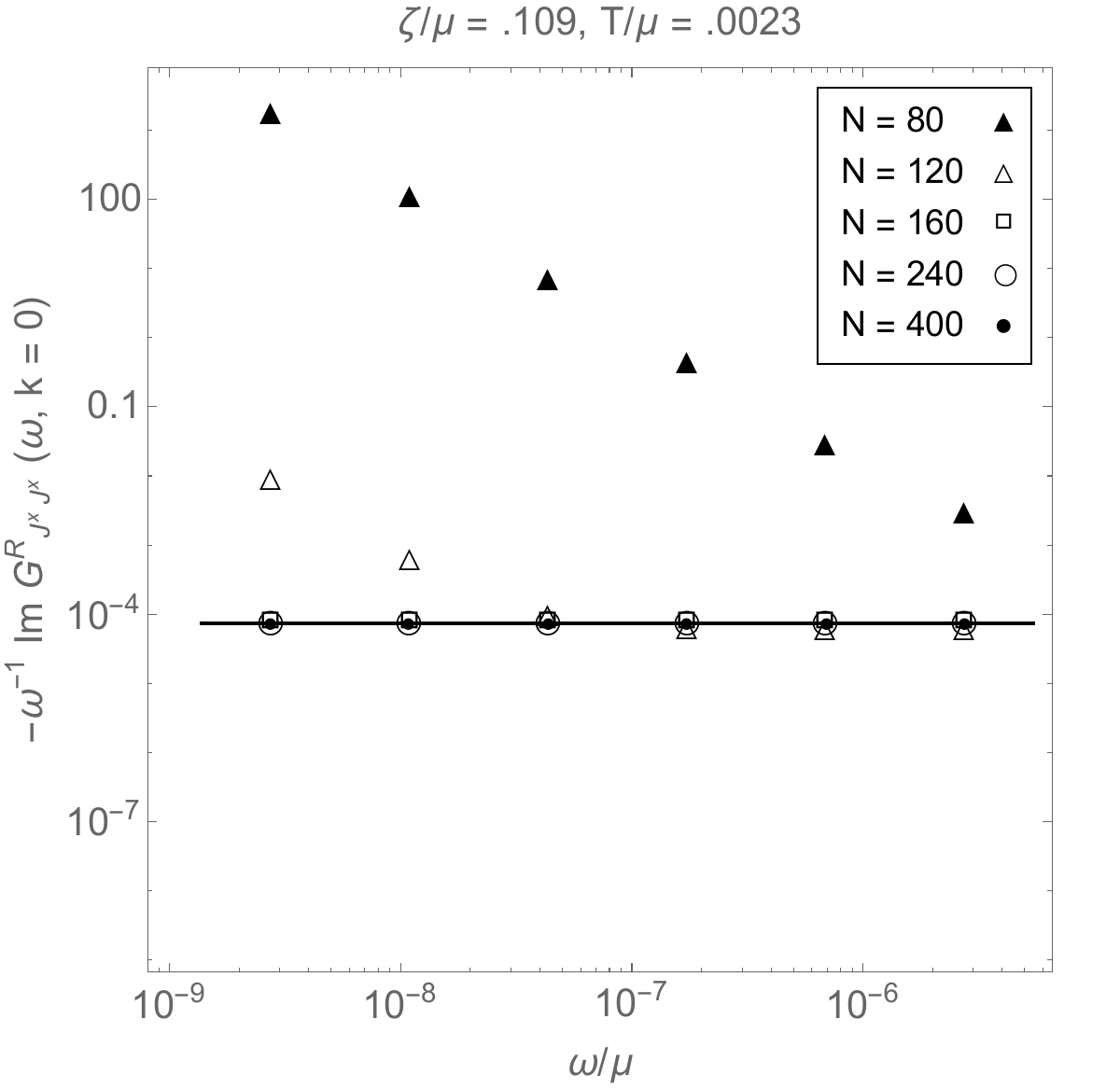}
\caption{\label{ConvergencePlots3}(Top left) The dissipative transport coefficients from the Kubo formulae are numerically extracted by extrapolation to $\omega \to 0$ (thick line) where the function is constant. At low temperatures and small grid sizes, this may not be possible and larger grid sizes are necessary. Here, we see that $N=120$ has good agreement with $N=400$ by sight. From Figure~\ref{ConvergencePlots1}, they differ by a relative magnitude of $\sim10^{-1}$.}
\end{figure}

In this Section we describe the numerical methods used to construct our solutions.
We work with a Fefferman-Graham type coordinate $u=1/r$ with a horizon located at $u=1$. We discretize the $u$ coordinate with a Chebyshev-Gauss-Lobatto collocation grid, and solve the resulting equations using a Newton-Raphson relaxation method. All fields are functions of the radial coordinate $u$ only. To allow for temperature variations at fixed $\mu$ and $\zeta$, we allow for a variable horizon location $u_h$ in our numerics. We work in $d=2$ with spacetime coordinates $x^\mu = \{t,x,y\}$. Choosing a superfluid velocity along the $x$ direction, without loss of generality, we can write the metric ansatz as
\begin{align}
ds^2 = \frac{1}{u^2}\left[-\frac{\chi^2(u)}{u_h^2}f(u)dt^2 + \frac{du^2}{f(u)}+\frac{\chi^2(u)}{u_h^2}dy^2 + u^2\left(2Y_1(u)dt\,dx + Y_2(u)dx^2\right)\right]
\end{align}
In writing the ansatz in this manner, we have chosen a frame in which the normal velocity, $u^\mu$, points only along the time direction. In other words, the boundary conditions on the metric are
\begin{align}
\lim_{u\to 0} g_{MN}dx^Mdx^N = \frac{du^2}{u^2}+\frac{1}{u^2}\left[-dt^2+dx^2+dy^2 + O(u^2)\right]
\end{align}
The Maxwell field in this case has components only along the $t$ and $x$ directions. Boundary conditions on this field are
\begin{align}
\lim_{u\to 0} A_t = \mu, \quad \lim_{u\to 0}A_{x} = -\zeta
\end{align}
Boundary conditions on the scalar field are
\begin{align}
\lim_{u\to 0}\psi'(u) = 0.
\end{align}
Boundary conditions at the black hole horizon are
\begin{align}
f(u_h) = 0, \quad Y_1(u_h) = 0, \quad A_t(u_h) = 0
\end{align}
and regularity of all other fields. The temperature and entropy associated with the black hole are 
\begin{align}
T = \frac{f'(u_h)\chi(u_h)}{4\pi u_h},\quad s = 4\pi\frac{\chi(u_h)}{u_h^2}\sqrt{\chi^2(u_h)+u_h^2Y_2(u_h)}
\end{align}
where we have set $G_N = 1/16\pi$.  Note that $u_h$ can be set equal to 1 via a scale transformation and as a consequence, the phase space of the background is fully specified by $\mu/T$ and $\zeta/T$. Nevertheless, we will keep $u_h$ around since it simplifies the derivatives with respect to the temperature, $T$.

The static susceptibilities are obtained via variations of the equilibrium solutions with respect to $\mu, T, \zeta$. We do not need to consider variations with respect to the normal velocity $u^\mu$ as these susceptibilities are determined via the other variations in terms of Onsager relations. To accurately obtain the variations, we discretize $\mu, u_h, \zeta$ over independent Chebyshev-Gauss-Lobatto grids and solve for the background fields, $g_{MN}, A_M, \psi$ as a function of these parameters. We then find the currents and conserved densities $\rho, n_s, \pi^x$ and $s$ using the holographic renormalization described in \cite{Sonner:2010yx} and in the main text, adapted to our specific ansatz. Finally, derivatives of these quantities with respect to $\mu, T, \zeta$ are obtained using the appropriate Chebyshev derivative matrices.

Having obtained a background solution, we then consider linearized perturbations. Since the instability will appear first for $k||\zeta$, we only discuss this particular perturbation. Working in a gauge where $A_r = 0$, $\delta G_{Mr} = 0$, the perturbed fields are
\begin{align}
\begin{split}
A_t &= A_{t,0} + a_t e^{-i\omega t+ikx}, 
\hspace{82pt} A_x = A_{x,0} + a_x e^{-i\omega t+ikx}, \\
\Psi &= \psi + (\sigma_r+i\sigma_i)e^{-i\omega t+ikx}, 
\hspace{64pt}  \bar{\Psi} = \psi + (\sigma_r-i\sigma_i)e^{-i\omega t+ikx}, \\
g_{tt} &= -\frac{f\chi^2}{u^2u_h}\left(1+h_{tt}e^{-i\omega t+ikx}\right), 
\hspace{46pt} g_{tx} = Y_1+\left(\frac{\chi^2}{u^2u_h^2}\right)h_{tx}e^{-i\omega t+ikx},\\
  g_{xx} &= \frac{\chi^2}{u^2u_h^2}+Y_2+\left(\frac{\chi^2}{u^2u_h^2}\right)h_{xx}e^{-i\omega t+ikx},
  \quad g_{yy} = \left(\frac{\chi^2}{u^2u_h^2}\right)(1+h_{yy}e^{-i\omega t+ikx})
 \end{split}
\end{align}
and all others can be set to zero. The linearized equations of motion are easily found, but their expressions are too cumbersome to show here.

The retarded Green's functions are found by solving the linearized equations of motion with Dirichlet boundary conditions at $u=0$ and satisfying ingoing boundary conditions at $u\to u_h$. Taking the set of linearized functions $X=\{a_t, a_x, \sigma_r, \sigma_i, h_{tt}, h_{tx}, h_{xx}, h_{yy}\}$, ingoing boundary conditions can be imposed as
\begin{align}
\label{numericalingoing}
X \to \tilde{X}e^{i\omega u_*(u)}
\end{align}
where $u_*(u)$ is the tortoise coordinate defined as
\begin{align}
du_*/du = \frac{u_h}{\chi(u)f(u)}, \quad u_*(0) = 0.
\end{align}
Note that $r_* \to \log(u_h-u)/{4\pi T}$ at the black hole horizon so that this agrees with the definition in the main text. Defining ingoing boundary conditions in this way, only $du_*/du$ and $d^2u_*/du^2$ appear in the linearized equations of motion, not $u_*$ itself, and often leads to lower powers of $\omega$ which is useful numerically when solving for the quasinormal modes. However, one must be careful to match the $u\to 0$ expansion of the left and right hand sides (\ref{numericalingoing}) to correctly extract the sources and currents.

As explained in the main text, the set of linearized equations contains both second order differential equations as well as first order differential equations that serve as constraints---in $d=2$, there are 4 second order equations and 4 first order. In order to find the Green's functions, we must therefore add four pure gauge solutions. These can be found in terms of terms of 4 functions, $\beta_t$, $\beta_x$, $\beta_u$ and $\Lambda$ under which
\begin{align}
\delta A_M &= \pounds_\beta A_M + \partial_M \Lambda, 
\quad \delta g_{MN} = \nabla_M \beta_N + \nabla_N \beta_M, \nonumber\\
\delta\Psi &= \beta^M\partial_M\psi+i\psi\Lambda,
\qquad \delta\bar{\Psi} = \beta^M\partial_M\psi - i\psi \Lambda.
\end{align}
We take these to have plane wave dependence and an asymptotic behavior as $u\to 0$,
\begin{align}
\beta^t(u,t,x) &= e^{-i\omega t+ikx}\eta_t,
\quad \beta^x(u,t,x) = e^{-i\omega t+ikx}\eta_x,\nonumber\\
\beta^u(u,t,x) &= u\sqrt{f(u)}, 
\qquad\;\;\; \Lambda(u,t,x) = e^{-i\omega t+ikx}\Lambda_0.
\end{align}
The retarded Green's functions, $\tilde{G}^{R}_{AB}$, are obtained by finding four independent sets of solutions, $X^I$ for $I=1...4$, corresponding to the boundary conditions 
\begin{align}
\left(\begin{array}{c}a_x(u_h)\\ h_{xx}(u_h)\\\sigma_r(u_h)\\\sigma_i(u_h)\end{array}\right)^I = \biggl\{\left(\begin{array}{c}1\\0\\0\\0\end{array}\right),\left(\begin{array}{c}0\\1\\0\\0\end{array}\right),\left(\begin{array}{c}0\\0\\1\\0\end{array}\right),\left(\begin{array}{c}0\\0\\0\\1\end{array}\right)\biggr\}^I.
\end{align}
The remaining boundary conditions on the fields are obtained by expanding the equations of motion near the black hole horizon and the UV boundary which results in a set of Robin boundary conditions relating the values of the functions on the hypersurface to their first derivatives. 

On the other hand, to find the poles of the Green's functions, we need to solve for the quasinormal modes. This can be done efficiently by requiring the gauge invariant combinations
\begin{align}
ka_t+\omega a_x\biggr|_{u=0}, \quad 2\omega k h_{tx} + \omega^2(h_{xx}-h_{yy})+k^2(h_{yy}-h_{tt})\biggr|_{u=0}
\end{align}
as well as a subset of the perturbations, for instance,
\begin{align}
h_{tt}\biggr|_{u=0}=\sigma_i\biggr|_{u=0}=\sigma_{r}\biggr|_{u=0}
\end{align}
to vanish in the UV. The remaining boundary conditions in the UV can be found in terms of a series expansion around $u=0$. The boundary conditions at the black hole horizon are similarly obtained in terms of a series expansion around $u = u_h$.

There are two methods we used to find the quasinormal modes. The most straightforward is as a generalized eigenvalue problem over the Chebyshev grid, see for instance \cite{Dias:2015nua}. When the linearized equations are evaluated on the numerical grid, we obtain a matrix equation
\begin{align}
\label{numericaleigenvalueeq}
(M^{(0)}_{ij}(k) + \omega M^{(1)}_{ij}(k) + \omega^2 M^{(2)}_{ij}(k) + ...)X^j = 0
\end{align}
Here, the index $i$ spans the equations as well as the grid points whereas $j$ spans the linearized perturbations as well as the grid points (the number of equations matches the number of fields, so these matrices are square). If $\omega^2$ (or even higher powers of $\omega$) was not present, by inverting $M^{(1)}$, it is clear that this is equivalent to an eigenvalue problem. As it happens, the presence of $\omega^n$ terms with $n\geq 2$ can be reframed as a standard eigenvalue problem in terms of generalized eigenvectors and matrices, 
\begin{align}
(\tilde{M}_{ij}^{(0)} + \omega \tilde{M}_{ij}^{(1)})\tilde{X}^j = 0
\end{align}
as follows. Take the largest power $\omega^n$ appearing in \eqref{numericaleigenvalueeq} and define the extended eigenvector
\begin{align}
\tilde{X} = \left(\begin{array}{c} X \\ \omega X \\  \omega^2 X\\ ... \\ \omega^{n-1}X \end{array}\right)
\end{align}
and matrix
\begin{align}
\tilde{M}^{(0)}(k) = \left(
				\begin{array}{ccccc} 
							   M^{(0)} & M^{(1)} & M^{(2)} & ... & M^{(n-1)}\\
							   0 & -\mathbb{I} & 0 & ... &0\\
							   0 & 0& -\mathbb{I} &...&0\\
							   0 & 0 & 0 &... & 0\\
							   0 & 0 & 0 & 0 &  -\mathbb{I}   
				\end{array}
		     \right), 
		     \quad
\tilde{M}^{(1)}(k) = \left(
				\begin{array}{ccccc} 
							   0 & 0& 0 & ... & M^{(n)}\\
							   \mathbb{I} & 0 & 0 & ... & 0\\
							   0 & \mathbb{I} & 0 & ... & 0\\
							   0 & 0 & ... & ... & 0\\
							   0 & 0 & ... &  \mathbb{I} & 0
				\end{array}
			      \right).
\end{align} 
In terms of the new matrices, $\tilde{M}^{(0)}$ and $\tilde{M}^{(1)}$, we are back to a standard eigenvalue problem.

Often, inversion is not necessary to find $\omega$, but the algorithms are still time consuming especially with a large number of grid points. Furthermore, when evaluating the $M^{(i)}$ in terms of numerical solutions, floating point errors can accumulate and fixed high-precision numerics are necessary, causing further slowing. For our purposes, we worked with fixed 60-point precision and monitored numerical accuracy by varying grid sizes. To mitigate this slowdown, another approach is to promote $\omega$ to a function of $u$ by introducing the additional equation of motion $\omega'(u) = 0$. The eigenvalue problem is then solved instead as a boundary value problem with a Newton-Raphson algorithm. This is a much faster approach when one has a good idea of the form of the quasinormal modes and spectrum and wants to increase accuracy. For instance, we used this method to monitor the convergence of our quasinormal modes as we increased the grid size from $N=40$ to $N=400$ points at a very low temperature, shown in Figures~\ref{ConvergencePlots1} and \ref{ConvergencePlots2}. The majority of numerical results reported in the main text were obtained at a grid size of $N=120$ points, though for low temperatures, we went up to $N=400$ points.

\newpage\newpage
\section{\label{probeAppendix} Probe superfluid}

In the main text, we considered the case of a superfluid with momentum and energy fluctuations. We can also consider the case where the dynamics of these conserved quantities are decoupled from the charge sector. This simple case is instructive so we have chosen to include it in this Appendix. 

The hydrodynamic modes are easily obtained by turning off fluctuations of the temperature and normal fluid velocities. We can then fix $u^\mu = \delta^\mu_t$. In this case, only three transport coefficients are left, namely $\sigma_0$, $\zeta_2$ and $\zeta_3$. We can write the dissipative currents in the collinear limit as
\begin{align}
\left(\begin{array}{c} \tilde{J}^x \\ \tilde{\mu} \end{array}\right) = -\left(\begin{array}{cc} \sigma_0  & \zeta_2 \zeta \\ \zeta_2 \zeta & \zeta_3 \end{array}\right)\left(\begin{array}{c} \partial_x \mu\\ \partial_x(\frac{n_s\zeta}{\mu})\end{array}\right).
\end{align}
Positivity of entropy production requires that $\sigma_0\zeta_3 - \zeta^2\zeta_2^2 \geq 0$ and in particular $\sigma_0,\zeta_3 \geq 0$ when $\zeta = 0$.

The spectrum of linearized fluctuations at small $k$ is then given by two sound modes with dispersion relations
\begin{align}
\label{modesprobe}
\omega_{\pm} = v_\pm\left(k -i\frac{\tilde{\Gamma}_\pm}{2}k^2+ \mathcal{O}(k^3)\right)
\end{align}
where
\begin{align}
\label{modesprobe2}
v_{\pm} =\frac{-\chi_{nh} \pm \sqrt{\chi_{nh}^2+\frac{ \chi_{nn}}{\chi_{\xi\xi}} }}{ \chi_{nn}},\qquad
\tilde{\Gamma}_{\pm} =  2\zeta\, \zeta_2 \pm \sqrt{\chi_{nh}^2+\frac{ \chi_{nn}}{\chi_{\xi\xi}} } \, \zeta_3 \pm \frac{\sigma_0}{ \sqrt{\chi_{nh}^2+\frac{ \chi_{nn}}{\chi_{\xi\xi}} }}\,.
\end{align}
In our conventions, $\chi_{nh}<0$ and $\chi_{nn}>0$ so that as we increase $\zeta_x$, $v_+$ increases. Most importantly, $v_+ = 0$ when
\begin{align} \label{thecriterion}
\partial_\zeta(\zeta n_s) = 0 \quad\quad \Longleftrightarrow \quad\quad \chi_{\xi\xi}\to\infty.
\end{align}
For larger $\zeta$, $v_+<0$ and $\Gamma_+<0$. This signals a dynamical instability. At the same time we see that the instability is also thermodynamic, for the diagonal susceptibility $\chi_{\xi \xi}$ diverges and changes sign. In \cite{Gouteraux:2022kpo}, we demonstrate that this same criterion coincides with the critical current in superconductors \cite{bardeen}. Furthermore, the coincidence of dynamical and thermodynamic instabilities has also been observed in \cite{schmitt2}.

We now match this to a holographic superfluid. To do so, we take the so-called probe limit, rescaling 
\begin{align}
A_M \to A_M/Q,\quad \Psi \to \Psi/Q
\end{align}
while taking $Q\to \infty$. In this limit, the complex scalar and gauge fields do not backreact on the geometry, so that we consider their propagation on a fixed background. In most cases, the equations of motion for holographic superfluids must be solved numerically. However, in $d=4$, there exists an analytic solution \cite{Herzog:2010vz,Bhattacharya:2011eea} when the mass sits at the Breitenlohner-Freedman bound $m^2L_{AdS}^2 = -4$ \cite{Breitenlohner:1982bm,Breitenlohner:1982jf} and the condensate is very small $\langle\mathcal{O}\rangle\ll1$. Defining $u=1/r$, we fix the background metric to be 
\begin{align}
ds^2 = \frac{L_{AdS}^2}{u^2}\left(-f(u)dt^2+\frac{du^2}{f(u)} + dx^2+dy^2+dz^2\right), \quad f(u) = 1-u^4\,.
\end{align}
The asymptotic expansion reads
\begin{align}
\Psi = \psi_su^2\log(u/\delta) - \langle \mathcal{O}\rangle u^2+...,\quad\quad A_\mu = - \xi_\mu + \frac{q^2}{2}\langle J_\mu\rangle u^2+...\,.
\end{align}
Here $\delta$ is a UV-cutoff that we introduce by hand. Choosing a superfluid velocity along the $x-$direction, without loss of generality, and introducing
\begin{align}
\epsilon \equiv -\sqrt{2}\langle \mathcal{O}\rangle, \quad \xi_x = \zeta,
\end{align}
the solution is found as an expansion in small $\epsilon$ and $\zeta$ (\cite{Herzog:2010vz} considers $\zeta = 0$). For our purposes, the important quantities obtained from these solutions are the densities
\begin{align}
\mu&=\frac{1}{e}\left\{2+\frac{\zeta^2}{2}+\frac{\log(2)-1}{4}\zeta^4+O(\zeta^6)+\left[\frac{1}{48}+\left(\frac{5}{144}-\frac{3\log(2)}{32}\right)\zeta^2+O(\zeta^4)\right]\epsilon^2+O(\epsilon^4)\right\}\nonumber\\
n_s &= \frac{1}{e}\left\{\left[\frac{1}{2}-\left(\frac{1-\log(2)}{4}\right)\zeta^2+O(\zeta^4)\right]\epsilon^2 + \left[\frac{1+4\log(2)}{192}+O(\zeta^2)\right]\epsilon^4 + O(\epsilon^6)\right\}\nonumber\\
n &= \frac{1}{e}\left\{4+\zeta^2-\frac{1-\log(2)}{2}\zeta^4+\left[\frac{7}{24}+\left(\frac{7}{36}-\frac{5\log(2)}{16}\right)\zeta^2+O(\zeta^4)\right]\epsilon^2+O(\epsilon^4)\right\}
\end{align}
from which we can obtain the static susceptibilities 
\begin{align}
\chi_{nn}  =14 + O(\epsilon^2,\zeta^2)\,, 
\quad \chi_{nh} =  -12\zeta + O(\epsilon^2\zeta, \zeta^3)\,,\quad \partial_\zeta n_s|_\mu = -24\zeta + O(\epsilon^2\zeta, \zeta^3)\,.
\end{align}
or
\begin{align}
\chi_{\xi\xi} = \frac{\mu}{n_s + \zeta \partial_\zeta n_s} = \frac{4+\zeta^2+\epsilon^2/24}{\epsilon^2-48\zeta^2}+O(\epsilon^4,\zeta^4,\epsilon^2\zeta^4)\,.
\end{align}
Next, we consider perturbations of this background of the following form,\begin{align}
\label{perturbationsprobe}
A_t &= A_{t,0}(u)+a_t(u)e^{-i\omega t+ikx},\qquad\qquad A_x = A_{x,0}(u)+a_x(u)e^{-i\omega t+ikx}\,,\nonumber\\
 \Psi &= \psi(u)+[\sigma_r(u)+i\sigma_i(u)]e^{-i\omega t+ikx}, \quad \bar{\Psi} =\psi(u)+ [\sigma_r(u)-i\sigma_i(u)]e^{-i\omega t+ikx}
\end{align}
where we have chosen the wavevector to point along the superfluid velocity. Importantly, while we can use a gauge transformation to set $\Psi=\bar{\Psi}$ in the background, the fluctuations will be independent. As $u\to 0$, the fluctuations can be expanded
\begin{align}
\begin{split}
a_t &\to \delta A_t^{(0)} + \frac{\delta J_t}{2}u^2+... \,, \\
a_x &\to \delta A_x^{(0)}+\frac{\delta J_x}{2}u^2+...\,,\\
\sigma_r&\to \sigma_r^{(0)}u^2\log(u/\delta) + \sigma_r^{(1)}u^2+...\,,\\
\sigma_i&\to \sigma_i^{(0)}u^2\log(u/\delta) + \sigma_i^{(1)}u^2+...\,.
\end{split}
\end{align}
At the horizon, we impose ingoing boundary conditions, 
\begin{align}
\begin{split}
a_t &\to (1-u)^{-i\omega/4}\left[a_t^h(1-u)+O(1-u)\right]\,,\\
a_x &\to (1-u)^{-i\omega/4}\left[a_x^h+O(1-u)\right]\,,\\
\sigma_r &\to (1-u)^{-i\omega/4}\left[\sigma_r^h+O(1-u)\right]\,,\\
\sigma_i &\to (1-u)^{-i\omega/4}\left[\sigma_i^h+O(1-u)\right]\,.
\end{split}
\end{align}
The equations of motion for the fluctuations are three second-order equations for $a_x, \sigma_r,$ and $\sigma_i$ and one first-order equation for $a_t$. As such, in addition to fixing the four ingoing boundary conditions, we are also able to fix $a_x^h, \sigma_r^h,$ and $\sigma_i^h$ at the horizon, though not $a_t^h$ which can be found as a function of the other three constants \cite{Amado:2009ts}. We choose three linearly independent sets of boundary conditions
\begin{align}
\{a_x^h,\sigma_r^h,\sigma_i^h\}^1 = \{1,0,0\},\quad \{a_x^h,\sigma_r^h,\sigma_i^h\}^2 = \{0,1,0\},\quad \{a_x^h,\sigma_r^h,\sigma_i^h\}^{3} = \{0,0,1\}.
\end{align}
which lead to three linearly independent solutions $X^i = \{a_t, a_x, \sigma_r, \sigma_i\}^I$ for $I=\{1,2,3\}$. Green's functions are obtained in the standard way from linear response \cite{KadanoffandMartin}--the Green's function for two currents $J^a$ and $J^b$ is given by the response of the current $J^a$ to the source $s^b$ of the current $J^b$,
\begin{align}
\delta J^a = -G^R_{J^aJ^b}(\omega,k)\delta s^b.
\end{align}
As opposed to the main text, we will see there are no contact terms. Here, the sources are the UV boundary values of the fluctuations, $\delta A_t^{(0)}, \delta A_x^{(0)}, \Sigma^{(0)},$ and $\bar{\Sigma}^{(0)}$. Clearly, to obtain all Green's functions, we want to independently vary each source. This is not possible with only three independent solutions, however, we are able to add a fourth solution via gauge invariance \cite{Amado:2009ts},
\begin{align}
X^{4} = \{-i\omega, ik, 0, \psi\}.
\end{align}
With this solution, any solution for the fluctuating fields can be written
\begin{align}
X = c_1 X^1 + c_{2}X^{2} + c_{3}X^{3} + c_{4}X^{4}
\end{align}
and in particular, we can find the solution corresponding to a particular choice of $\delta A_t^{(0)}, \delta A_x^{(0)}, \sigma_r^{(0)},$ and $\sigma_i^{(0)}$ via this method. In the end, we find that as $u\to 0$,
\begin{align}
\delta J_t &=  4k\frac{k\delta A_t^{(0)} +\omega \delta A_x^{(0)}}{\mathcal{P}}\biggl\{-480\omega^2-4i\omega\biggl[37\epsilon^2+24k\bigl(k+6i\zeta\bigr)\biggr]\nonumber\\
&\hspace{180pt}
+\biggl[\epsilon^2+4k^2\biggr]\biggl[7\epsilon^2+12\bigl(k^2-4\zeta^2\bigr)\biggr]+...\biggr\}\nonumber\\
\delta J_x &=  - 4\omega\frac{k\delta A_t^{(0)} + \omega \delta A_x^{(0)}}{\mathcal{P}}\biggl\{-480\omega^2-4i\omega\biggl[37\epsilon^2+24k\bigl(k+6i\zeta\bigr)\biggr]\nonumber\\
&\hspace{180pt}
+\biggl[\epsilon^2+4k^2\biggr]\biggl[7\epsilon^2+12\bigl(k^2-4\zeta^2\bigr)\biggr]+...\biggr\}\nonumber\\
\sigma_r^{(1)}+i\sigma_i^{(1)} &= -24\sqrt{2}\epsilon\frac{k\delta A_t^{(0)}+\omega\delta A_x^{(0)}}{\mathcal{P}}\biggl\{k\biggl[\epsilon^2+4k^2\biggr] - 8\omega\biggl[\zeta+2ik\biggr]+...\biggr\}\nonumber\\
\sigma_r^{(1)}-i\sigma_i^{(1)} &=\frac{i\sqrt{2}\epsilon\delta A_x^{(0)}}{\mathcal{P}}\biggl\{288\omega^2\biggl[k+2i\zeta\biggr]+k\biggl[\epsilon^2+4k^2\biggr]\biggl[\epsilon^2+12\bigl(k^2-4\zeta^2\bigr)\biggr]\nonumber\\
&\hspace{120pt}-12i\omega\biggl[\epsilon^2\bigl(3k+4i\zeta\bigr)+16k\bigl(k^2+2ik\zeta-2\zeta^2\bigr)\biggr]+...\biggr\}\nonumber\\
&\qquad+\frac{ i\sqrt{2}\epsilon\delta A_t^{(0)}}{\mathcal{P}}\biggl\{-960 i\omega^2+8\omega\biggl[7\epsilon^2+24k(2k+3i\zeta)\biggr]-48k\biggl[\epsilon^2+4k^2\biggr]\zeta+...\biggr\}
\end{align}
where
\begin{align}
\mathcal{P} &=-1920i\omega^3 +16\omega^2(7\epsilon^2+12k(7k+12i\zeta))+24ik\omega(16k(k+i\zeta)(k+2i\zeta)+\epsilon^2[3k+8i\zeta])\nonumber\\
&\quad-2k^2(\epsilon^2+4k^2)(\epsilon^2+12[k^2-4\zeta^2]).
\end{align}
In the limit $k\ll \epsilon\ll1$ and $k\ll \zeta \ll1$, the zeroes of $\mathcal{P}$ give a gapped non-hydrodynamic mode and two hydrodynamic sound modes,
\begin{align}
\label{modesprobeanalytic}
\begin{split}
\omega_\pm &= v_\pm k\left(1 - i\frac{\tilde{\Gamma}_\pm}{2} k^2+...\right)\,,\\
\omega_g &= -i\frac{7}{120}\epsilon^2 -\frac{18}{35}k\zeta -ik^2\left(\frac{89}{245}+\frac{2496}{343}\frac{\zeta^2}{\epsilon^2}\right)+...
\end{split}
\end{align}
where 
\begin{align}
\label{modesprobeanalytic2}
v_\pm = \frac{6\zeta}{7}\pm\frac{\sqrt{14\epsilon^2-96\zeta^2}}{28}, \quad \tilde{\Gamma}_\pm = \pm\frac{462\epsilon^2+48\zeta(8\zeta\mp9\sqrt{14\epsilon^2-96\zeta^2})}{49\epsilon^2\sqrt{14\epsilon^2-96\zeta^2}}.
\end{align}

The two sound modes obtained here match those of \cite{Bhattacharya:2011eea}. To confirm that this indeed obeys the probe superfluid hydrodynamics, we also need some Kubo formulae to find the dissipative coefficients \cite{Herzog:2011ec}, 
\begin{align}
\sigma_0 = -\lim_{\omega\to 0}\frac{1}{\omega} \text{Im}, G_{J^xJ^x}(\omega,0),\quad
\zeta_2 = \frac{1}{\zeta}\lim_{\omega\to 0}\text{Re}\, G_{J^x\phi}(\omega,0),\quad \zeta_3 = \lim_{\omega\to 0}\omega\,\text{Im}\, G_{\phi\phi}(\omega,0).
\end{align}
From the expressions for the asymptotics above, we find that 
\begin{align}
\zeta_2 = -\frac{216}{49\epsilon^2},\quad \zeta_3 = \frac{52}{49\epsilon^2},\quad \sigma_0 = 1+\frac{1440}{49}\frac{\zeta^2}{\epsilon^2}.
\end{align}
Using these definitions, we can see that the hydro expressions for the sound modes (\ref{modesprobe}) and (\ref{modesprobe2}) match the holographic expressions (\ref{modesprobeanalytic}) and (\ref{modesprobeanalytic2}). In particular, we can see that when 
\begin{align}
\zeta_c = \epsilon/\sqrt{48}
\end{align}
the velocity $v_- = 0$ and $\tilde{\Gamma}_- = 7\epsilon/\sqrt{3}$. In other words, for $\zeta<\zeta_c$, $v_-\tilde{\Gamma}_- <0$, whereas for $\zeta>\zeta_c$, $v_-\tilde{\Gamma}_->0$ indicating a dynamical instability. Furthermore, it is easily checked that $n_s + \zeta(\partial n_s/\partial\zeta)|_{\zeta_c} = 0$. 

A convenient feature of the analytic superconductor is that we can obtain the modes exactly when $\zeta = \zeta_c$. Away from $k=0$, the unstable mode reads
\begin{align}
\label{criticalvelocitymode}
\omega_+ = 0k + 0k^2 + \frac{\sqrt{3}}{2\epsilon}k^3 - i\frac{7}{4\epsilon^2}k^4+...
\end{align}
However, from hydrodynamics, we may have expected that $\omega_- = 0$ exactly due to the diverging static susceptibility matrix. The resolution is that the dynamic susceptibility does not diverge at $\zeta_c$,
\begin{align}
\chi_{\xi\xi}^{-1}(k) \to 0 + \frac{3\epsilon^2k^2}{4(\epsilon^2+12k^2)}+O(k^4)\,.
\end{align}
Using this expression in (\ref{modesprobe}) and (\ref{modesprobe2}), one recovers (\ref{criticalvelocitymode}). In other words, the critical velocity is momentum dependent, which for $k\ll \epsilon$ reads
\begin{align}
\zeta_c(k) =\epsilon\left( \frac{1}{4\sqrt{3}}+\frac{\sqrt{3}k^2}{2\epsilon^2}+O(k^4)\right).
\end{align}
This value gives $\chi_{\xi\xi}^{-1} = \omega_- = 0$ for all $k$. Nevertheless, increasing $\zeta$ slightly, $\chi_{\xi\xi}^{-1}(k=0)$ changes sign and indeed $\zeta_c(k=0)$ indicates the global instability. Finally, the probe superfluid also exhibits the two-stream instability for $\zeta > \sqrt{\frac{7}{48}}\epsilon > \zeta_c$, where it is observed that the velocities of the sound modes become complex and one lies in the upper half plane. This is well beyond the critical velocity.

\bibliographystyle{JHEP}
\bibliography{local}

\end{document}